\documentclass{aa} 
\usepackage{comment}
\usepackage{mathtools}
\usepackage{hyperref}
\hypersetup{
    colorlinks=true,
    linkcolor=blue,
    citecolor=blue,
    urlcolor=green,
    }
\usepackage{natbib}
\usepackage{graphicx}
\usepackage{txfonts}
\usepackage{ulem}
\usepackage{threeparttable}
\usepackage{mathrsfs}
\usepackage{amsmath}
\begin{document}

\title{From 100 MHz to 10 GHz: Unveiling the spectral evolution of the X-shaped radio galaxy in Abell 3670}

   \author{L. Bruno
          \inst{1,2},
          M. Brienza
          \inst{2},
          A. Zanichelli
          \inst{2},
          M. Gitti
          \inst{1,2},
          F. Ubertosi
           \inst{1,3},
          K. Rajpurohit
          \inst{4,2},
          T. Venturi
          \inst{2,3,5},
          D. Dallacasa
          \inst{1,2}
          }

   \institute{
   Dipartimento di Fisica e Astronomia (DIFA), Universit\`a di Bologna, via Gobetti 93/2, 40129 Bologna, Italy
         \and
    Istituto Nazionale di Astrofisica (INAF) - Istituto di Radioastronomia (IRA), via Gobetti 101, 40129 Bologna, Italy      
        \and
    Istituto Nazionale di Astrofisica (INAF) - Osservatorio di Astrofisica e Scienza dello Spazio (OAS) di Bologna, Via P. Gobetti 93/3, 40129, Bologna, Italy
         \and
    Center for Astrophysics | Harvard \& Smithsonian, 60 Garden Street, Cambridge, MA 02138, USA
    \and
    Center for Radio Astronomy Techniques and Technologies, Rhodes University, Grahamstown 6140, South Africa
    \\
    \email{luca.bruno4@unibo.it}
 }

 
  \abstract
   {X-shaped radio galaxies (XRGs) are characterised by two pairs of misaligned lobes, namely the active lobes hosting radio jets and the wings. None of the formation mechanisms proposed so far can exhaustively reproduce the diverse features observed among XRGs. The emerging evidence is the existence of sub-populations of XRGs forming via different processes. }
   {The brightest cluster galaxy in Abell 3670 (A3670) is a dumbbell system hosting the XRG MRC 2011-298. The morphological and spectral properties of this interesting XRG have been first characterised through Karl G. Jansky Very Large Array (JVLA) data at 1-10 GHz. In the present work, we followed-up MRC 2011-298 with the upgraded Giant Metrewave Radio Telescope (uGMRT) at 120-800 MHz to further constrain its properties and origin.  }
   {We carried out a detailed spectral analysis sampling different spatial scales. Integrated radio spectra, spectral index maps, radio colour-colour diagrams, and radiative age maps of both the active lobes and prominent wings have been employed to test the origin of the source. }
   {We confirm a progressive spectral steepening from the lobes to the wings. The maximum radiative age of the source is $\sim 80$ Myr, with the wings being older than the lobes by $\gtrsim 30$ Myr in their outermost regions. }
   {The observed properties are in line with an abrupt reorientation of the jets by $\sim 90$ deg from the direction of the wings to their present position. This formation mechanism is further supported by the comparison with numerical simulations in the literature, which additionally highlight the role of hydrodynamic processes in the evolution of large wings such as those of MRC 2011-298. Potentially, the coalescence of supermassive black holes could have triggered the spin-flip of the jets. Moreover, we show that the S-shape of the radio jets is likely driven by precession with a period $P\sim 10$ Myr.}

   
\titlerunning{The X-shaped radio galaxy in A3670}
\authorrunning{Bruno et al.}
   \maketitle
%

\section{Introduction}

Extended radio galaxies are characterised by a pair of radio lobes hosting relativistic jets. Depending on the absence or presence of hotspots at the tip of the jets, radio galaxies are primarily classified into Fanaroff-Riley classes I (FRI) and II (FRII), respectively, with the latter having higher radio powers \citep{fanaroff&riley74}. Within these two broad categories, different morphologies have been observed. For instance, the radio jets can be bent by ram pressure \citep{gunn&gott72} when the host galaxy moves at high velocity throughout the environmental medium, producing mirror-symmetric (U-shaped, V-shaped) morphologies \citep[e.g.][]{miley72,owen&rudnick76,pinkney93,pfrommer&jones11,ternidegregory17,missaglia19}. On the other hand, the nature of X-shaped radio galaxies (XRGs) is still debated. Similarly to classical radio galaxies, XRGs exhibit bright primary (`active') lobes hosting bipolar jets, but they additionally show a pair of fainter secondary lobes, called `wings', which are misaligned by a large angle ($\sim 70-90$ deg) with respect to the primary lobes and, as far as we know from current data, always lack jets \citep[e.g.][]{leahy&williams84}.

Observational studies at different wavelengths have outlined some typical properties of XRGs, their host, and environment, but they also revealed many targets with less common features. The majority ($\sim 70-90\%$) of XRGs exhibit jets of FRII-type, while a significantly smaller fraction of XRGs belong to the FRI class \citep[e.g.][]{yang19,bera20}. The radio power of XRGs is often found to be intermediate between that dividing FRI and FRII classes \citep[e.g.][]{dennett-thorpe02,cheung09,landt10}. The size of the wings is typically comparable to or smaller than the size of the active lobes \citep[e.g.][]{saripalli&subrahmanyan09,saripalli&roberts18,joshi19}, but XRGs with more extended wings have also been reported \citep[e.g.][]{wang03,bruno19,hardcastle19,ignesti20}. Typically, XRGs exhibit a steepening of the spectral index from the lobes to the wings \citep[e.g.][]{bruno19,cotton20,patra23}, thus indicating that the wings are radiatively older than the lobes; nevertheless, a few XRGs with opposite spectral index trends are known \citep[e.g.][]{lal&rao04,lal19}. Interestingly, a connection with the properties of the host and environment was shown for a number of XRGs. The hosts of XRGs are early-type galaxies typically characterised by a high ellipticity ($\varepsilon = 1-b/a  \gtrsim 0.2$, where $a$ and $b$ are the projected major and minor axis, respectively) of both the optical stellar component \citep{capetti02} and the hot (interstellar, intragroup, intracluster medium) X-ray atmosphere \citep{hodges-kluck10}. The wings are predominantly aligned with the minor axis of the host and its X-ray atmosphere, while the active lobes tend to be aligned with the major axis \citep[e.g.][]{capetti02,saripalli&subrahmanyan09,hodges-kluck10,gillone16,joshi19}. Likewise standard FRIIs, XRGs mostly inhabit isolated or poorly populated environments such as galaxy groups \citep{joshi19}. Additional studies focused on possible signatures of mergers associated with XRGs, such as young stellar populations, presence of dust, and statistically higher supermassive black hole (SMBH) masses \citep[e.g.][]{landt10,mezcua11,mezcua12,gillone16,joshi19}, but conflicting results have been obtained among these works, thus indicating that merger is not a general event associated to all XRGs.

Several theoretical models have been proposed to explain the observed diverse properties of XRGs \citep[see][for reviews]{gopal-krishna12,giri24}. At present, the two main competing models are the reorientation and hydrodynamical scenarios. In reorientation models, the wings trace old plasma emitted in the past, while the main lobes consist of young plasma emitted after the reorientation of the jets along a new direction \citep[e.g.][]{wirth82,rees78,dennett-thorpe02,merritt&ekers02,liu04,gergely&biermann09}. Either a merger of SMBHs or inhomogeneous mass accretion can induce an abrupt reorientation of the jets (`spin-flip') by large angles ($\lesssim 90$ deg), while jet precession is responsible for a slow reorientation. In hydrodynamical models, the backflow plasma from the hotspots falls back towards the core, but can be diverted along the path of the lowest pressure in highly asymmetric environments. If the FRII-type jets propagate along the major axis of the host/atmosphere, the backflow can produce wings along the minor axis due to the pressure gradient \citep[e.g.][]{leahy&williams84,kraft05,capetti02,hodgeskluck11}. Both models can reproduce some observational features of XRGs, but fail to replicate other properties. Notably, reorientation models can naturally explain XRGs with wings longer than the active lobes, whereas hydrodynamical scenarios are perfectly in line with the observed multi-wavelength alignments. Further insight into the origin of XRGs has come from numerical simulations, which can test strengths and limits of each model \citep[e.g.][]{hodgeskluck11,rossi17,giri22b,giri22,giri23,nolting23,dominguez-fernandez24}. Owing to the variety of observational features and results of such simulations, the emerging evidence is that a universal scenario reproducing all XRGs is highly disfavoured, and multi-frequency data are necessary to constrain the model that better fits the properties of a specific target. 

An interesting example among XRGs is the radio source MRC 2011-298 in the galaxy cluster Abell 3670 (A3670). This exhibits S-shaped jets of FRI-type, wings that are prominently larger than the active lobes, and a spectral steepening towards the wings, as derived from Karl Jansky Very Large Array (JVLA) observations at 1-10 GHz \citep{bruno19}. In the present work we followed-up A3670 with deep upgraded Giant Metrewave Radio Telescope (uGMRT) observations at 120-850 MHz to further investigate its spectral properties over a wide frequency range, and constrain its origin. 

The paper is organised as follows: in Sect. \ref{sect: The galaxy cluster Abell 3670} we describe A3670 and its BCG hosting the XRG; in Sect. \ref{sect: Observations and data reduction} we present the radio data and summarise their processing; in Sect. \ref{sect: Results} we report on our results, showing radio images, radio spectra, and spectral index and age maps; in Sect. \ref{sect: Discussion} we discuss the origin of the target in the light of our results; in Sect. \ref{sect: Summary and conclusions} we summarise our work. Throughout this paper we adopt a standard $\Lambda$CDM cosmology with $H_0=70\;\mathrm{km\; s^{-1}\; Mpc^{-1}}$, $\Omega_{\rm M}=0.3$ and, $\Omega_{\rm \Lambda}=0.7$. We define the spectral index $\alpha$ through $S_{\rm \nu}\propto \nu^{-\alpha}$, where $S_{\rm \nu}$ is the flux density at the frequency $\nu$. At the brightest cluster galaxy (BCG) redshift of $z=0.1366$, the luminosity distance is $D_{\rm L}=644$ Mpc and $1''=2.42$ kpc.

\section{The galaxy cluster Abell 3670}   
\label{sect: The galaxy cluster Abell 3670}

\begin{figure*}
	\centering
\includegraphics[width=0.8\textwidth]{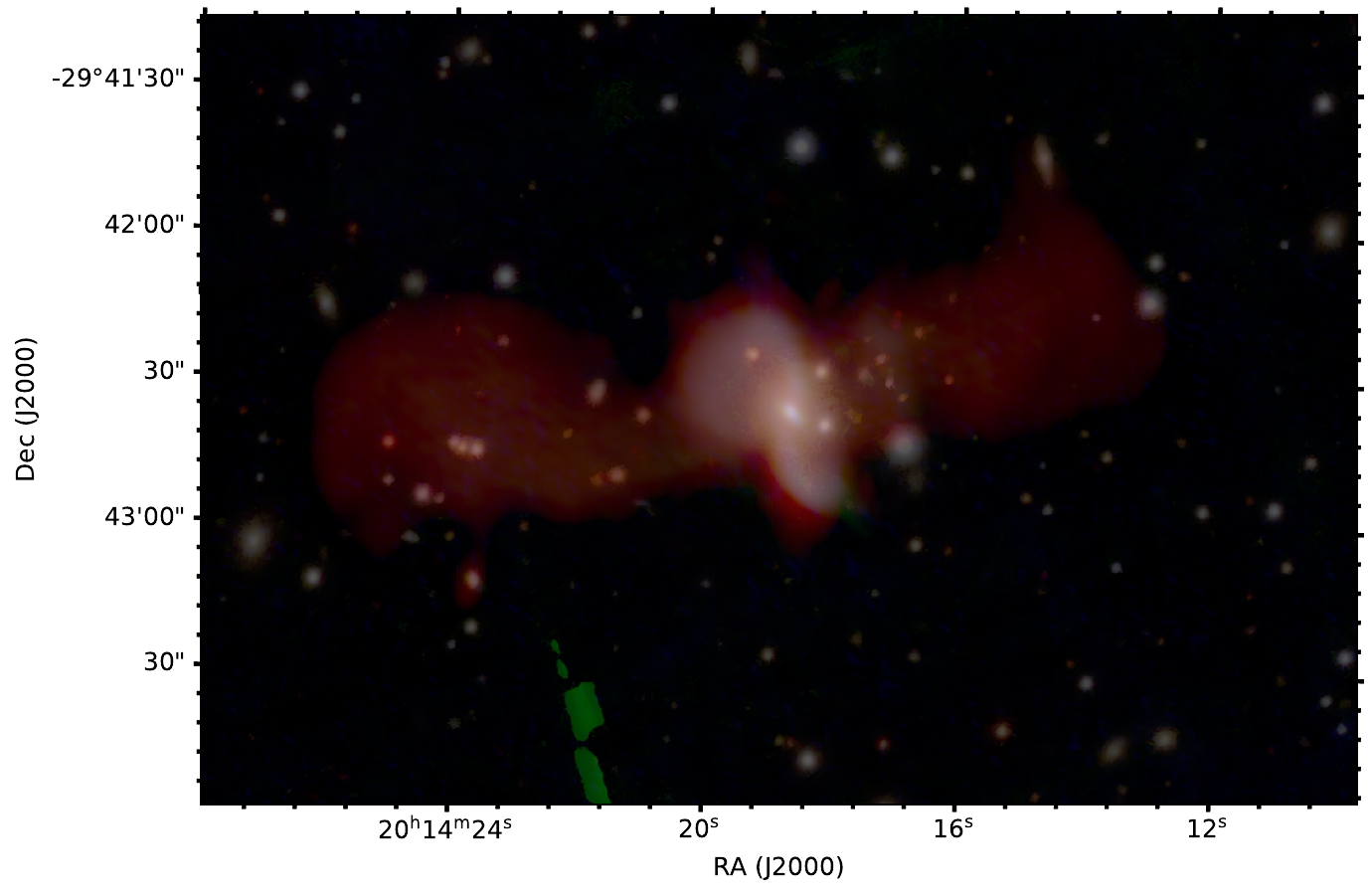}
\includegraphics[width=0.26\textwidth]{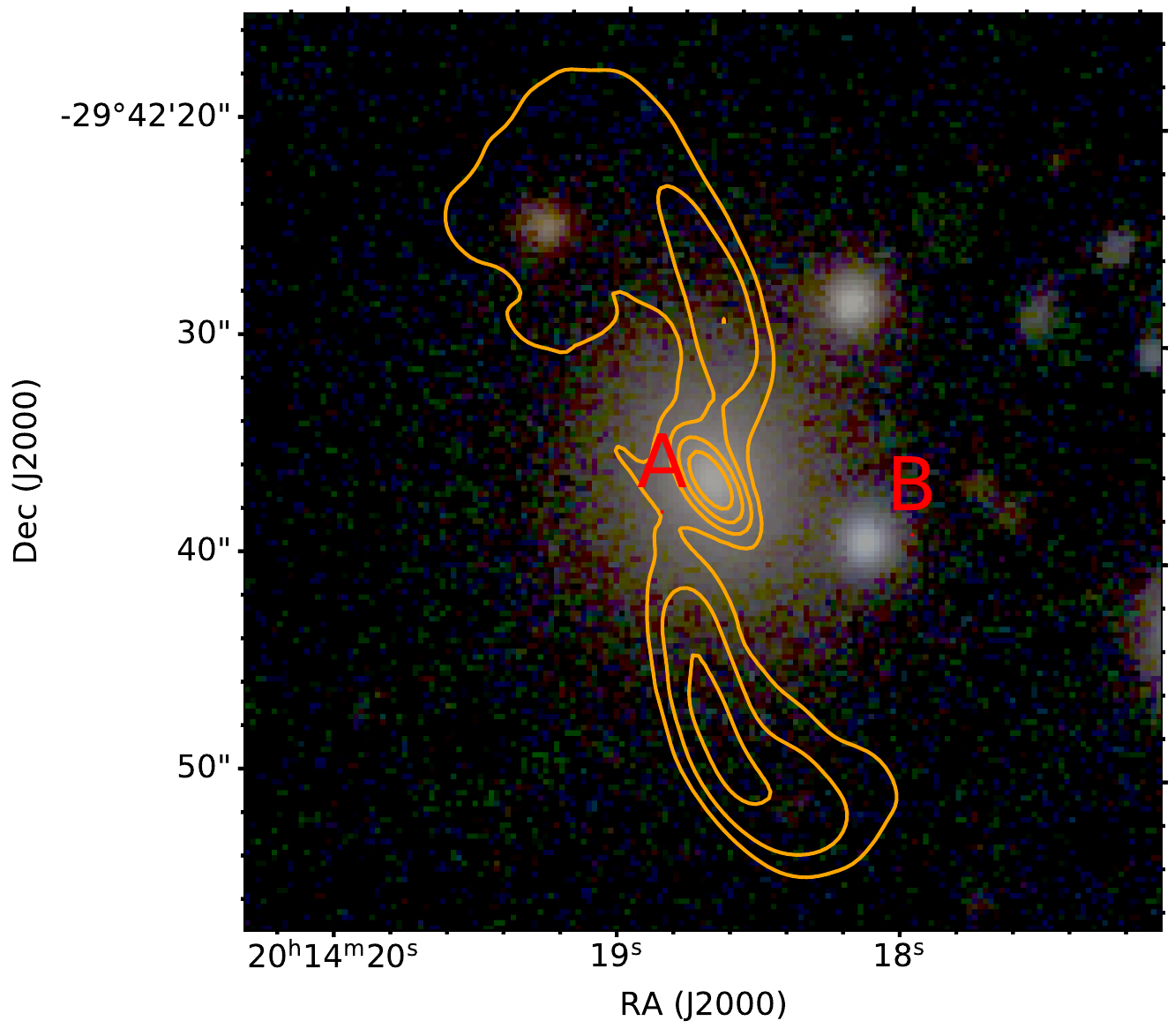}
\includegraphics[width=0.36\textwidth]{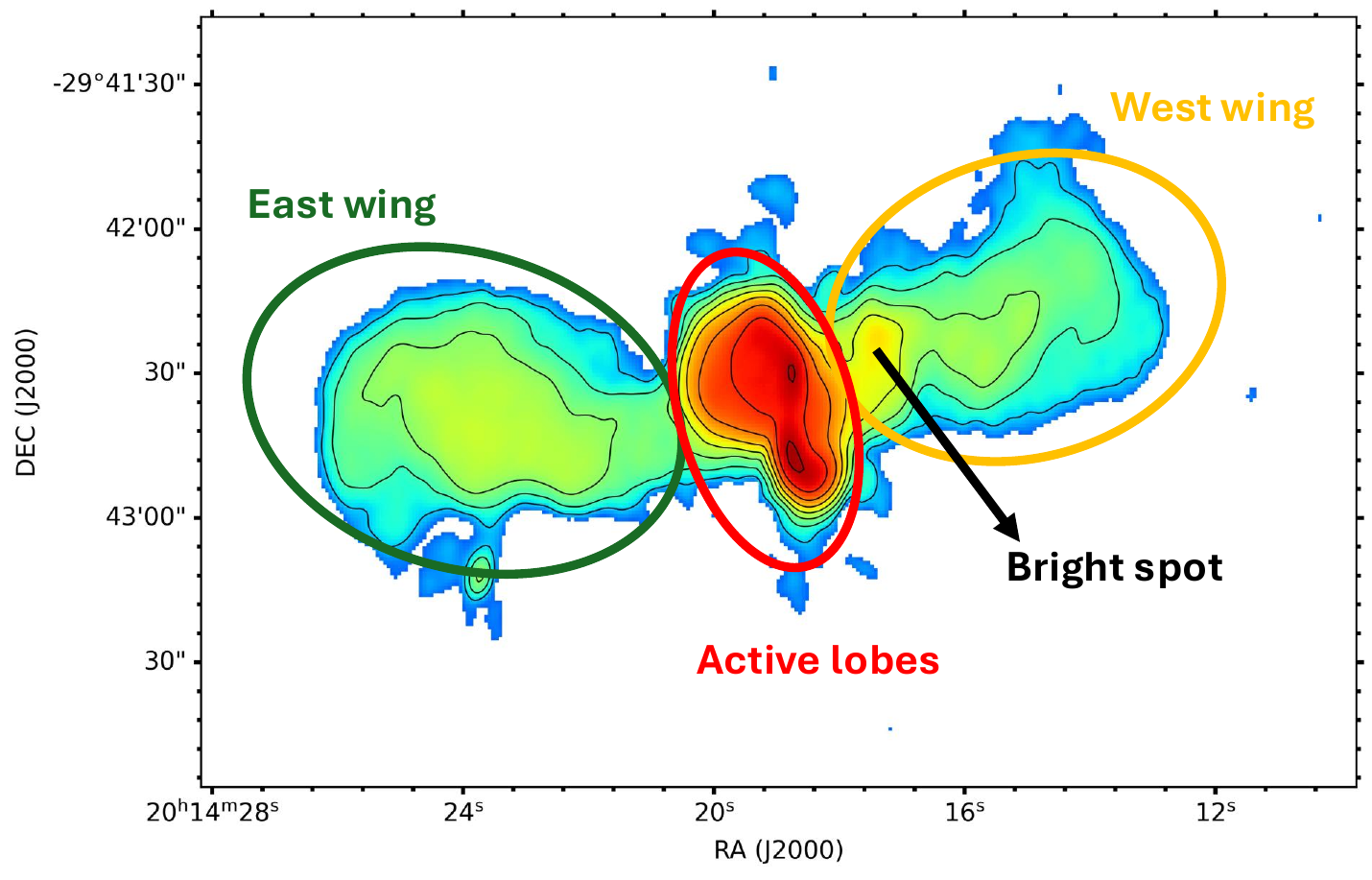}
\includegraphics[width=0.36\textwidth]{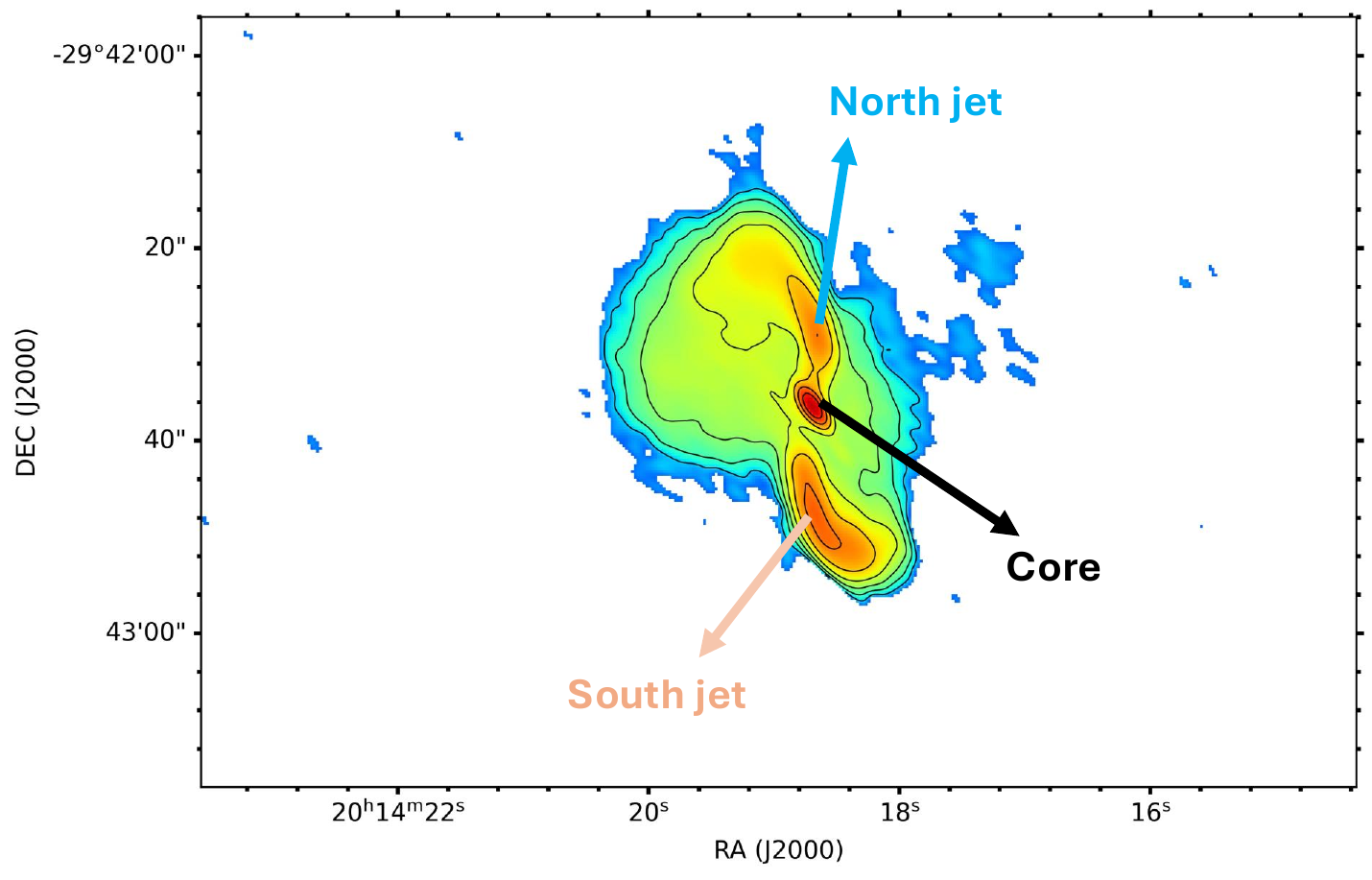}
\caption{\textit{Top}: Pan-STARSS optical (combined r, g, i filters) image of A3670 with overlaid (reddish colours) composite band-4, C-band, X-band radio images of MRC 2011-298 (from Fig. \ref{fig: mappefullres}). \textit{Bottom left}: zoom-in of the Pan-STARSS image with overlaid 9 GHz contours (starting from $64\sigma$) indicating the two optical nuclei (`A' and `B'). \textit{Bottom middle and right}: 700 MHz and 9 GHz images labelling the regions discussed in the text.  }
	\label{fig: radio+panstarss}
\end{figure*}

A3670 (${\rm RA}_{\rm J2000} =$ $20^{\rm h}14^{\rm m}18^{\rm s}$, ${\rm Dec}_{\rm J2000} =$  $-29^{\rm o}44'51''$) is a galaxy cluster at redshift $z=0.142$ \citep{coziol09} in the southern hemisphere. According to the classification of \cite{abell58,abell89}, A3670 is a richness class 2 cluster, meaning that it is a poor system hosting 80-130 member galaxies. On large scales, \cite{chow-martinez14} reported on a candidate supercluster (centred at ${\rm RA}_{\rm J2000}=$ $20^{\rm h}20^{\rm m}46^{\rm s}$, ${\rm Dec}_{\rm J2000}=$  $-30^{\rm o}37'48''$) formed by A3670 and A3678 (${\rm RA}_{\rm J2000}=$ $20^{\rm h}27^{\rm m}13^{\rm s}$, ${\rm Dec}_{\rm J2000}=$  $-31^{\rm o}31'05''$) with a separation of 38 Mpc.

Key information on the thermal properties, dynamical state, and mass of A3670 is completely missing, as the cluster is neither detected by sky surveys nor pointed observations are available at X-ray and sub-mm wavelengths. Observations recorded during slewing manoeuvres of XMM-Newton have been used to produce a shallow survey of the sky in the X-rays \citep{saxton08SLEW}. While A3670 is not detected in the slew survey, this provides a useful upper limit ($2\sigma$ confidence) on its soft-band (0.2-2 keV) flux of $F_{\rm \left[0.2-2\right]}=8.35\times 10^{-13} \; {\rm erg \; cm^{-2} \; s^{-1}}$. This value can be used to infer upper limits on the mass\footnote{We consider $M_{500}$ as the mass within a radius $R_{500}$ enclosing $500\rho_{\rm c}(z)$, where $\rho_{\rm c}(z)$ is the critical density of the Universe at a given redshift.}, size, and temperature of the target by considering self-similar scaling relations with the luminosity \citep[e.g.][]{lovisari20,lovisari21}. We obtained an upper limit to the X-ray luminosity of $L_{\rm \left[0.2-2\right]}\lesssim 2\times 10^{43} \; {\rm erg \; s^{-1}}$, which yields $M_{500}\sim1\times 10^{14} \; M_{\rm \odot}$, $R_{500}\sim 675$ kpc, and $kT\sim 2$ keV. These values are typical of extremely low mass clusters, towards the regime of groups, and explain the elusive nature of A3670 in previous surveys.

The brightest cluster galaxy (${\rm RA}_{\rm J2000}=$ $20^{\rm h}14^{\rm m}18.672^{\rm s}$, ${\rm Dec}_{\rm J2000}=$  $-29^{\rm o}42'36.360''$, $z=0.1366$) in A3670 is a giant ellipsoid of stellar mass $M_*=3.1\times 10^{11} \; M_{\odot}$ that hosts MRC 2011-298, the XRG first reported by \cite{gregorini94}, and afterwards confirmed by \cite{bruno19}. In Fig. \ref{fig: radio+panstarss} we show optical images from the Panoramic Survey Telescope \& Rapid Response System (Pan-STARSS; \citealt{flewelling20PANSTARRS}) with overlaid radio images (see details in Sect. \ref{sect: Radio morphology}), and panels labelling the regions that we will discuss. Owing to the presence of two bright optical cores (`A' and `B') of similar magnitude embedded by a common stellar halo, the BCG is classified as a dumbbell galaxy \citep{gregorini92,andreon92}. Such configuration may be the result of a galaxy merger event \citep[e.g.][]{valentijn&casertano88,gregorini94}. The two optical nuclei are separated by less than 20 kpc in projection, and only the primary core A is radio emitting \citep{bruno19}. As typical of galaxies hosting XRGs, the BCG in A3670 shows high ellipticity $\varepsilon=0.28$ \citep{makarov14}, and the major and minor axes are aligned with the radio jets and wings, respectively \citep{bruno19}.




\section{Observations and data reduction}
 \label{sect: Observations and data reduction}

\begin{table*}
\centering
\caption[]{Details of the radio data analysed in this work.}
\label{datiRADIO}
\begin{tabular}{cccccccc}
\hline
\noalign{\smallskip}
Instrument & Band name & Frequency & Observation date & On-source time & Project code & PI \\
 & & (MHz) & & (h) & & \\
\noalign{\smallskip}
\hline
\noalign{\smallskip}
uGMRT & band-2 & 120-220 & 20-Mar.-2022 & 8  & $41\_020$ & L. Bruno \\
uGMRT & band-3 & 250-450 & 19-Mar.-2022 & 8  & $41\_020$ & L. Bruno\\
uGMRT & band-4 & 550-850 & 1,3-Jun.-2021 & 9  & $40\_050$ & L. Bruno\\
${\rm JVLA_{[CnB]}}$ & L-band & 1000-2000 & 10-Jan.-2015 & 0.35  & 14B-027 & M. Gitti \\
ASKAP & L-band & 1224-1512 & 25,26,27-Dec.-2020 & 0.75 & RACS-mid & - \\
${\rm JVLA_{[B]}}$ & S-band & 2000-4000 & 1-Jul.-2017; 15-Feb.-2022 & - & VLASS & - \\
${\rm JVLA_{[CnB]}}$ & C-band & 4500-6500 & 11-Jan.-2015 & 0.7  & 14B-027 & M. Gitti \\
${\rm JVLA_{[DnC]}}$ & C-band & 4000-8000 & 20-Sep.-2014 & 0.35  & 14B-027 & M. Gitti \\
${\rm JVLA_{[CnB]}}$ & X-band & 8000-10000 & 9-Jan.-2015 & 0.6  & 14B-027 & M. Gitti \\
\noalign{\smallskip}
\hline
\end{tabular}
\end{table*}

In this Section we present new uGMRT observations of MRC 2011-298 and summarise the data reduction procedures that we used. Furthermore, we summarise the extraction algorithm of Karl G. Jansky Very Large Array Sky Survey \citep[VLASS;][]{lacyVLASS20} pointings, and the reprocessing of JVLA observations first presented in \cite{bruno19}. Finally, we describe public images of the Australian SKA Pathfinder (ASKAP; \citealt{hotan21}) surveys that we will use. Details on all radio data are reported in Table \ref{datiRADIO}.  

\subsection{uGMRT data}
\label{sect: uGMRT data}

A3670 was observed with the uGMRT wide-band receiver (GMRT Wideband Backend, GWB) in band-2 (120-220 MHz), band-3 (250-450 MHz), and band-4 (550-850 MHz) for 8, 8, and 9 hours, respectively. Data were recorded in channels of width 0.34, 1.1, and 1.6 MHz each for band-2, band-3, and band-4, respectively. Furthermore, the outdated narrow-band receiver (GMRT Software Backend, GSB) was active during the observations. Each observation includes pointings ($\sim 10$ min) on 3C380 and 3C48, which were used as flux density scale calibrators.

We processed the data by means of the Source Peeling and Atmospheric Modeling ({\tt SPAM}\footnote{\url{http://www.intema.nl/doku.php?id=huibintemaspampipeline}}; \citealt{intema09}) pipeline. {\tt SPAM} first calculates flux density and bandpass gain solutions for the calibrator, which are then transferred to the target. Ionospheric effects are corrected by performing rounds of direction-dependent calibration by means of bright sources in the field of view. We first processed the GSB data; the final GSB images were exploited to build a catalogue of the sources in the field of view through the Python Blob Detector and Source Finder ({\tt PyBDSF}\footnote{\url{https://pybdsf.readthedocs.io/en/latest/}}; \citealt{mohan&rafferty15}). This catalogue was used to model the sky for the subsequent direction-dependent calibration of the GWB data. For a proper calibration of the GWB data with {\tt SPAM}, the total bandwidth needs to be split into narrower sub-bands. Therefore, band-2, band-3, and band-4 datasets were split into 6 sub-bands of 16, 33, and 50 MHz, respectively, and each of them was processed independently. The calibrated sub-bands are finally simultaneously imaged with proper multi-frequency algorithms (see below).

The automatic data processing with {\tt SPAM} was not able to completely remove artefacts around the target, which are likely due to its high dynamic range. To mitigate these artefacts in band-3 and band-4, we performed additional cycles of phase plus amplitude self-calibration with the National Radio Astronomy Observatory (NRAO) Common Astronomy Software Applications ({\tt CASA}; \citealt{mcmullincasapaper07}) v. 6.4, by building model images through wide-field, multi-frequency, and multi-scale synthesis with {\tt WSClean} \citep{offringa14,offringa17} v. 2.10. This calibration strategy was successful for band-4 data. However, we rejected self-calibration of band-3 data because of excessively high flagging levels of short baselines, which are crucial for the purposes of our work. We stress that amplitude gain corrections computed from self-calibration are expected to vary the flux density of the target by a few percent at most (less than typical calibration uncertainties, see Sect. \ref{sect: Flux density measurements}), therefore, spectral measurements are not biased by the combination of self-calibrated and non-self-calibrated datasets. 

Being $\theta$ the restoring beam of the radio images, we report final noise levels near MRC 2011-298 (at a distance of $\sim 3'$) and far from it (at a distance of $\sim 6'$) of $\sigma_{\rm near}^{\rm b-2} \sim 0.8 \; {\rm  mJy \; beam^{-1}}$ and $\sigma_{\rm far}^{\rm b-2} \sim 0.7 \; {\rm  mJy \; beam^{-1}}$ at $\theta=24''\times 11''$, $\sigma_{\rm near}^{\rm b-3} \sim 0.25 \; {\rm mJy \; beam^{-1}}$ and $\sigma_{\rm far}^{\rm b-3} \sim 0.12 \; {\rm mJy \; beam^{-1}}$ at $\theta=12''\times 9''$, and $\sigma_{\rm near}^{\rm b-4} \sim 25 \; {\rm \mu Jy \; beam^{-1}}$ and $\sigma_{\rm far}^{\rm b-4} \sim 12 \; {\rm \mu Jy \; beam^{-1}}$ at $\theta=6''\times 3''$, in band-2 , band-3, and band-4, respectively (see also Table \ref{tab: image full res}). Throughout the next Sections, we will consider $\sigma_{\rm near}$ as the reference noise.

\subsection{JVLA data}   
\label{sect: JVLA data}

\cite{bruno19} first presented JVLA observations of MRC 2011-298. We reprocessed these data with more sophisticated calibration and imaging strategies to obtain higher quality images. MRC 2011-298 was observed at 1-2 GHz (L-band, CnB array configuration), 4-8 GHz (C-band, both CnB and DnC array configurations), and 8-10 GHz (X-band, CnB array configuration) for $\sim30$ min on-source each. In all the observations, 3C48 and J2003-3251 were used as flux density scale and phase calibrators, respectively. The data in CnB and DnC configuration were recorded in 16 and 34 spectral windows, respectively, each split into 64 channels. 
    
We carried out a standard data reduction with {\tt CASA}, which includes iterations of RFI excision and delay, bandpass, amplitude, and phase calibration. Differently from \cite{bruno19}, here we refined the self-calibration by making use of multi-frequency and multi-scale models produced with {\tt WSClean}. For each observation, one round of phase only and one round of phase plus amplitude self-calibration were sufficient to improve the quality of our data. Finally, we combined the visibilities of the CnB and DnC configuration data of the C-band observations into a single dataset.

We reached noise levels of $\sigma^{\rm L} \sim 55 \; {\rm \mu Jy \; beam^{-1}}$, $\sigma^{\rm C} \sim 8 \; {\rm \mu Jy \; beam^{-1}}$, and $\sigma^{\rm X} \sim 6.5 \; {\rm \mu Jy \; beam^{-1}}$ for L-band ($\theta=11''\times 8''$), C-band ($\theta=4''\times 2''$), and X-band ($\theta=3''\times 1''$), respectively (Table \ref{tab: image full res}). We therefore reduced the noise by a factor $\sim 1.2$ and $\sim 1.5$ for L-band and C-band with respect to the images presented in \cite{bruno19} at similar resolutions.

\subsection{VLASS data}
\label{sect: VLASS data}

The VLA Sky Survey \citep[VLASS;][]{lacyVLASS20} is observing the sky above declination $-40^{\rm o}$ at 2-4 GHz (S-band) in B configuration. A number of snapshots of $\sim 10$ s are mosaicked to reach a typical noise level of $  120 \; {\rm \mu Jy \; beam^{-1}}$ at the nominal resolution of $ 2.5''$.

We considered two epochs of VLASS observations in the direction of A3670. The public VLASS images are prominently affected by artefacts around the target, which increase the local noise up to $\sigma \sim 180 \; {\rm \mu Jy \; beam^{-1}}$. To improve the quality of these images we adopted the extraction procedure\footnote{\url{https://github.com/erikvcarlson/VLASS_Scripts}} described in \cite{carlson22}. We first extracted the calibrated visibilities of the two epochs covering the field of interest, thus reducing the size of the datasets from $\sim 500$ GB to a few GB. The two extracted \textit{uv}-datasets were then combined into a single dataset, and imaged with {\tt tclean} in {\tt CASA} with proper mosaicking, multi-frequency, multi-scale, and weighting options. This procedure allowed us to remove the artefacts and reduce the local noise by a factor of 1.6, thus reaching $\sigma \sim 110 \; {\rm \mu Jy \; beam^{-1}}$ (Table \ref{tab: image full res}).

\subsection{RACS data}
\label{sect: RACS data}

The Australian SKA Pathfinder (ASKAP; \citealt{hotan21}) is currently surveying the sky below declination $+40^{\rm o}$ at 700-1800 MHz. In this context, the Rapid ASKAP Continuum Survey (RACS;  \citealt{mcconnell20RACS-DR1,duchesne23RACS-DR2}) is providing shallow (15 min/tile) observations of the sky in different sub-bands that will be used as starting calibration models for deeper surveys.   

The field of A3670 was covered by both RACS-low (744-1032 MHz) and RACS-mid (1224-1512 MHz). For the purposes of our work, we only retrieved RACS-mid images from the public archive\footnote{\url{https://data.csiro.au/domain/casda}}. Owing to the dense \textit{uv}-coverage at short spacings of RACS observations, we will use these RACS-mid images to confirm the flux density of extended components measured from our L-band JVLA data, which may suffer from losses due to a sparser inner \textit{uv}-coverage. The noise level at 1367 MHz and $10''$-resolution in the direction of the target is particularly low, being $\sigma \sim 150 \; {\rm \mu Jy \; beam^{-1}}$ (that is a factor of $\sim1.3$ better than the median noise of the survey). While we stacked three available images, we did not obtain appreciable increases in terms of signal-to-noise ratio with respect to each single image.

\subsection{Flux density measurements}
\label{sect: Flux density measurements}

Throughout this work, uncertainties on the reported flux densities $S$ are computed as: 
\begin{equation}
\Delta S= \sqrt{ \left( \sigma^2 \cdot N_{\rm beam} \right) + \left(  \xi_{\rm cal} \cdot S \right) ^2} \; \; \; ,
\label{eq: erroronflux}
\end{equation}
where $\sigma$ is the noise of the image close to the target, $N_{\rm beam}$ is the number of independent beams within the considered region, and $\xi_{\rm cal}$ is the calibration error. For JVLA, we assumed $\xi_{\rm cal}=5\%$ in L-band and $\xi_{\rm cal}=3\%$ in S-band, C-band, and X-band \citep{perley&butler13}. For uGMRT data, we assumed $\xi_{\rm cal}= 7\%$ in band-2, $\xi_{\rm cal}= 6\%$ in band-3, $\xi_{\rm cal}= 5\%$ in band-4 \citep{chandra04}.

For the purposes of our work, a sanity check on the flux density scales and possible losses of our data is recommended. By measuring the radio spectra of $\sim 10$ compact sources spread across the field of view, we can exclude systematic offsets on the flux density scales. An additional issue that needs to be carefully discussed is the sampling of the \textit{uv}-coverage. Indeed, the significantly shallower observing time of JVLA datasets with respect to uGMRT datasets implies a much sparser \textit{uv}-coverage at GHz frequencies (see Fig. \ref{fig: uvplane} in Appendix \ref{sect: uvplane}). Therefore, flux density losses due to missing spacings in our JVLA observations have to be estimated to avoid misinterpretation of spectral measurements. To this aim, we used the Mock UV-data Injector Tool ({\tt MUVIT}\footnote{\url{https://github.com/lucabruno2501/MUVIT}}; \citealt{bruno23}) to simulate mock visibilities into our JVLA datasets. Following the procedure described in \cite{bruno23}, we first produce a model image of a mock source; the mock visibilities obtained by predicting the model image are added to the observed \textit{uv}-data, and finally, the real plus mock visibilities are imaged. Losses can be estimated by comparing the injected flux density ($S_{\rm inj}$) with the actual measured flux density of the mock source. As a simplistic assumption, we modelled the surface brightness profile of the mock source with either a spherically-symmetric or elliptically-symmetric Gaussian function in the form 
\begin{equation}
I(r)=I_{0}e^{-\frac{r^2}{2\hat{\sigma}^2}}  \; \; \; ,
\end{equation}
where $r=\sqrt{x^2+y^2}$, $I_{0}$, and $\hat{\sigma}$ are the radial distance, corresponding peak, and standard deviation, respectively. We assumed that the mock source has a radius of  $R=3\hat{\sigma}$ and a flux density given by $S_{\rm inj}=2\pi I_{0}  \hat{\sigma}^2$. In Sect. \ref{sect: Integrated spectra} we will constrain the recovery fraction of our JVLA observations for different values of $R$, $S_{\rm inj}$, and \textit{uv}-range.

\section{Results}
\label{sect: Results}

\subsection{Radio morphology}
\label{sect: Radio morphology}

\begin{figure*}
	\centering

\includegraphics[width=0.495\textwidth]{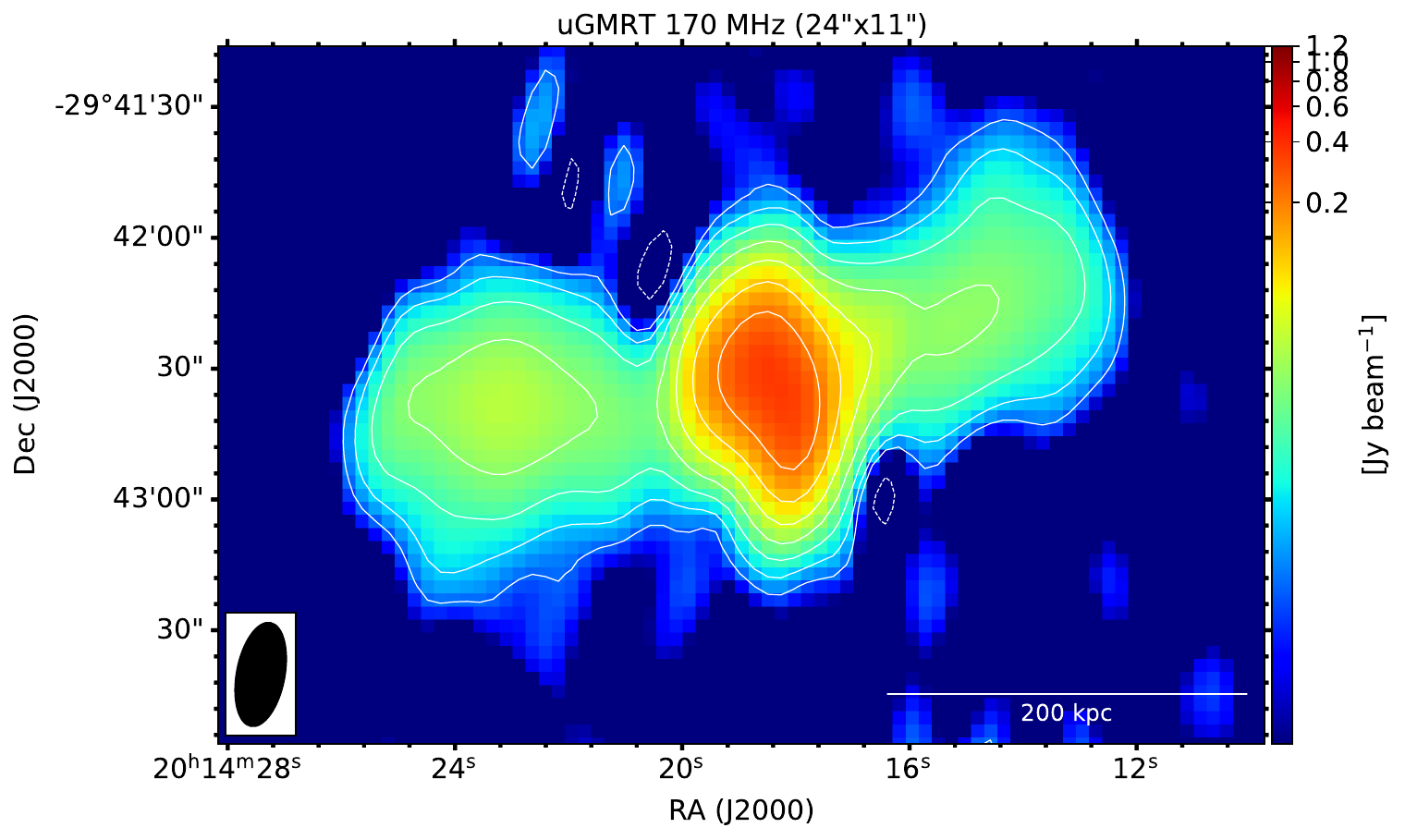}
\includegraphics[width=0.495\textwidth]{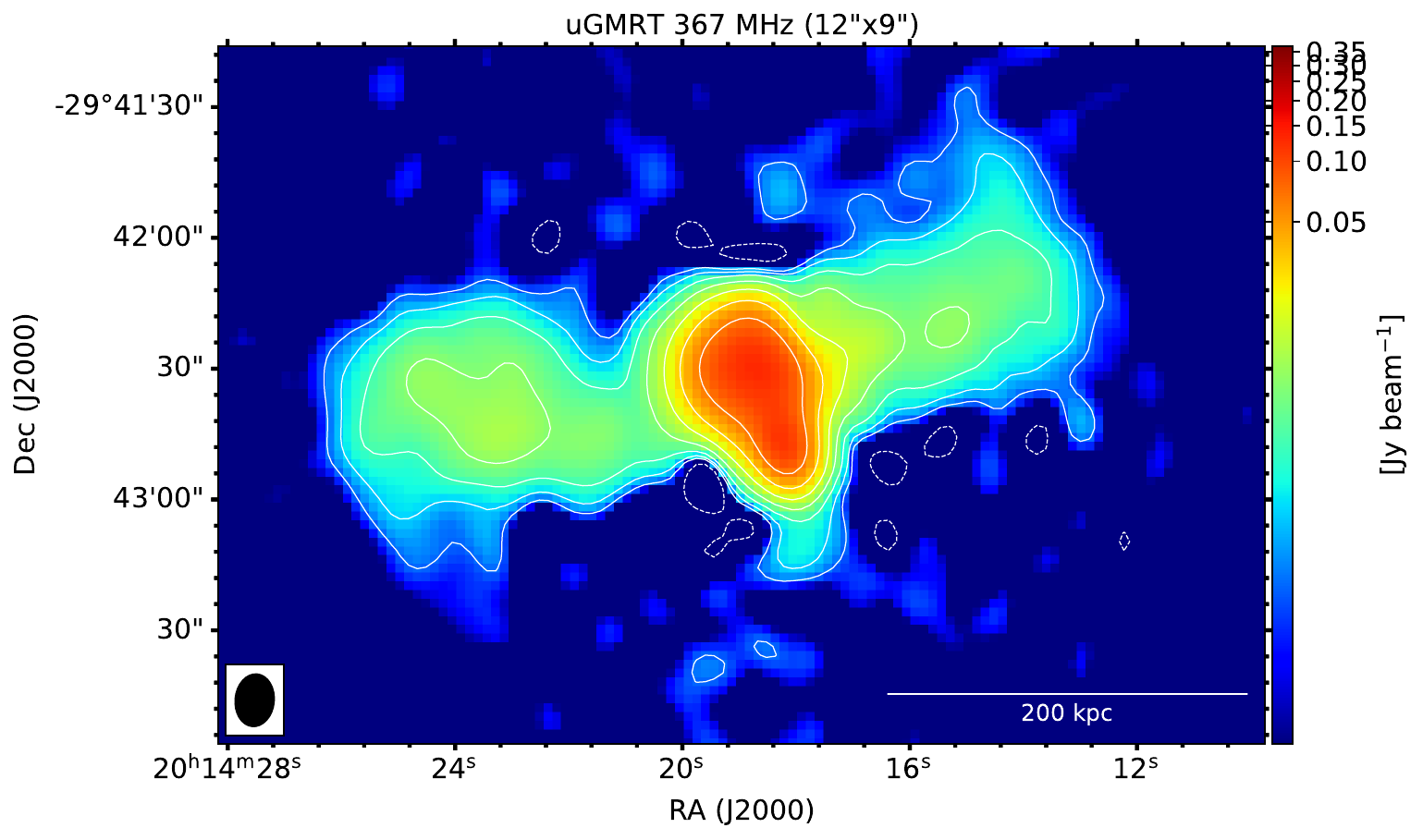}
\includegraphics[width=0.495\textwidth]{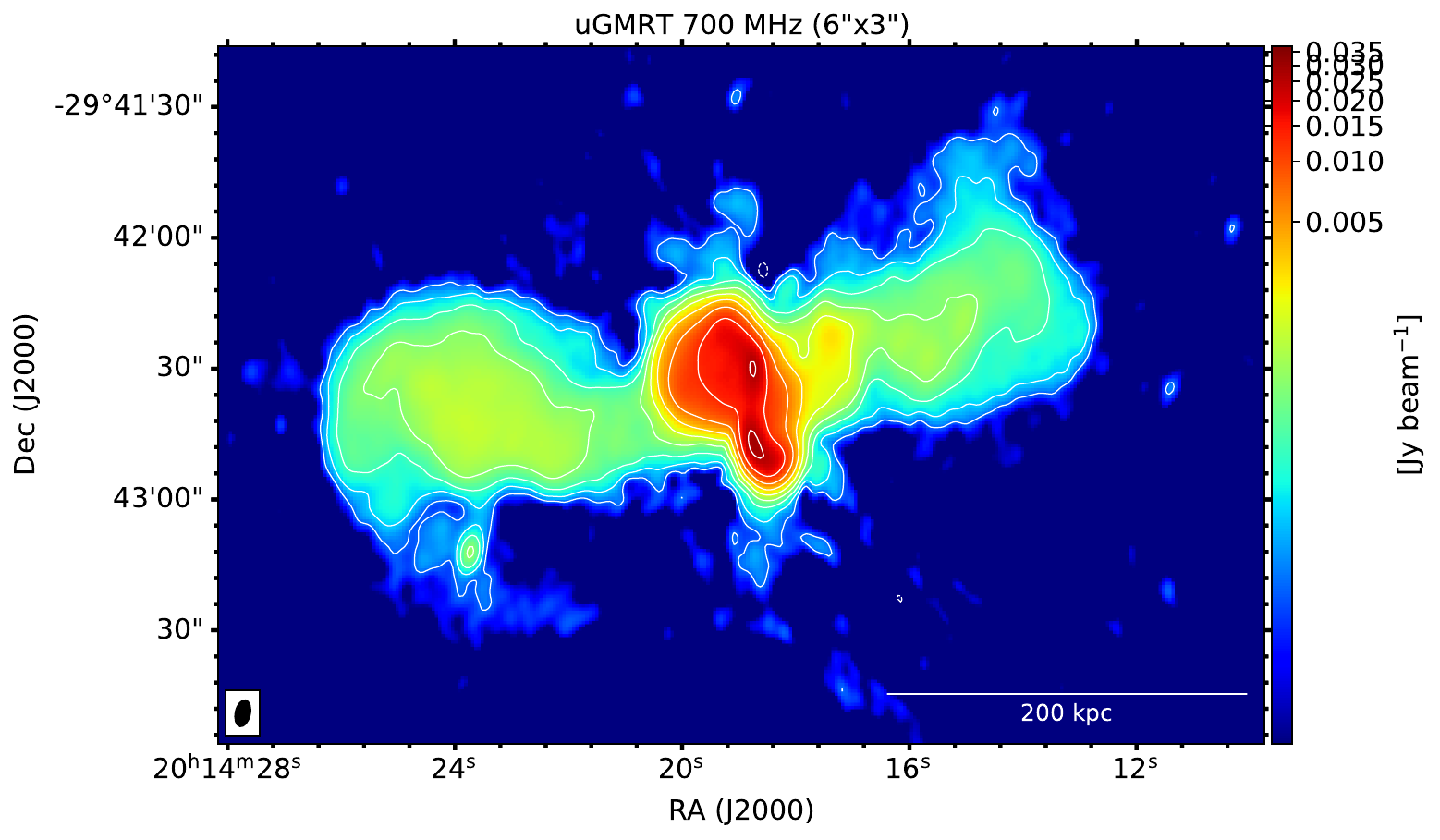}
\includegraphics[width=0.495\textwidth]{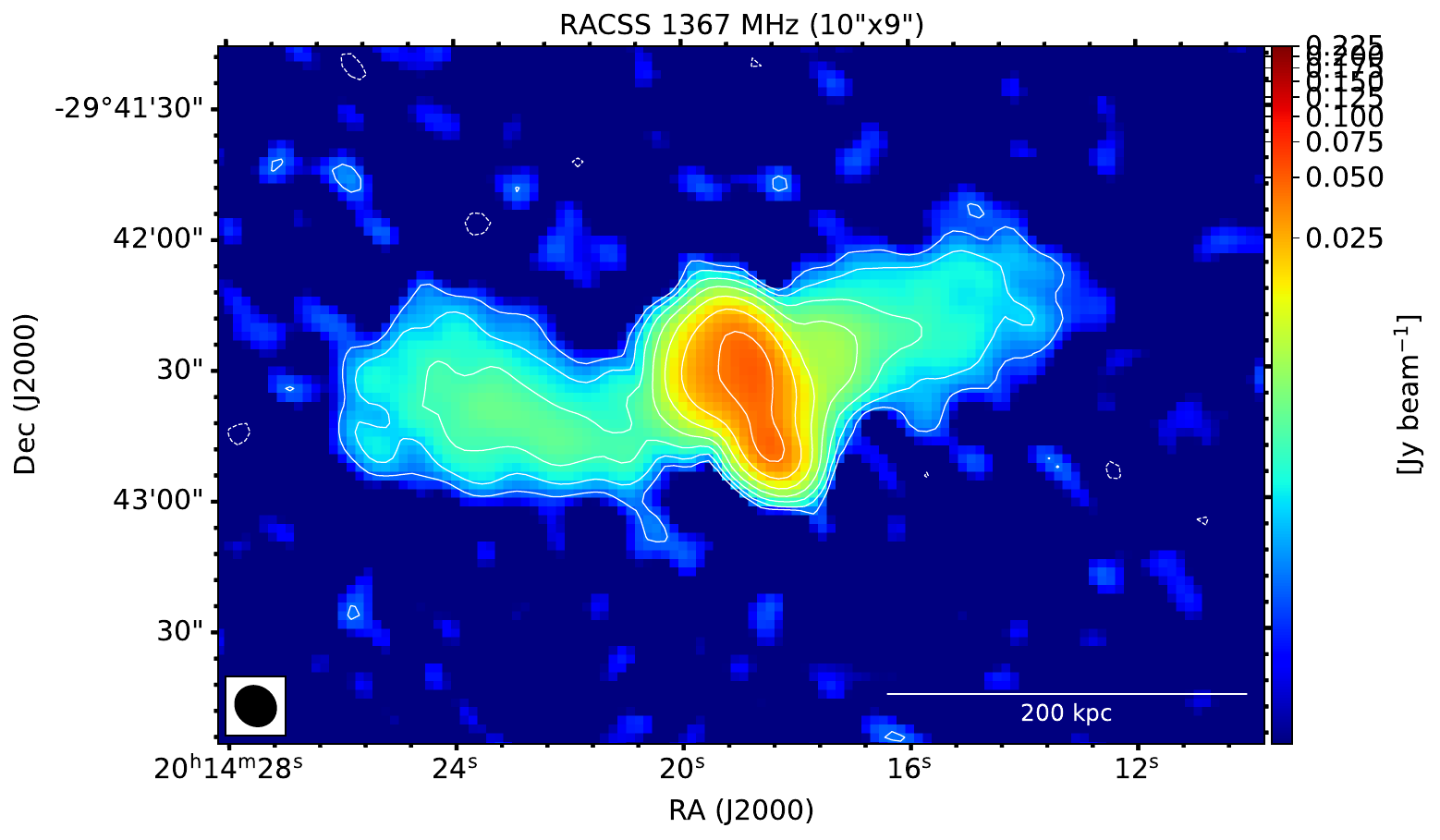}
\includegraphics[width=0.495\textwidth]{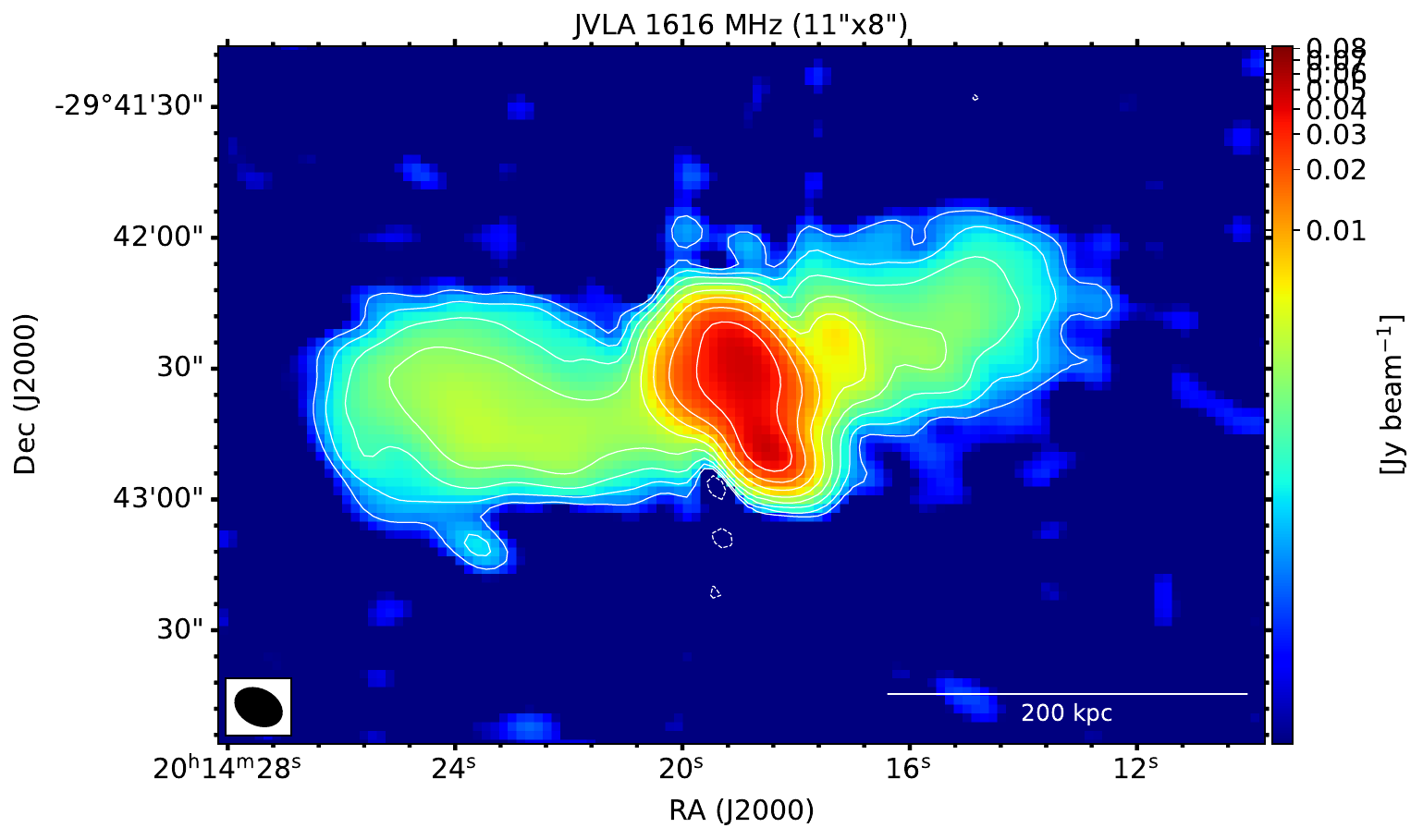}
\includegraphics[width=0.495\textwidth]{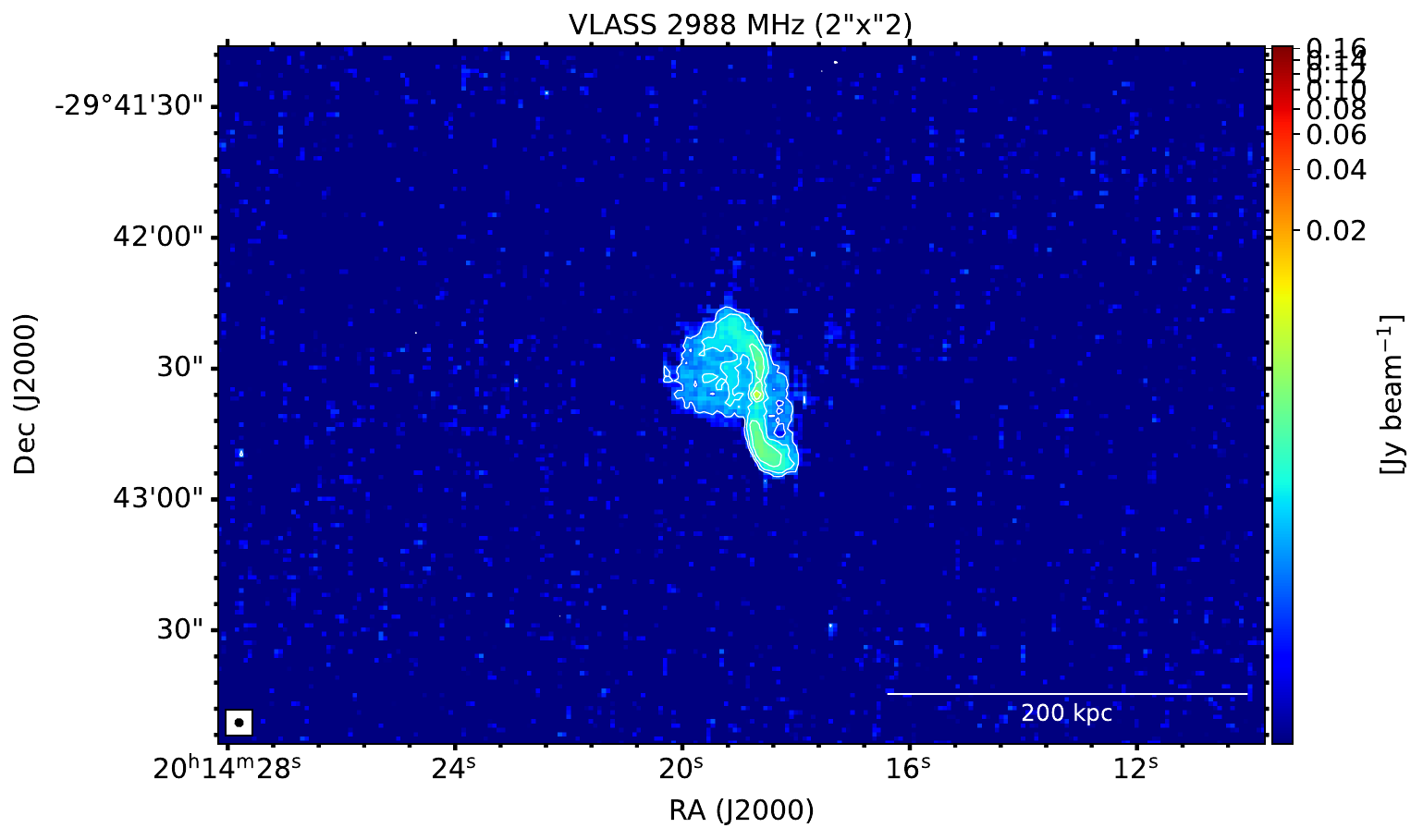}
\includegraphics[width=0.495\textwidth]{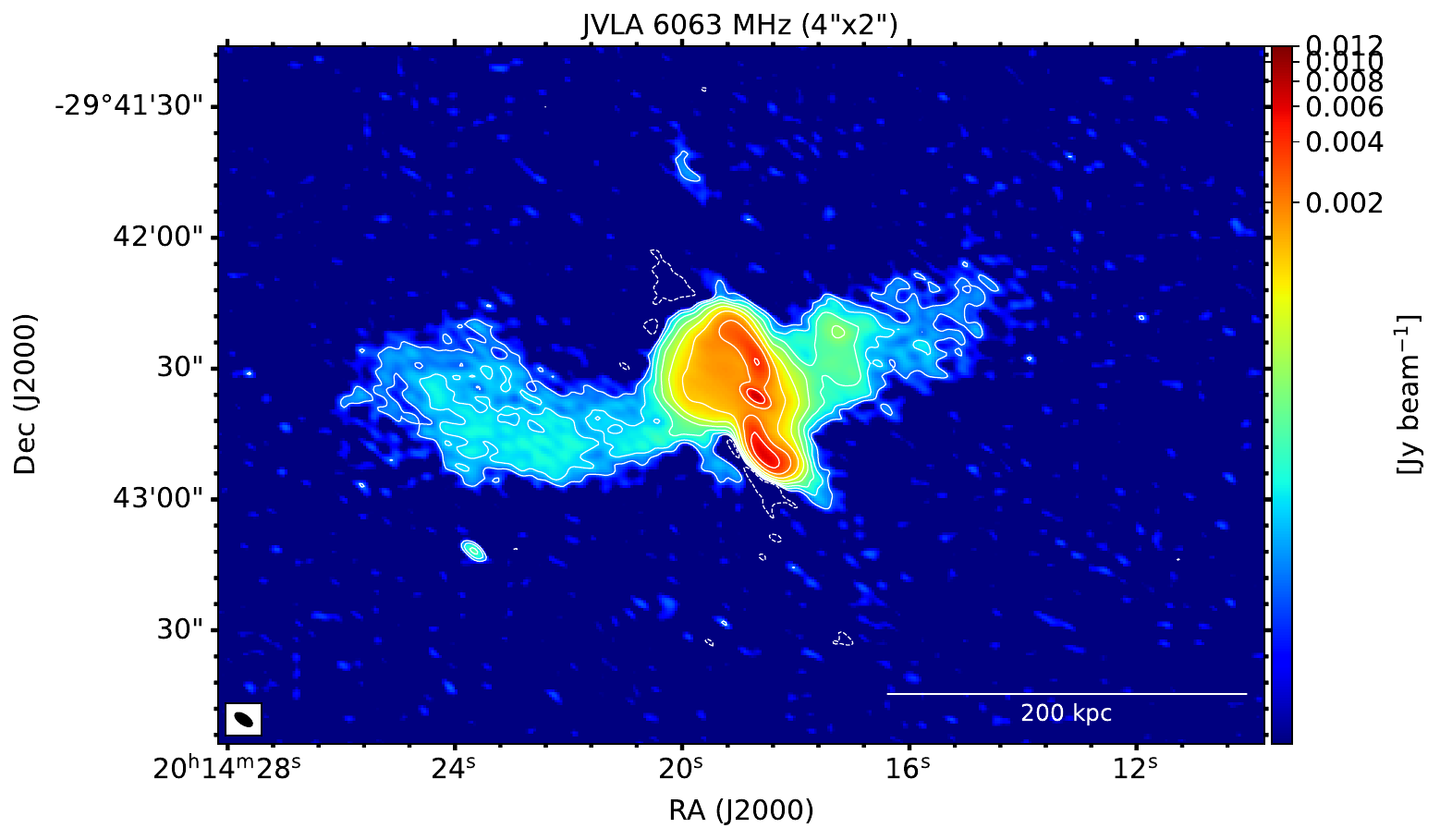}
\includegraphics[width=0.495\textwidth]{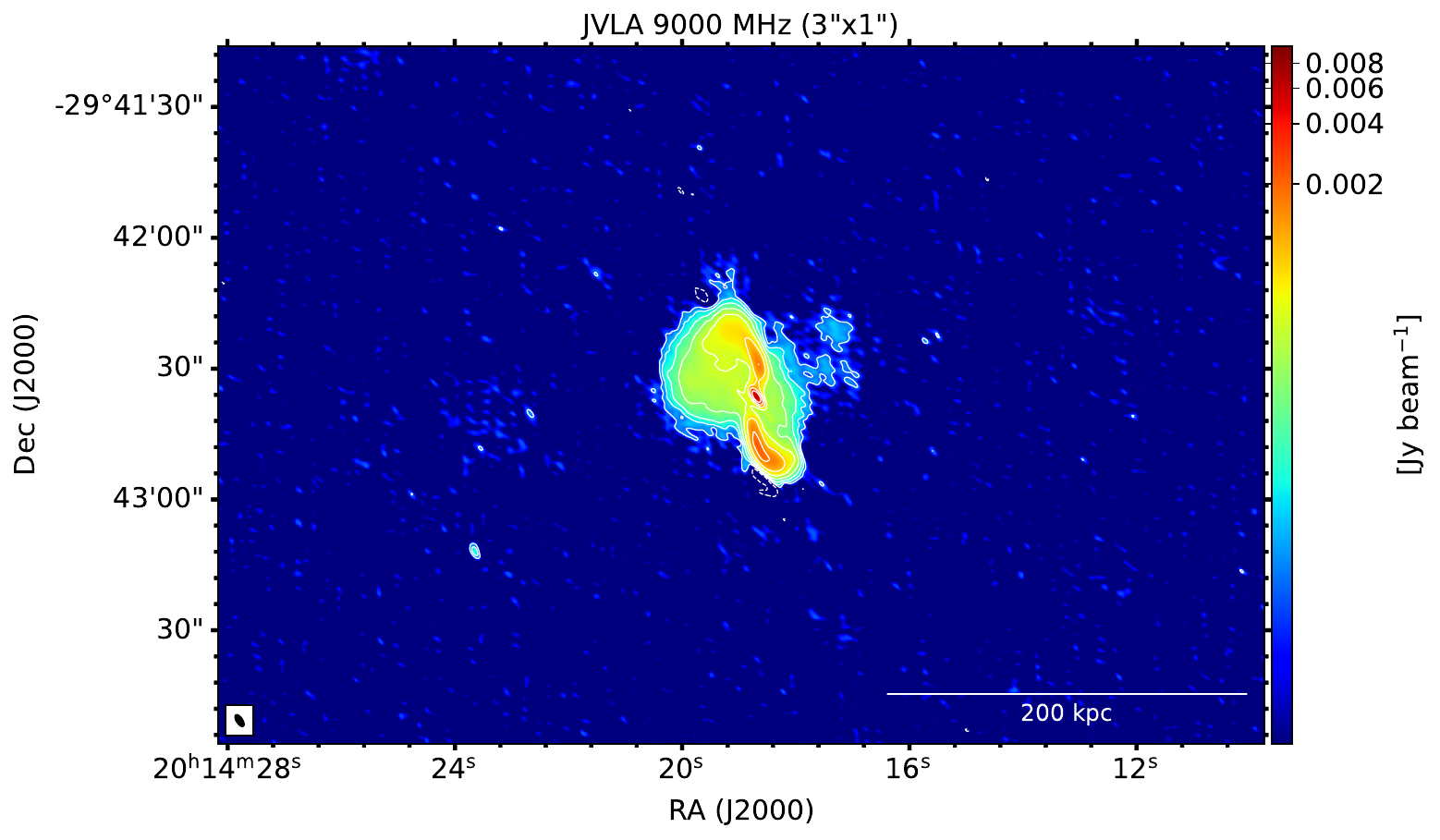}

	\caption{Radio images of MRC 2011-298. The overlaid contour levels are $[\pm4, \;8, \;16,\; 32,\; ...]\times \sigma$. {\it From top left to bottom right}: uGMRT at 170 MHz ($\theta=24''\times11''$, $\sigma = 0.8 \; {\rm mJy \; beam^{-1}}$), uGMRT at 367 MHz ($\theta=12''\times9''$, $\sigma = 0.25 \; {\rm mJy \; beam^{-1}}$), uGMRT at 700 MHz ($\theta=6''\times3''$, $\sigma = 25 \; {\rm \mu Jy \; beam^{-1}}$), RACS-mid at 1367 MHz ($\theta=10''\times9''$, $ \sigma = 0.15 \; {\rm mJy \; beam^{-1}}$), JVLA at 1616 MHz ($\theta=11''\times8''$, $ \sigma = 55 \; {\rm \mu Jy \; beam^{-1}}$), VLASS at 2988 MHz ($\theta=2''\times2''$, $\sigma = 0.11 \; {\rm mJy \; beam^{-1}}$), JVLA at 6063 MHz ($\theta=4''\times2''$, $\sigma = 8 \; {\rm \mu Jy \; beam^{-1}}$), JVLA at 9000 MHz ($\theta=3''\times1''$, $\sigma = 6.5 \; {\rm \mu Jy \; beam^{-1}}$). }
	\label{fig: mappefullres}
\end{figure*}

 \begin{table}
 \fontsize{8.5}{8.5}\selectfont
      \centering
   	\caption[]{Summary of the parameters of radio images shown in Fig. \ref{fig: mappefullres}. Cols. 1-5: Instrument, central frequency ($\nu$), restoring beam ($\theta$), beam position angle ($P.A.$), noise ($\sigma$) close to the target, and robust (R) parameter of the \cite{briggs95} 
 baseline weighting scheme.}
   	\label{tab: image full res}
   	\begin{tabular}{cccccc}
   	\hline
   	\noalign{\smallskip}
   	 Instrument & $\nu$ & $\theta$ & $P.A.$ & $\sigma$ & R \\
   	&  (MHz) & ($'' \; \times \; ''$) & (deg) & (${\rm \mu Jy \; beam^{-1}}$) &  \\
   	\noalign{\smallskip}
   	\hline
   	\noalign{\smallskip}
    uGMRT &	170  & $24\times11$ & 169 & 800 & -0.2 \\
    uGMRT &	367 & $12\times9$ & 175 & 250 & -0.2 \\
    uGMRT & 700 & $6\times3$ & 168 & 25 & -0.2 \\
    JVLA &   1616 & $11\times8$ & 63 & 55 & 0.2 \\
    VLASS & 2988 & $2\times2$ & 0 & 110 & 0 \\
    JVLA & 6063 & $4\times2$ & 55 & 8 & -0.2 \\
     JVLA & 9000 & $3\times1$ & -146 & 6.5 & 0.2 \\
   	\noalign{\smallskip}
   	\hline
   	\end{tabular}
   \end{table}

\begin{figure*}
	\centering
  \includegraphics[width=0.4\textwidth]{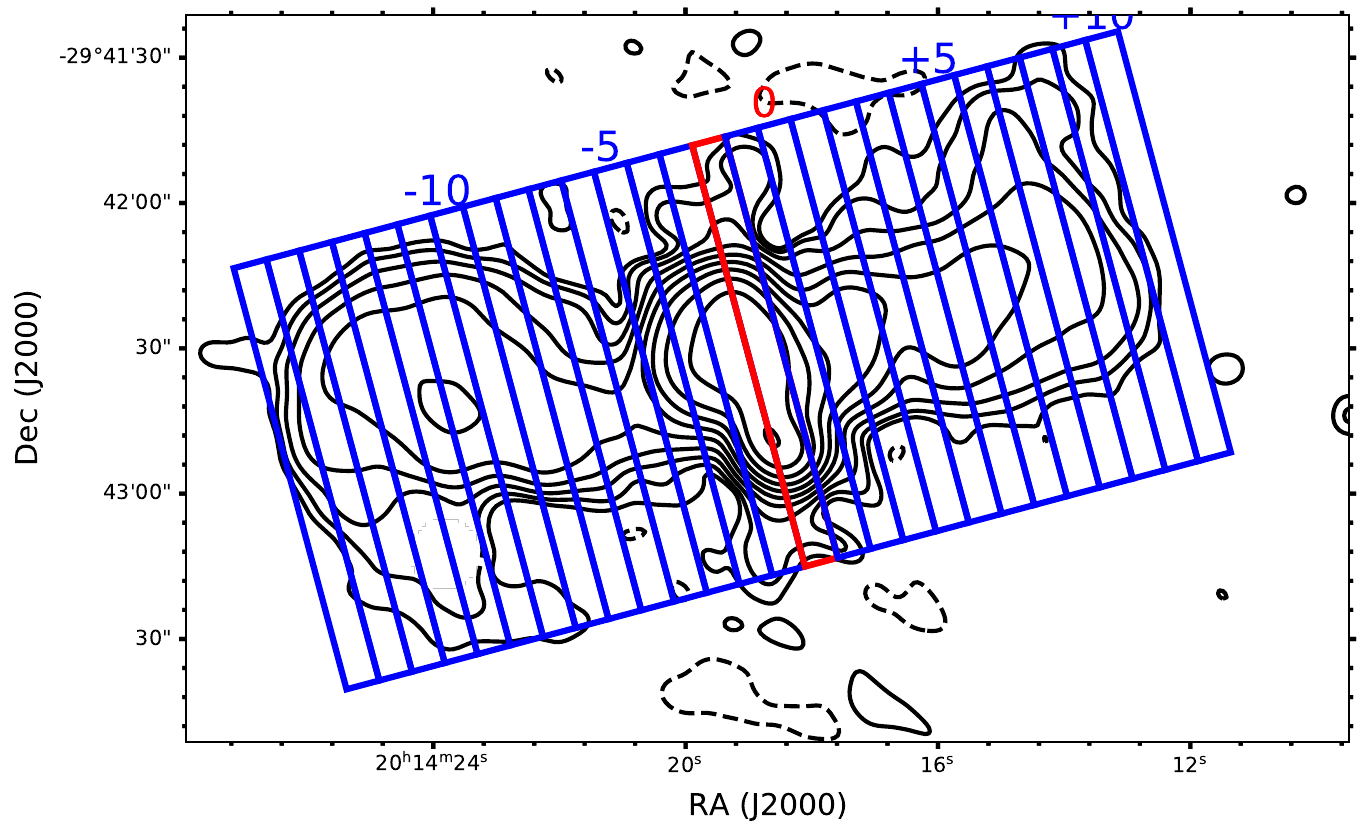}
\includegraphics[width=0.28\textwidth]{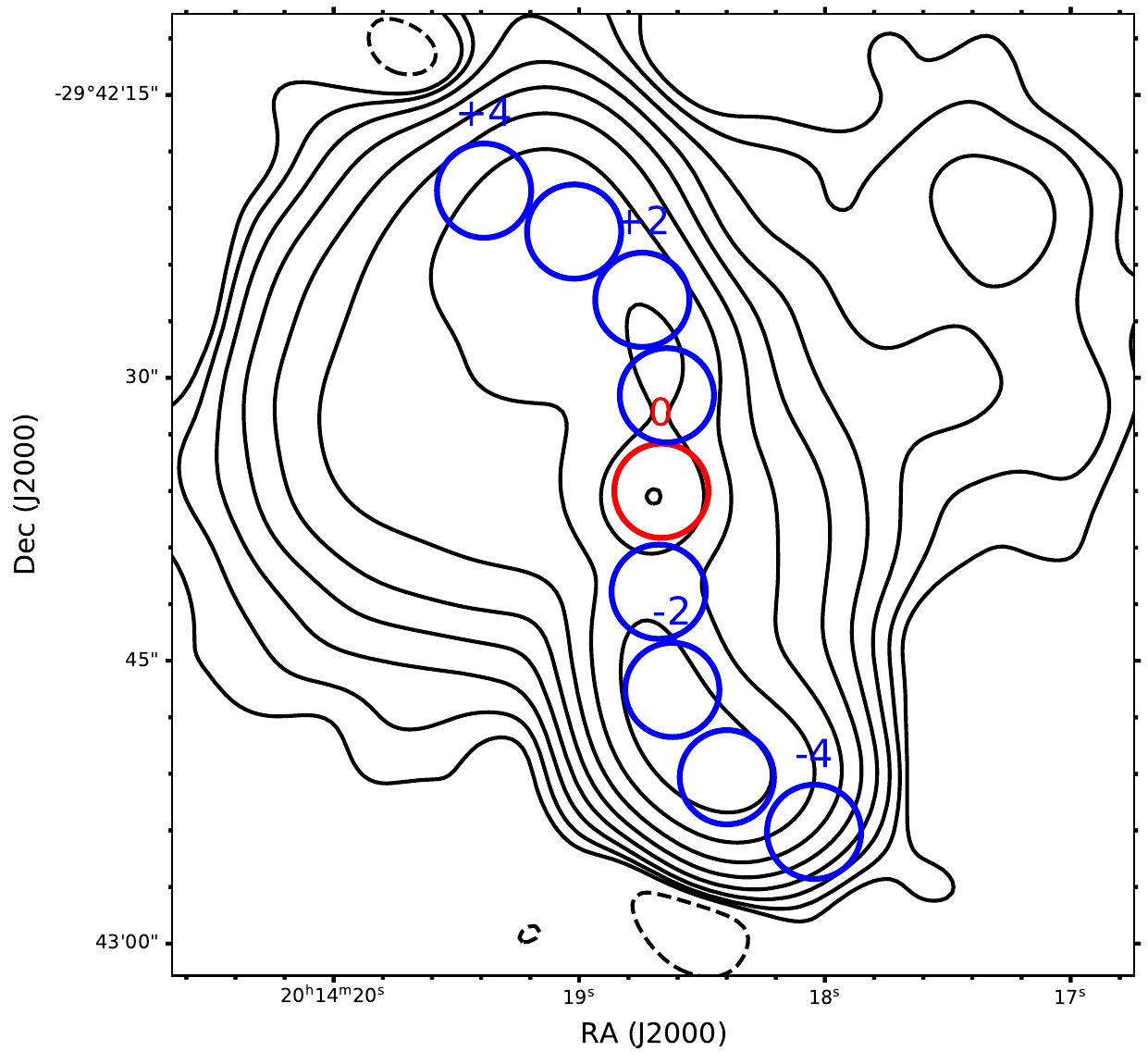}
 \includegraphics[width=0.49\textwidth]{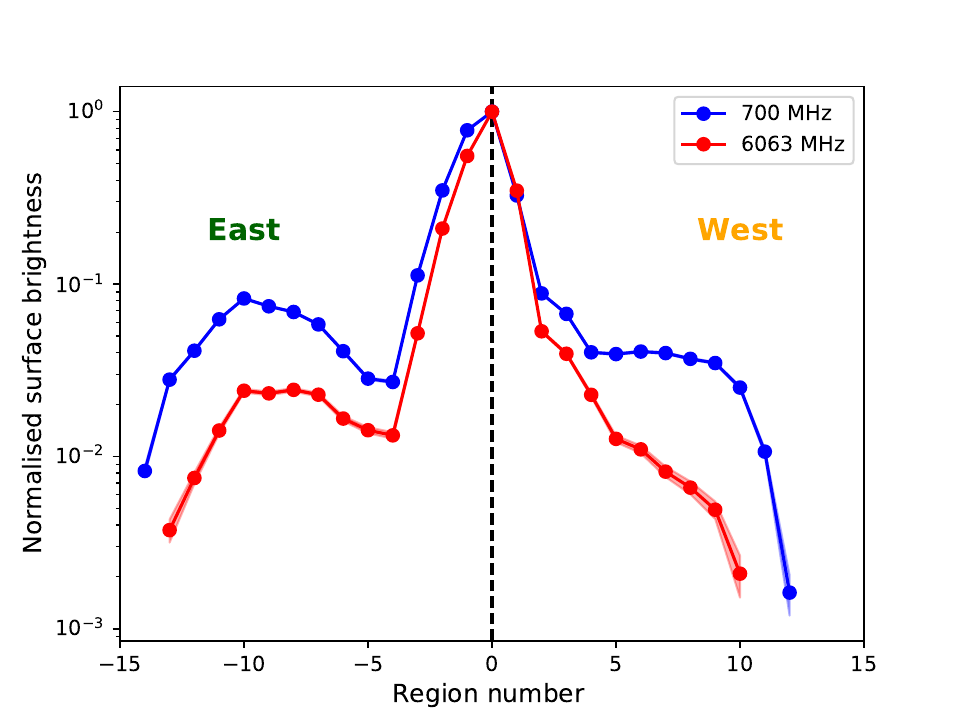}
\includegraphics[width=0.49\textwidth]{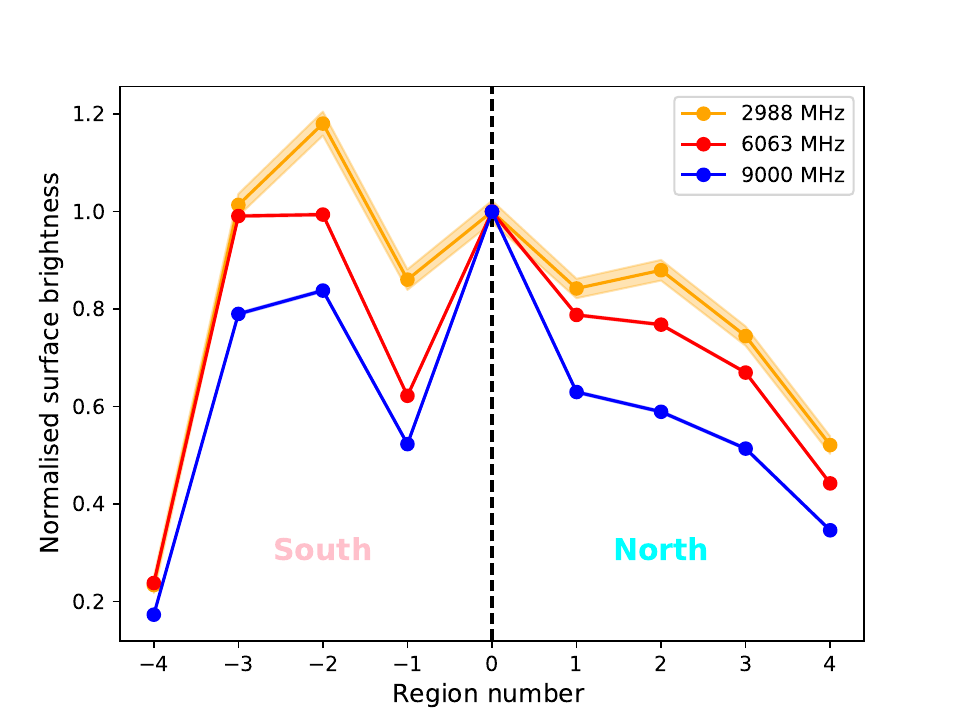}
	\caption{Radio surface brightness profiles of MRC 2011-298. \textit{Top}: Sampling regions overlaid on the radio contours at 700 (left) and 9000 (right) MHz. The identifier of the sampling regions increases towards north and west, and decreases towards south and east with respect to region `0' (in red), which includes the radio core. \textit{Bottom left}: Profiles of the active lobes and wings measured at 700 (blue) and 6063 (red) MHz in boxes of beam-size ($7''$) width, and normalised by the brightness of region `0' (dashed black line). \textit{Bottom right}: Profiles of the jets measured at 2988 (orange), 6063 (red), and 9000 (blue) MHz in beam-size ($4.5''$) circles, normalised by the brightness of region `0' (dashed black line).}
	\label{fig: profili brillanza}
\end{figure*}

In Fig. \ref{fig: mappefullres} we present the radio images of the XRG in A3670 from 170 to 9000 MHz. Table \ref{tab: image full res} summarises the resolution and noise level of these images. The features that we will discuss are labelled in Fig. \ref{fig: radio+panstarss}.

The active lobes along the north-south axis are the brightest sub-region of the target. They can be approximated as an ellipse of length (along N-S) $\sim 60''$ and width (along E-W) $\sim 40''$, corresponding to $\sim 150 \; {\rm kpc} \times 100 \; {\rm kpc}$ at the cluster redshift. The very high-resolution ($\leq 4''$, that is $\leq 10$ kpc) provided by S-band, C-band, and X-band JVLA images allows us to distinguish the core and the jets within the lobes. As claimed in \cite{bruno19}, MRC 2011-298 is a FRI-type XRG owing to the absence of hotspots at the edge of the jets. The jets exhibit an inversion-symmetric S-shape with respect to the core, which is suggestive of a precession motion. The southern jet is slightly brighter than the northern jet, likely due to a combination of projection effects and precession.

The wings are fainter than the lobes at all frequencies, and extend along the east-west axis. By considering the sensitive 700 MHz image as reference, the eastern and western wings can be represented as ellipses of length (along E-W) and width (along N-S) $\sim 80''\times 60'' \sim 200 \; {\rm kpc} \times 150 \; {\rm kpc}$ and $\sim 70''\times 50'' \sim 175 \; {\rm kpc} \times 125 \; {\rm kpc}$, respectively. Therefore, the whole structure is approximately $\sim 3'\times 1' \sim 450 \; {\rm kpc} \times 150 \; {\rm kpc}$, and the total axis ratio is $\sim 3$ (the single wing to lobe ratio is $\sim1.2-1.3$). XRGs having wings larger than the main lobes as MRC 2011-298 are not common \citep[e.g.][]{saripalli&subrahmanyan09,saripalli&roberts18,joshi19}, but it is unclear whether projection effects bias their actual number. Although the apparent lengths of the eastern and western wings of MRC 2011-298 are similar, their morphology is not symmetric. The western wing exhibits a bright spot (Fig. \ref{fig: radio+panstarss}) in its inner regions, which is detected at each frequency (except in VLASS, where wings are completely undetected) with a $\geq 4 \sigma$ significance. Interestingly, the uGMRT images reveal a deflection of the wings at their tips, which provides a global Z-shape that is not observed at higher frequencies; we will further discuss this feature in Sect. \ref{sect: Jet precession}.

In Fig. \ref{fig: profili brillanza} we report the surface brightness profiles of the target as measured from different images convolved to the same resolution of $7''$ and $4.5''$ for the E-W and N-S analysis, respectively. Each data point is normalised to the brightness of the reference region `0' that includes the core. 

The bottom left panel of Fig. \ref{fig: profili brillanza} reports the global profiles along the east-west axis at 700 MHz (blue) and 6063 MHz (red). The absolute peak is associated with the emission of the lobes, while secondary peaks are associated with the wings. The transition from the lobe to the eastern wing is sharp, whereas the presence of the bright spot produces a shallower transition to the western wing. The normalised surface brightness profiles of the active lobes are consistent at 700 MHz and 6063 MHz, while the brighter profile of the wings at lower-$\nu$ is indicative of spectral steepening (see also Sects. \ref{sect: Integrated spectra}, \ref{sect: Resolved spectral analysis}). 

The bottom right panel of Fig. \ref{fig: profili brillanza} reports the normalised brightness profiles of the jets at 3 GHz (orange), 6 GHz (red), and 9 GHz (blue). The orientation of the sampling circles track the S-shape of the jets, but the measured profiles indicate that the brightness of the jets is not symmetric. The brightness of the southern jet exhibits a drop (region `-1'), which is followed by an enhancement (region `-2'), and then decreases, while the profile of the northern jet gradually decreases with the distance from the core. The comparison of the brightness indicates the jets do not completely lie on the plane of sky (under the assumption of intrinsic symmetry of brightness and advance speed), but the southern and northern jets are approaching and receding us, respectively.

\subsection{Integrated spectral analysis}
\label{sect: Integrated spectra}

\begin{table}
 \fontsize{8.8}{8.8}\selectfont
\centering
   		\caption{Summary of the sets of images used to measure flux densities for integrated spectra. Col. 1: set identifier. Cols. 2, 3: considered frequencies and \textit{uv}-range. Col. 4: image resolution after convolution with common beam. }
    \label{tab: parametriperspettri}   
    \begin{tabular}{cccc}
   	\hline
   	\noalign{\smallskip}
   	Set & $\nu$ & \textit{uv}-range & $\theta$ \\  
	& (MHz) & (${\rm k}\lambda$) & (arcsec)  \\  
   	\noalign{\smallskip}
  	\hline
   	\noalign{\smallskip} 
 A & 170, 367, 700, 6063 & [0.5, 17] & 17  \\
 B & 170, 367, 700, 1616, 6063 & [0.9, 17] & 16  \\
 C & 170, 367, 700, 1616, 2988, 6063, 9000 & [2, 17] & 15  \\
 D & 700, 2988, 6063, 9000 & [2, 66] & 4 \\
 E & 4743, 5575, 6371, 7383 & [0.75, 36] & 5 \\
   	\noalign{\smallskip}
   	\hline
   	\end{tabular}
   	  	\centering
   \end{table}

\begin{table*}
\centering
   		\caption{Flux density measurements of MRC 2011-298 shown in Fig. \ref{fig: spettro integrato}. The considered sub-region and set of images are listed in Table \ref{tab: parametriperspettri}. }
    \label{tab: flussiperspettri}   
    \begin{tabular}{ccccccccc}
   	\hline
   	\noalign{\smallskip}
   	Set & Component & $S_{170}$ & $S_{367}$ & $S_{700}$ & $S_{1616}$ & $S_{2988}$ & $S_{6063}$ & $S_{9000}$  \\  
	& & (mJy) & (mJy) & (mJy) & (mJy) & (mJy) & (mJy) & (mJy)   \\  
   	\noalign{\smallskip}
  	\hline
   	\noalign{\smallskip}

A & Total & $1588.4 \pm 111.4$ & $1002.6 \pm 60.2$ & $692.9 \pm 34.6$ & - & - & $144.4 \pm 4.3$ & - \\
B & Total & $1525.9 \pm 107.1$ & $973.3 \pm  58.5$ & $687.7 \pm 34.4$ & $373.1 \pm 18.7$ & - & $143.4 \pm 4.3$ & - \\
C & Total  & $1190.0 \pm 83.6$ & $858.4 \pm  51.6$ & $604.0 \pm 30.2$ & $348.1 \pm 17.4$ & $211.5\pm 7.0$ & $137.7 \pm 4.1$ & $79.6 \pm 2.4$ \\
A & Lobe & $1143.3 \pm 80.1$ & $750.6 \pm 45.0$ & $535.9 \pm 26.8$ & - & - & $130.5 \pm 3.9$ & - \\
B & Lobe & $1130.3 \pm 79.2$ & $745.8 \pm 44.8$ & $ 535.7 \pm 26.8$ & $310.5 \pm 15.5$ & - & $130.5 \pm 3.9$ & - \\
C & Lobe & $1014.7 \pm 71.1$ & $705.3 \pm 42.3$ & $509.4 \pm 25.5$ & $302.3  \pm 15.1$ & $210.0 \pm 6.4$ & $127.8 \pm 3.8$ & $79.7 \pm  2.4$ \\
A & East wing  & $267.2 \pm 19.0$ & $149.0 \pm 9.1$ & $100.2 \pm  5.0$ & - & - & $9.3\pm 0.3$ & - \\
B & East wing  & $239.3 \pm 17.3$ & $131.7 \pm 8.1$ & $96.2 \pm  4.8$ & $42.8 \pm 2.2$ & - & $8.9 \pm 0.3$ & - \\
C & East wing & $145.4 \pm 10.7$ & $107.4 \pm 6.5$ & $63.9 \pm  3.2$ & $32.9 \pm 1.7$ & - & $7.3 \pm 0.3$ & - \\
A & West wing & $160.1 \pm 11.6$ & $91.9 \pm 5.7$ & $52.9 \pm  2.7$ & - & - & $3.9 \pm 0.2$ & - \\
B & West wing  & $156.6 \pm 11.5$ & $94.8 \pm 5.8$ & $53.9 \pm 2.7$ & $19.9 \pm 1.1$ & - & $3.7 \pm 0.2$ & - \\
C & West wing  & $75.6 \pm 6.3$ & $55.9 \pm 3.5$ & $34.0 \pm 1.7$ & $11.6 \pm 0.7$ & - & $2.0 \pm 0.1$ & - \\
D & Core & - & - & - & - & $10.6 \pm 0.4$ & $8.6 \pm 0.3$ & $7.7 \pm 0.2$ \\
D & North Jet & - & - & $76.2 \pm 3.8$ & - & $30.4 \pm 1.0$ & $21.1 \pm 0.6$ & $13.4 \pm 0.4$ \\
D & South Jet  & - & - & $83.2 \pm 4.2$ & - & $36.6 \pm 1.1$ & $26.4 \pm 0.8$ & $17.3 \pm 0.5$ \\
   	\noalign{\smallskip}
   	\hline
   	\end{tabular}
   	  	\centering
   \end{table*}

\begin{figure*}
	\centering

\includegraphics[width=0.40\textwidth]{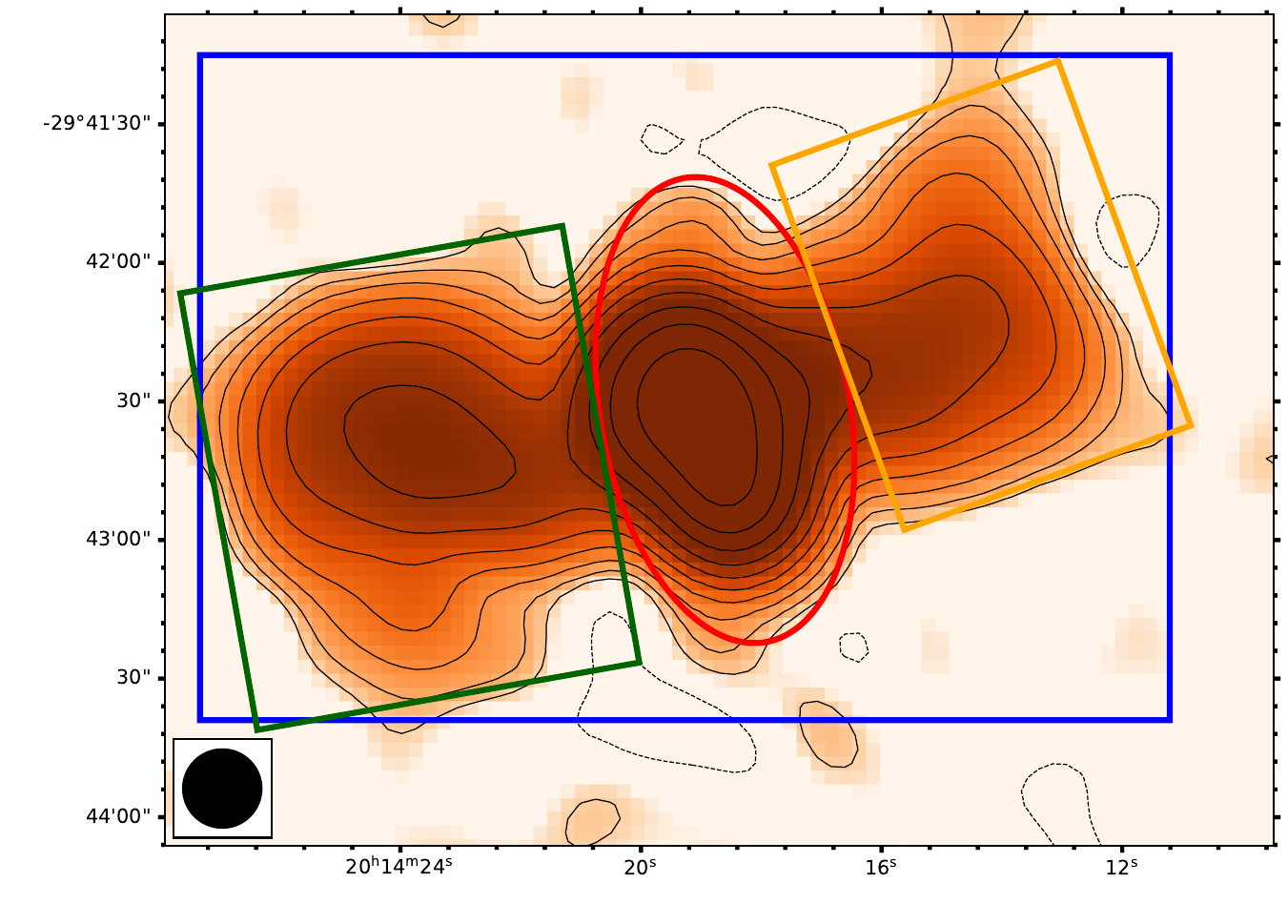} 
\includegraphics[width=0.34\textwidth]{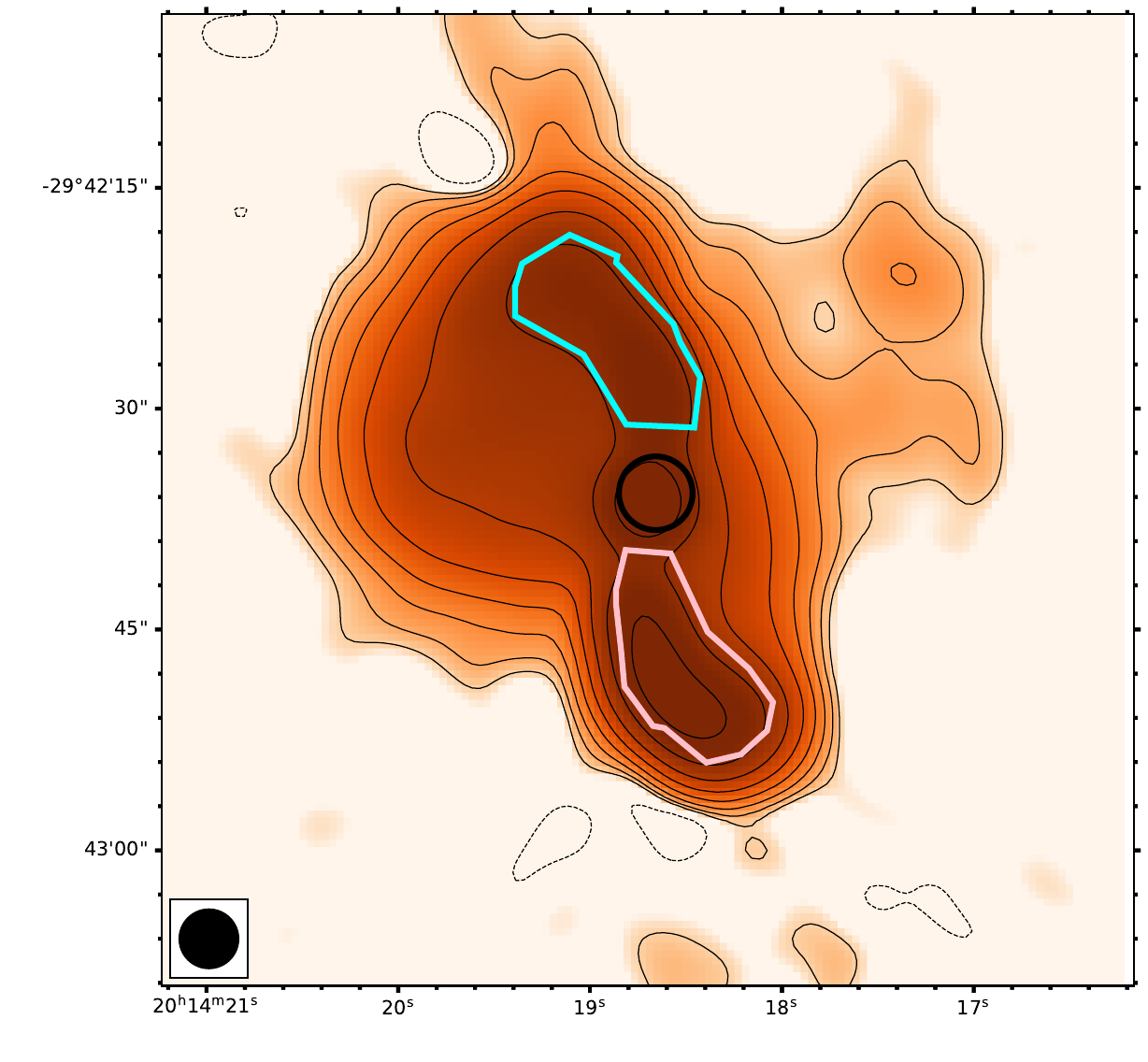} 
\includegraphics[width=0.49\textwidth]{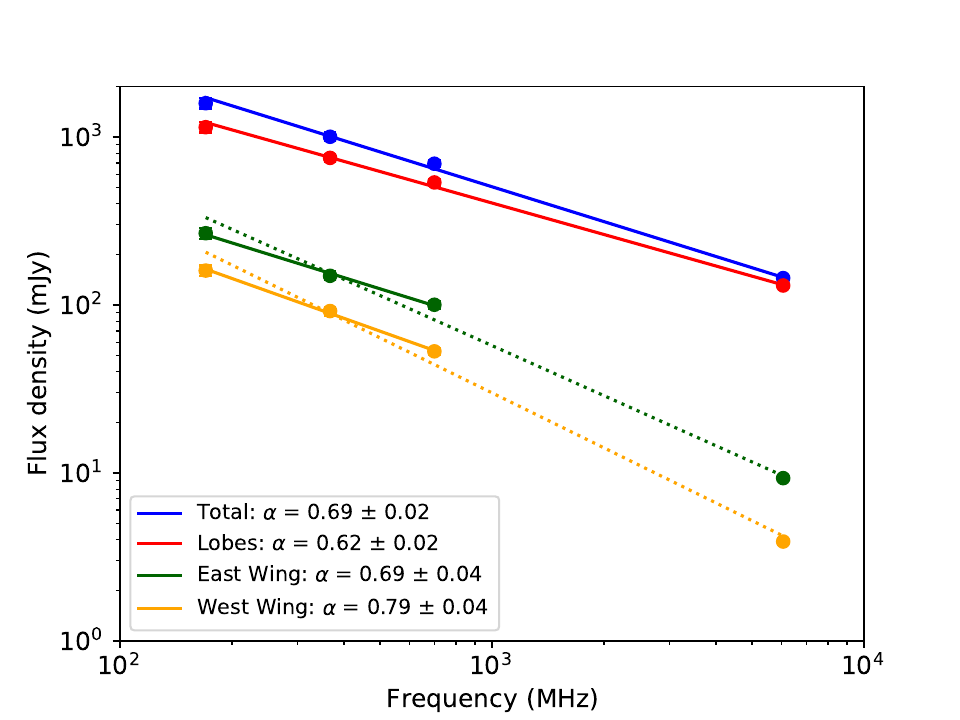}
\includegraphics[width=0.49\textwidth]{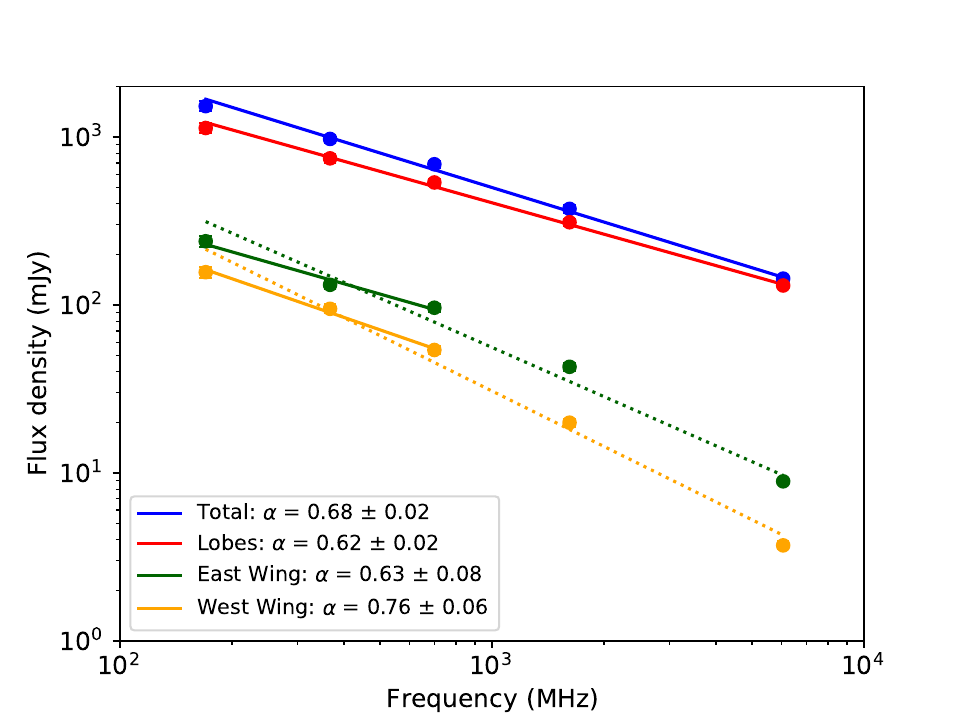}
\includegraphics[width=0.49\textwidth]{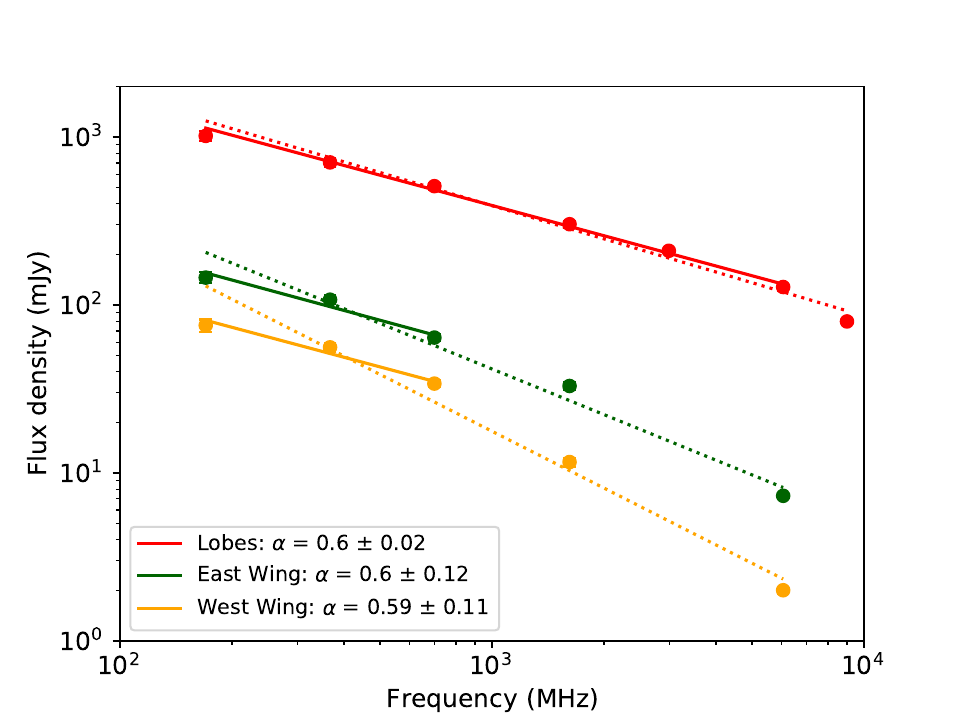}
\includegraphics[width=0.49\textwidth]{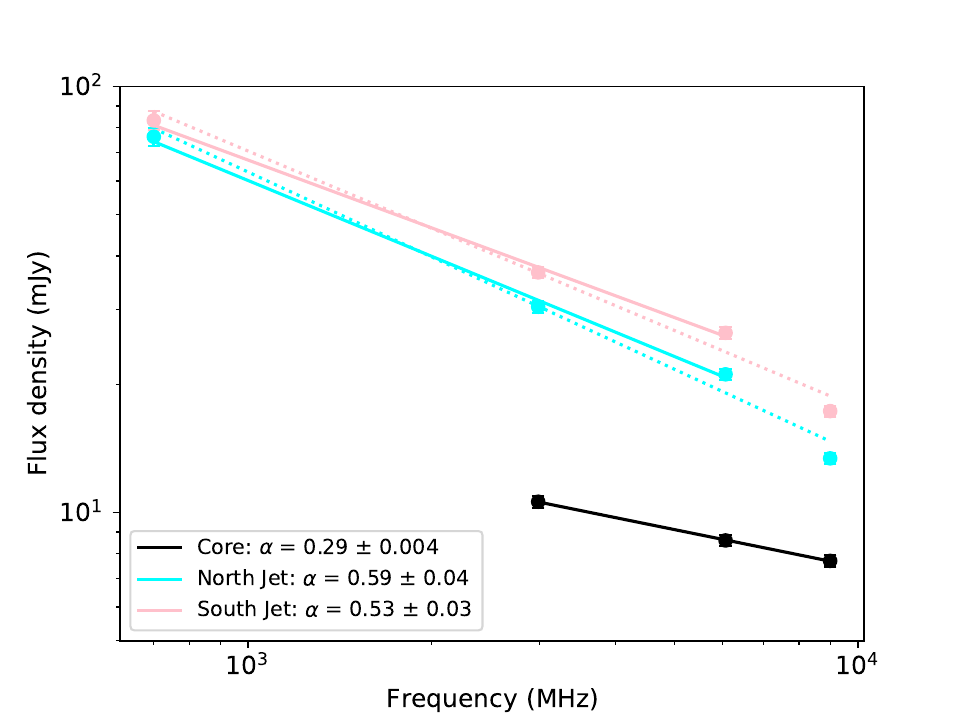}
	\caption{Radio spectrum of MRC 2011-298. Upper panels report the 700 MHz (left) and 9000 MHz (right) images with overlaid an example of the regions used to measure the flux densities of the whole source (blue box), lobes (red ellipse), east wing (green box), west wing (orange box), core (black circle), north jet (cyan polygonal), and south jet (pink polygonal). Middle and lower panels show the spectra obtained for sets of images A, B, C, and D (Table \ref{tab: flussiperspettri}). Dashed and solid lines are the fitted power-laws to all data points and to subsets of data points, respectively. Results of the fits are summarised in Table \ref{tab: fitspix}.  }
	\label{fig: spettro integrato}
\end{figure*}

\begin{table}
\centering
   		\caption{Integrated spectral index of MRC 2011-298. Col. 1: Considered area. Col. 2: Colour and line style of the fitted power-law as in Figs. \ref{fig: spettro integrato}, \ref{fig: inbandspec}. Col. 3: Frequency range. Col. 4: Fitted spectral index. Col. 5: Reduced $\chi$-squared.}
    \label{tab: fitspix}
     \resizebox{0.45\textwidth}{!}{
    \begin{tabular}{ccccc}
   	\hline
   	\noalign{\smallskip}
   	Area & Power-law & $\nu$ & $\alpha$ & $\chi^2_{\rm red}$  \\  
	  &  & (MHz) &  &   \\  
   	\noalign{\smallskip}
  	\hline
   	\noalign{\smallskip}
  Total (A) & Solid blue & 170-6063 & $0.69 \pm  0.02$ & 1.6  \\
  Lobes (A) & Solid red & 170-6063 & $0.62 \pm 0.02$ & 1.2  \\
  East Wing (A) & Dotted green & 170-6063 & $0.99 \pm 0.06$ &  13.8 \\
  East Wing (A) & Solid green & 170-700 & $0.69 \pm 0.04$ &  0.5 \\
  West Wing (A) & Dotted orange & 170-6063 & $1.09 \pm 0.08$ & 13.6  \\
  West Wing (A) & Solid orange & 170-700 & $0.79 \pm 0.04$ & 0.4  \\
  Total (B) & Solid blue & 170-6063 & $0.68 \pm  0.02$ & 1.7  \\
  Lobes (B) & Solid red & 170-6063 & $0.62 \pm 0.02$ & 1.0  \\
  East Wing (B) & Dotted green & 170-6063 & $0.97 \pm 0.07$ &  18.1 \\
  East Wing (B) & Solid green & 170-700 & $0.63 \pm 0.08$ &  1.8 \\
  West Wing (B) & Dotted orange & 170-6063 & $1.10 \pm 0.08$ & 13.2  \\
  West Wing (B) & Solid orange & 170-700 & $0.76 \pm 0.06$ & 1.1  \\
  Lobes (C) & Dotted red & 170-9000 & $0.66 \pm 0.04$ & 10.2  \\
  Lobes (C) & Solid red & 170-6063 & $0.60 \pm 0.02$ & 1.7  \\
  East Wing (C) & Dotted green & 170-6063 & $0.90 \pm 0.08$ &  16.6 \\
  East Wing (C) & Solid green & 170-700 & $0.60 \pm 0.12$ &  3.8 \\
  West Wing (C) & Dotted orange & 170-6063 & $1.12 \pm 0.11$ & 26.9  \\
  West Wing (C) & Solid orange & 170-700 & $0.59 \pm 0.11$ & 2.9  \\
  Core (D) & Solid black & 2988-9000 & $0.290 \pm 0.004$ & 0.01  \\
  North Jet (D) & Dotted cyan & 700-9000 & $0.66 \pm 0.11$ & 11.4   \\
  North Jet (D) & Solid cyan & 700-6063 & $0.59 \pm 0.04$ & 1.8   \\
  South Jet (D) & Dotted pink & 700-9000 & $0.60 \pm 0.07$ & 10.3   \\
  South Jet (D) & Solid pink & 700-6063 & $0.53 \pm 0.03$ & 1.5   \\
  East Wing (E) & Solid green & 4743-7383 & $2.29 \pm 0.09$ &  0.6 \\
  West Wing (E) & Solid orange & 4743-7383 & $1.66 \pm 0.12$ & 0.8  \\
   	\noalign{\smallskip}
   	\hline
   	\end{tabular}
    }
   	  	\centering
   \end{table}

\begin{figure}
	\centering

\includegraphics[width=0.48\textwidth]{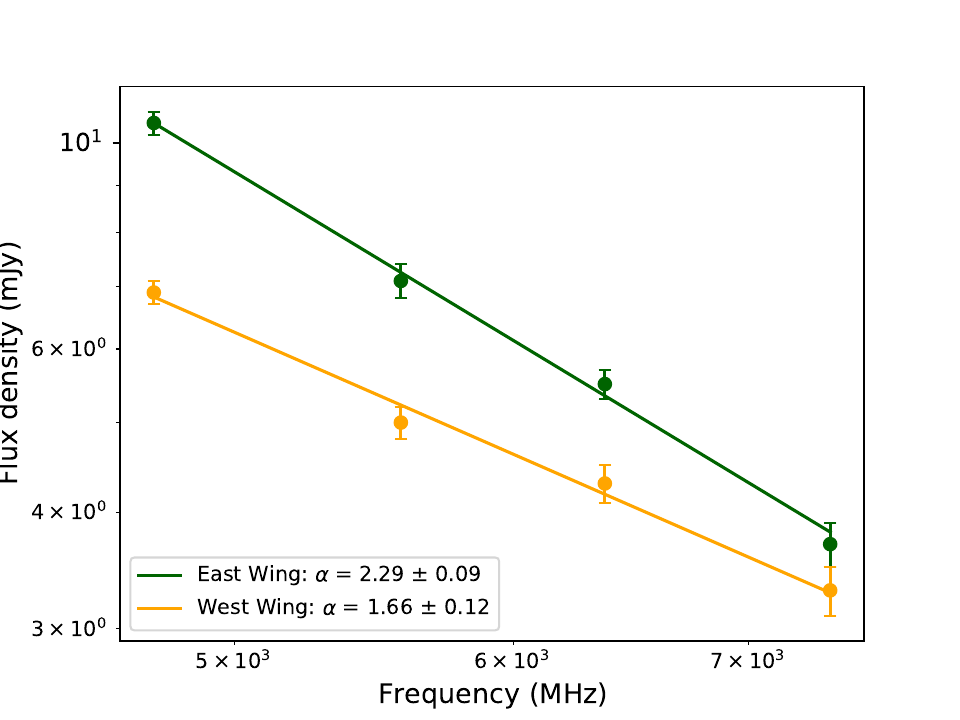} 
	\caption{In-band spectra of the wings at C-band. Results of the fits are summarised in Table \ref{tab: fitspix}.  }
	\label{fig: inbandspec}
\end{figure}

In this Section we derive the integrated spectrum of each component of MRC 2011-298. To this aim, we produced three sets of images (`A', `B', `C') with fixed weighting scheme (robust ${\rm R}=0$ as for \citealt{briggs95}) and same maximum baseline length, but different minimum baseline length, which allowed us to measure the flux density of the lobes and wings at comparable resolutions ($\sim 15''$), and investigate flux density losses associated with the \textit{uv}-coverage. An additional set of images (`D') was produced at higher resolution to measure the spectrum of the core and jets. Details on these images are summarised in Table \ref{tab: parametriperspettri}. For sets A, B, and C, flux densities were measured in regions encompassing the $4\sigma$ contour level at 700 MHz (see Fig. \ref{fig: spettro integrato}). We fitted the flux density measurements with single power-laws. When such power-laws are not representative of all measurements, we considered only data points at low frequencies that can be reasonably reproduced by a single power-law to constrain the spectral break. We report our measurements and results of the fits in Tables \ref{tab: flussiperspettri}, \ref{tab: fitspix}, respectively; the corresponding spectra are shown in Fig. \ref{fig: spettro integrato}. In Tables \ref{tab: parametriperspettri}, \ref{tab: fitspix} we anticipate information on set of images `E', which will be used and discussed in Sect. \ref{sect: Constraining the break frequency}.

The total flux density of the target is dominated by the active lobes. The average spectral index between 170-6063 MHz is $\alpha \sim 0.7$, as we obtained $\alpha_{170-6063}^{\rm A, tot}=0.69\pm 0.02$ and $\alpha_{170-6063}^{\rm B, tot}=0.68\pm 0.02$ for sets A and B, respectively (blue line). Nevertheless, different spectral indices are found within the components of MRC 2011-298. A single power-law cannot properly reproduce the spectrum of the lobes up to 9000 MHz (dashed red line for set C). A better description of the data points from 170 to 6063 MHz (solid red line) is provided by a power-law of slope $\alpha_{170-6063}^{\rm C,lobe}=0.60\pm 0.02$, which is consistent with the spectra inferred from sets A and B ($\alpha_{170-6063}^{\rm A,lobe}=\alpha_{170-6063}^{\rm B,lobe}=0.62\pm 0.02$).

The flux densities of the wings show prominent deviations from a single power-law at GHz frequencies. Although a single power-law can fit data points at 170-700 MHz, scatter and uncertainties progressively increase for sets B and C. The comparison of the flux densities of the wings (Table \ref{tab: flussiperspettri}) provides a clear example of the role of the \textit{uv}-range to recover extended components. For instance, we notice that a fraction $\sim 40-50\%$ of the flux density is lost as a consequence of the chosen \textit{uv}-range of $2-17 \; {\rm k\lambda}$ (set C); when including shorter baselines, such effect is mitigated and flux density measurements from sets A and B are consistent within errors, as well as the fitted spectral indices. Therefore, we rejected the spectra of the wings obtained from set C, and we will consider those obtained from set A as reference. The corresponding spectral indices are $\alpha_{170-6063}^{\rm A,ew}=0.69\pm 0.04$ (solid green line) and $\alpha_{170-6063}^{\rm A,ww}=0.79\pm 0.04$ (solid orange line) for the eastern wing and western wing, respectively (Fig. \ref{fig: spettro integrato}). 

By means of the high-resolution ($4''$) images provided by set D, we obtained the spectra of the core and jets. Flux densities of the core are the peak values obtained by fitting a Gaussian component to its surface brightness, while flux densities of the jets were measured in regions encompassing the $256\sigma$ level of the X-band image (see Fig. \ref{fig: spettro integrato}). While the core is not distinguished at 700 MHz, we derived a flat spectral index $\alpha_{2988-9000}^{\rm D,core}=0.290\pm 0.004$ at 2988-9000 MHz (solid black line). As highlighted in Fig. \ref{fig: profili brillanza}, the southern (approaching) jet is brighter than the northern (receding) jet. Single power-laws do not reproduce all data points; therefore, we fitted data points up to 6063 MHz, and obtained $\alpha_{700-6063}^{\rm D,nj}=0.59\pm 0.04$ for the northern jet (solid cyan line) and $\alpha_{700-6063}^{\rm D,sj}=0.53\pm 0.03$ for the southern jet (solid pink line). The two spectral index measurements are consistent within errors. 

As a useful comparison with XRGs in the literature, we computed the total radio power at 1.4 GHz as $P_{\rm 1400}= 4 \pi D_{\rm L}^{2}S_{\rm 1400}(1+z)^{{\alpha-1}}=(1.9\pm 0.1)\times 10^{25} \; {\rm W \; Hz^{-1}}$, by considering the average spectral index $\alpha_{170-6063}^{\rm A,tot}$. This value is consistent with typical radio powers measured for XRGs \citep[e.g.][]{cheung09,landt10}, intermediate between those of FRI and FRII galaxies.

\subsection{Constraining the break frequency}
\label{sect: Constraining the break frequency}

The integrated spectra that we obtained in Sect. \ref{sect: Integrated spectra} show hints of curvature above 6 GHz for the active lobes and jets, and above 700 MHz for the wings. However, flux density losses associated with the sparse \textit{uv}-coverage of our JVLA data need to be discussed in detail. Indeed, obtaining unbiased constraints on the break frequency $\nu_{\rm b}$ is crucial, as it provides a direct measure of the radiative age of a radio source (for a fixed magnetic field value, as further discussed in Sect. \ref{sect: Spectral age and magnetic field}). To this aim, we followed the procedure described in Sect. \ref{sect: Flux density measurements}. First, we assumed values of $S_{\rm inj}$ as obtained by extrapolation at higher frequencies of our spectra under the hypothesis that no break is present. Second, based on the approximate geometry and angular size of the different components, we produced models of an elliptically-symmetric source of semi-axes $ 27''\times42'' $ ($\hat{\sigma} = 9''\times14'' $) and a spherically-symmetric source of radius $R= 45''$ ($\hat{\sigma} = 15''$) mimicking the lobes and the wings, respectively. Finally, the corresponding mock plus real visibilities were imaged with the same \textit{uv}-coverage as in set C for the lobes and set B for the wings (Table \ref{tab: parametriperspettri}). We stress that a simple Gaussian profile is not physically representative of the lobes of radio galaxies, but our procedure still provides useful estimates of flux density losses due to missing short baselines, regardless of the actual morphology and surface brightness profile of extended objects.

We measured the flux density of the mock emission and compared to $S_{\rm inj}$ to constrain the losses. At X-band, we found a loss fraction of $\xi_{\rm loss}\sim 15\%$ for the main lobes. We notice that if losses were $\xi_{\rm loss}\sim 20\%$, the spectrum of the lobes would be well described by a single power-law up to 9 GHz. On the other hand, even the spectrum of the jets deviates from a single power-law at 9 GHz, although they are less affected by losses being smaller and brighter. Therefore, the observed curvature in the spectra of the active lobes and jets is likely genuine, and we do not expect a break at $\nu \gg 9$ GHz.

At C-band, we found a loss fraction for the wings $\xi_{\rm loss}\sim 8\%$, which is incompatible with $\xi_{\rm loss}\sim 60\%$ required to match the  single power-law at 170-700 MHz. The combined BnC and CnD configuration datasets in C-band provide an adequately sampled \textit{uv}-coverage at short spacings and negligible flux density losses, which decrease to $\xi_{\rm loss}\sim 3\%$ when the shortest baselines (as for set A) are included. The high sensitivity of our C-band dataset allowed us to measure the in-band spectral index and to search for evidence of a break at 4-8 GHz. We imaged four sub-bands of width $\sim 1$ GHz each with a common \textit{uv}-range (0.75-36 k$\lambda$), convolved the images to a resolution of $5''$, and obtained the spectrum that is shown in Fig. \ref{fig: inbandspec} (set E in Tables \ref{tab: parametriperspettri}, \ref{tab: fitspix}). The spectra of the wings can be described by very steep single power-laws at 4.7-7.4 GHz ($\alpha=2.29 \pm 0.09$ for the eastern wing, $\alpha=1.66 \pm 0.12$ for the western wing), thus confirming that the corresponding break frequency is $\nu_{\rm b}<4.7$ GHz.

At L-band, we estimate a high $\xi_{\rm loss}\sim 20\%$ for the wings, which prevents us from deriving a trustworthy in-band spectrum. Moreover, by correcting the flux density of the wings by this amount, $S_{1616}$ is perfectly aligned with the 170-700 MHz power-law. As a further sanity check, we produced a 1360 MHz image from a single spectral window, and compared the measured flux densities (without any correction) with those from the 1367 MHz RACS-mid image (Sect. \ref{sect: RACS data}). Notably, in spite of a much denser inner \textit{uv}-coverage of RACS and different noise sensitivities, the flux densities are consistent within $5\%$. This evidence provides a solid lower limit on $\nu_{\rm b}\geq1.4$ GHz; furthermore, this may suggest that the observed deviations from the power-law at 1616 MHz are not completely driven by missing flux density, but that the spectrum intrinsically steepens at $\sim 1.5$ GHz. 

In summary, we found $\nu_{\rm b}\gtrsim 9$ GHz for the active lobes and jets, and we constrained the break frequency of the wings within the range 1.4-4.7 GHz. Although tighter constraints require deeper L-band and S-band observations of the wings, it is likely that $\nu_{\rm b}\sim 1.5$ GHz, as we have independent indications from RACS on the sufficient \textit{uv}-coverage of our L-band JVLA data. In the next Sections we will provide refined, spatially-resolved, spectral measurements, and derive the radiative age of MRC 2011-298.

\subsection{Resolved spectral analysis}
\label{sect: Resolved spectral analysis}

\begin{table}
\centering
   		\caption{Summary of the parameters used to produce images for the spectral index maps in Fig. \ref{fig: spixmap}. Col. 1: considered frequencies. Col. 2: \textit{uv}-range. Col. 3: common circular beam used to convolve each image.}
    \label{tab: spixparam}   
    \begin{tabular}{ccc}
   	\hline
   	\noalign{\smallskip}
   	$\nu$ & \textit{uv}-range & $\theta$ \\  
	 (MHz) & (${\rm k}\lambda$) &  (arcsec) \\  
   	\noalign{\smallskip}
  	\hline
   	\noalign{\smallskip}
  170, 367 & [0.2, 17] & 18 \\
  170, 700 & [0.2, 17] & 18 \\
  367, 700 & [0.2, 30] & 12 \\
  367, 1616 & [0.9, 21] & 12 \\
  700, 6063 & [0.5, 66] & 6 \\
  1616, 6063 & [0.9, 21] & 9 \\
  700, 2988 & [1.6, 66] & 4 \\
  6063, 9000 & [1.9, 85] & 3 \\
   	\noalign{\smallskip}
   	\hline
   	\end{tabular}
   	  	\centering
   \end{table}

\begin{figure*}
	\centering

\includegraphics[width=0.49\textwidth]{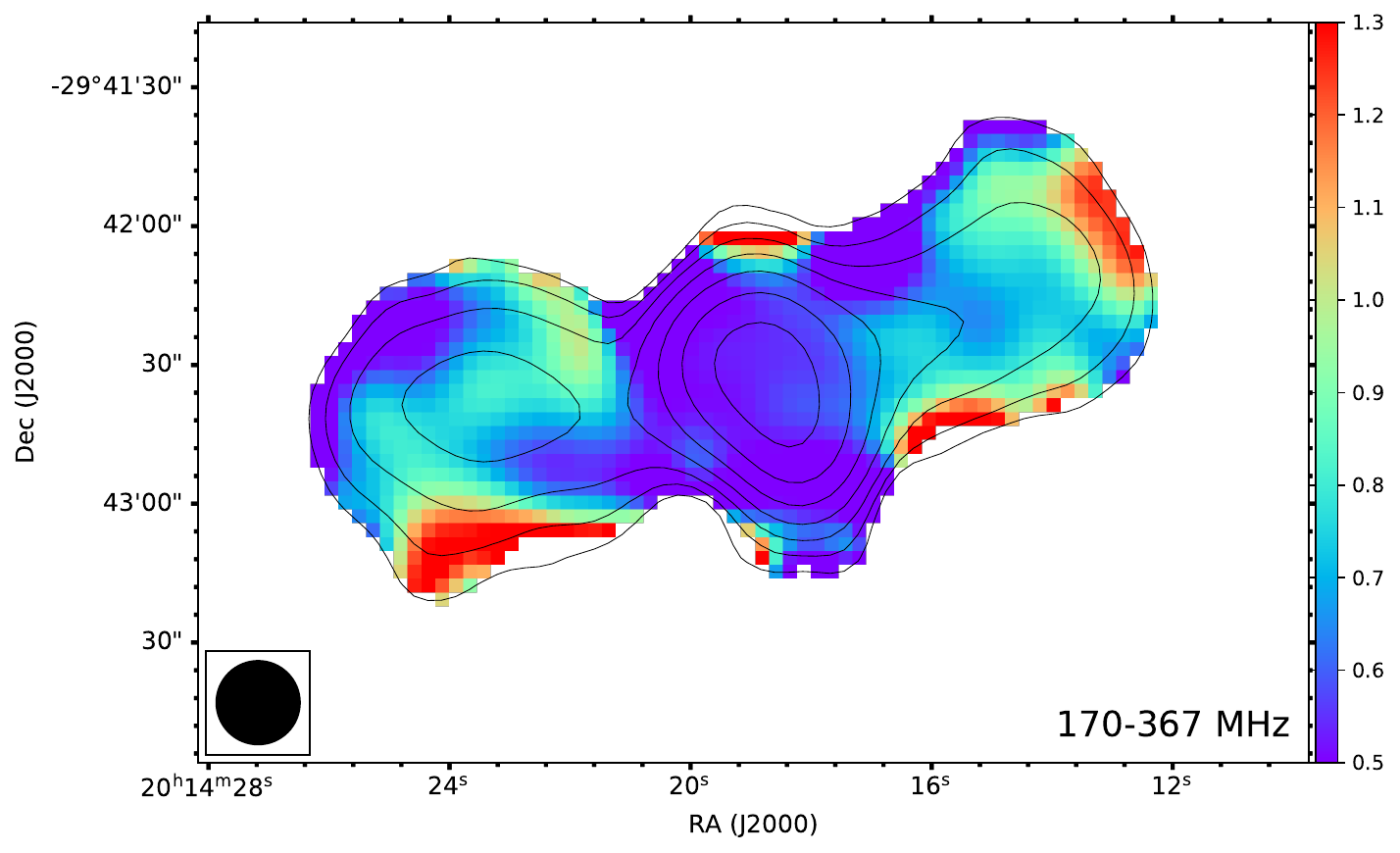}
\includegraphics[width=0.49\textwidth]{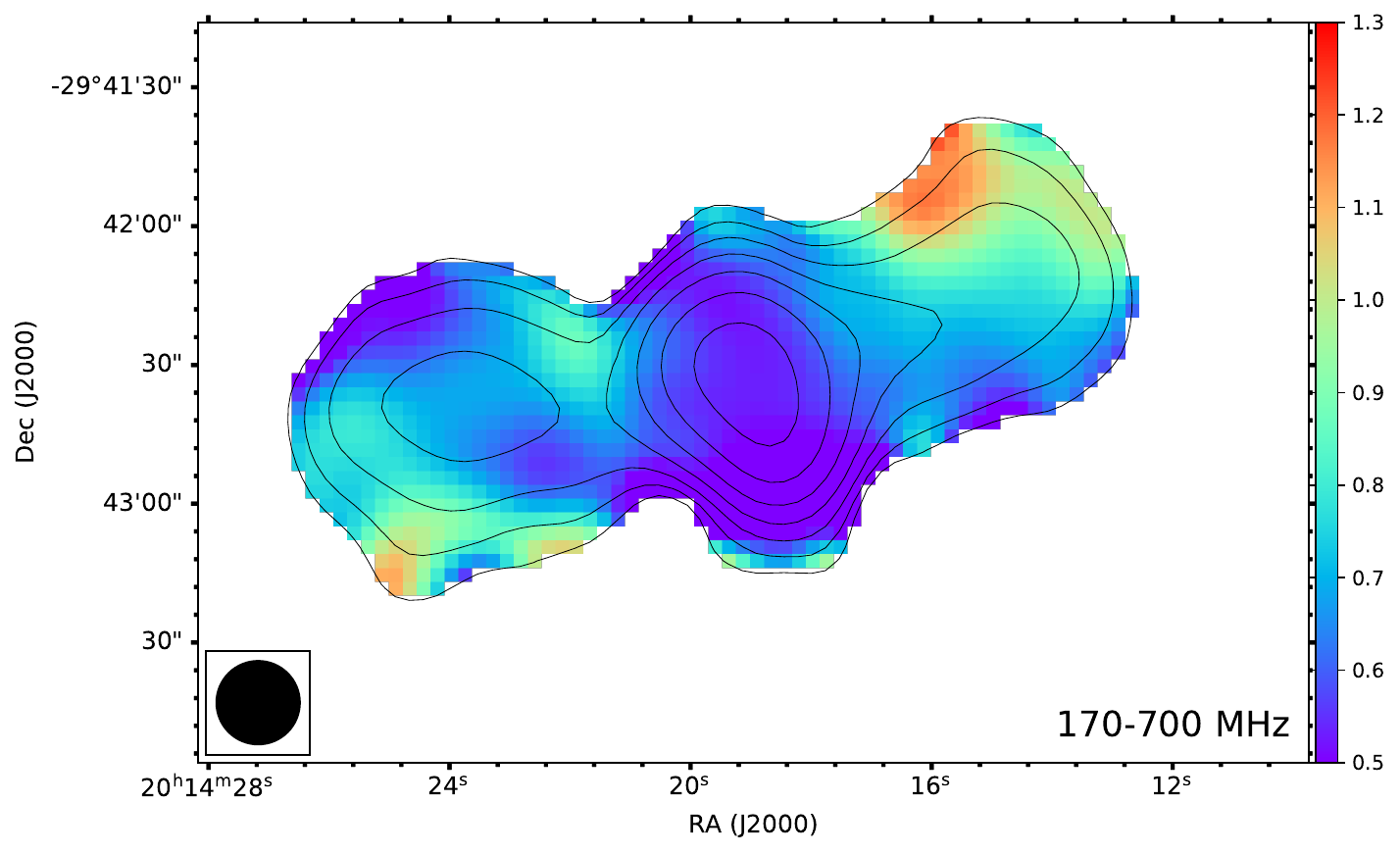}
\includegraphics[width=0.49\textwidth]{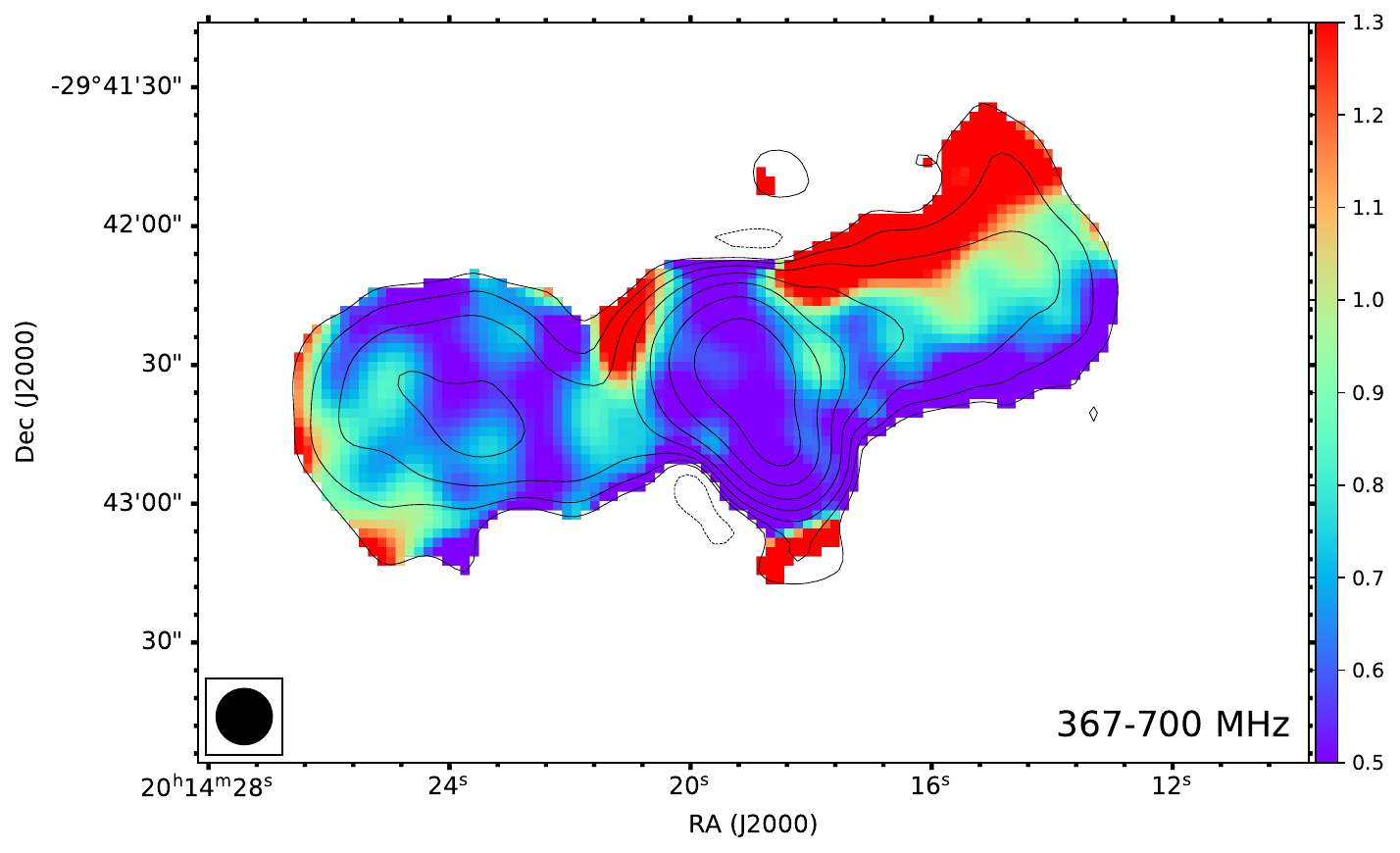}
\includegraphics[width=0.49\textwidth]{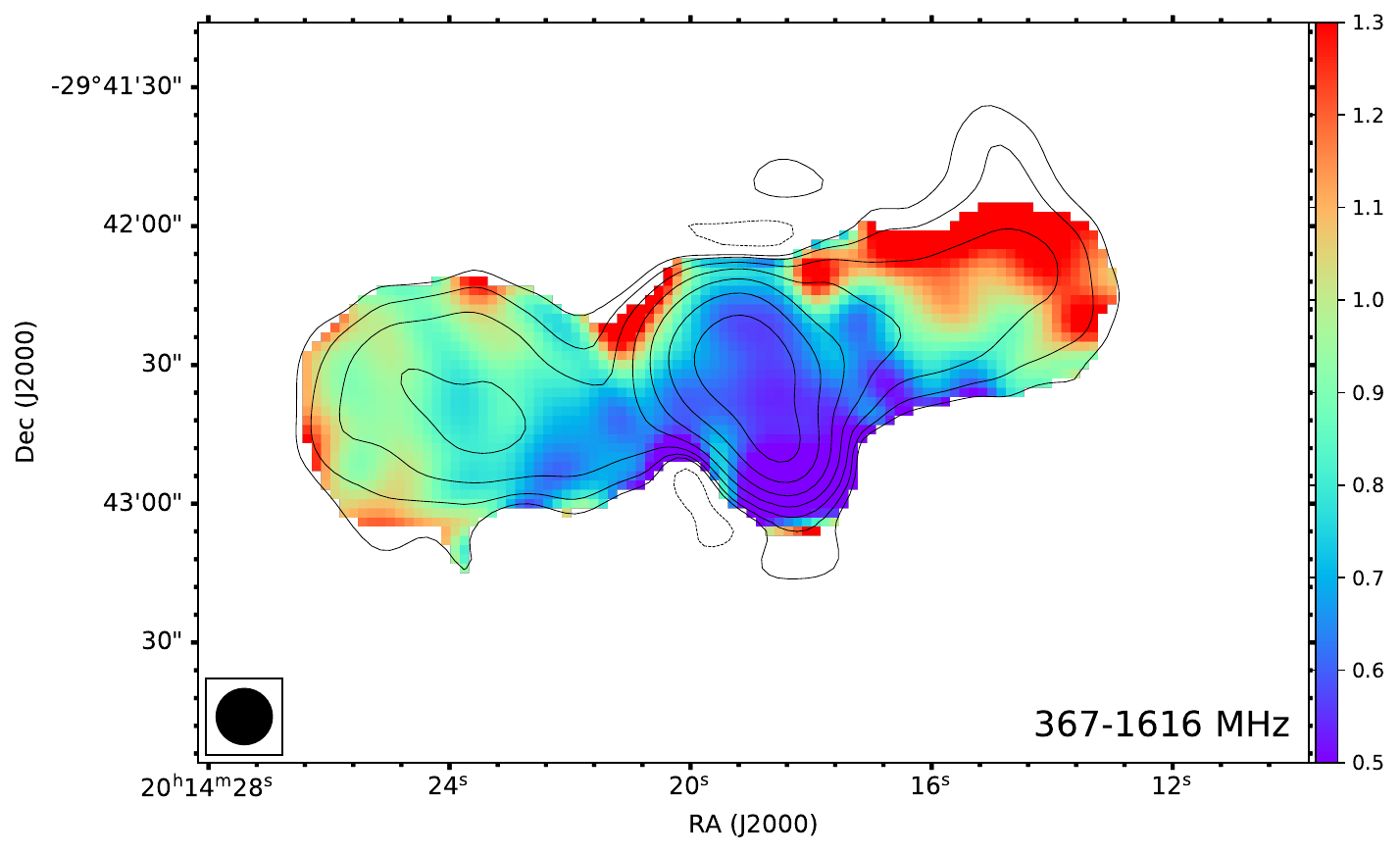}
\includegraphics[width=0.49\textwidth]{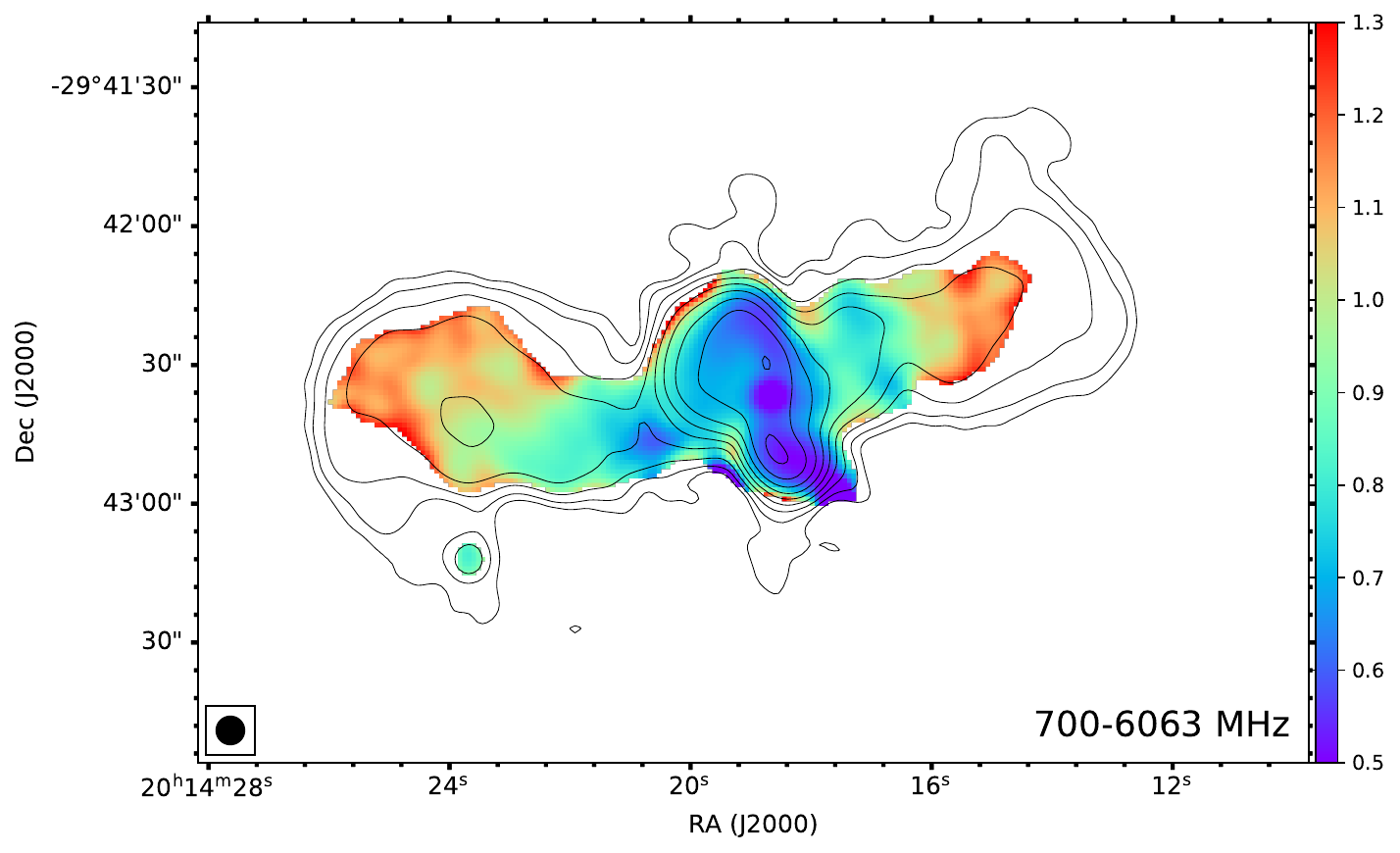}
\includegraphics[width=0.49\textwidth]{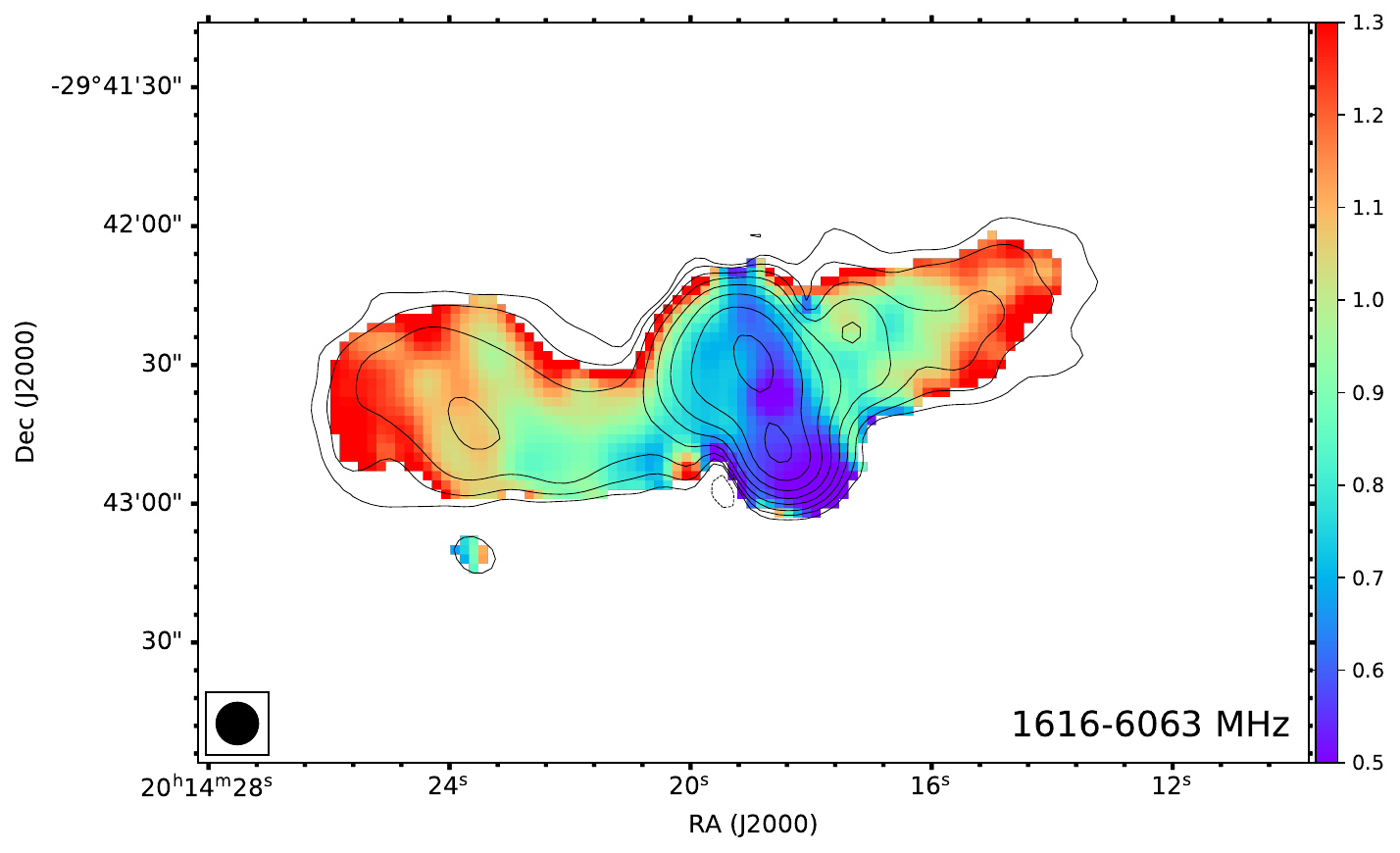}
\includegraphics[width=0.40\textwidth]{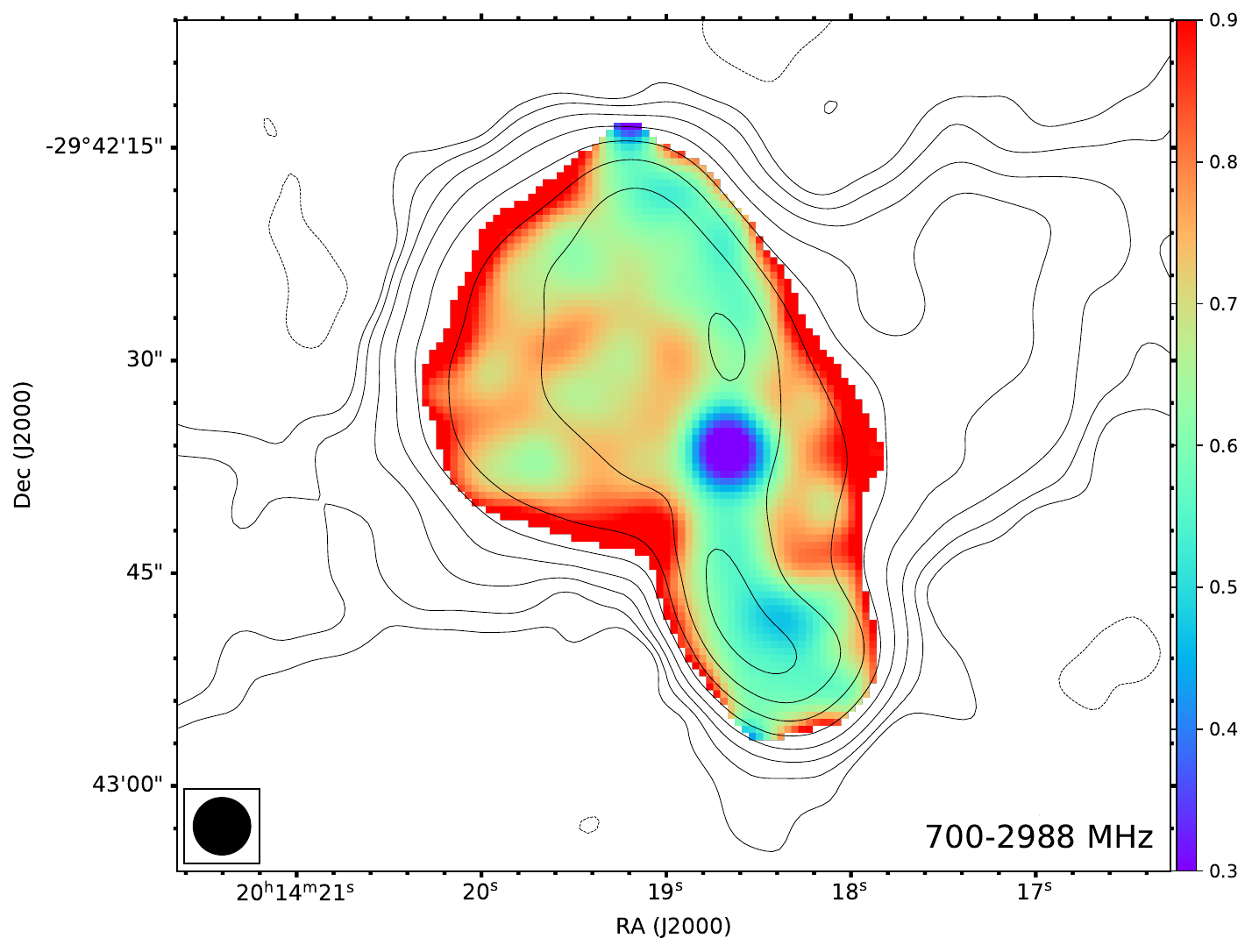}
\includegraphics[width=0.40\textwidth]{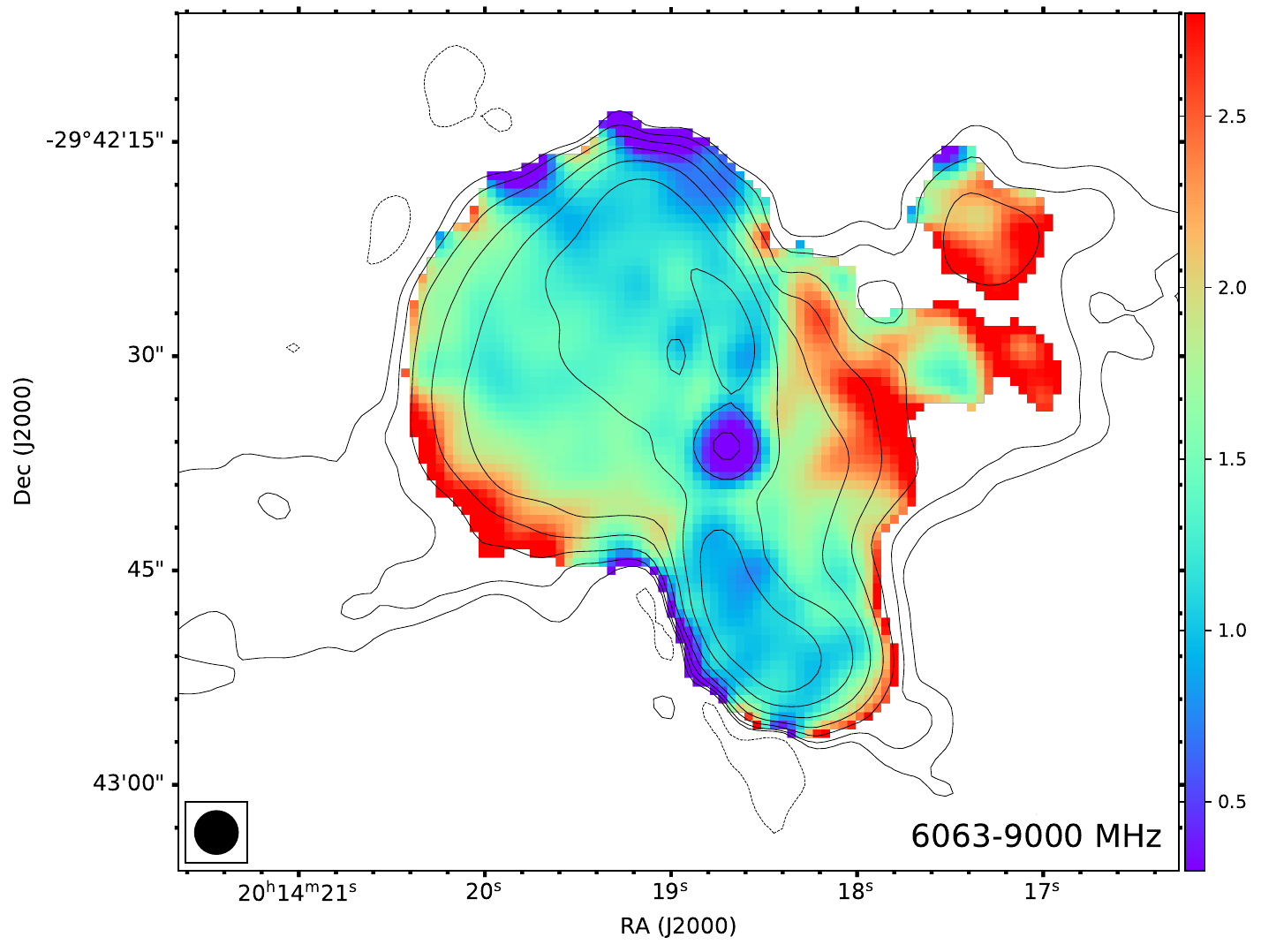}
	\caption{Spectral index maps of MRC 2011-298 for different frequency pairs and resolution. The overlaid contour levels are $[\pm5, \;10, \;20,\; 40,\; ...]\times \sigma$ from the lower frequency image. {\it From top left to bottom right}: 170-367 MHz map at $18''$, 170-700 MHz map at $18''$, 367-700 MHz map at $12''$, 367-1616 MHz map at $12''$, 700-6063 MHz map at $6''$, 1616-6063 MHz map at $9''$, 700-2988 MHz map at $4''$, 6063-9000 MHz map at $3''$. The corresponding error maps are shown in Fig. \ref{fig: errspixmap}.}
	\label{fig: spixmap}
\end{figure*}

\begin{figure*}
	\centering

\includegraphics[width=0.40\textwidth]{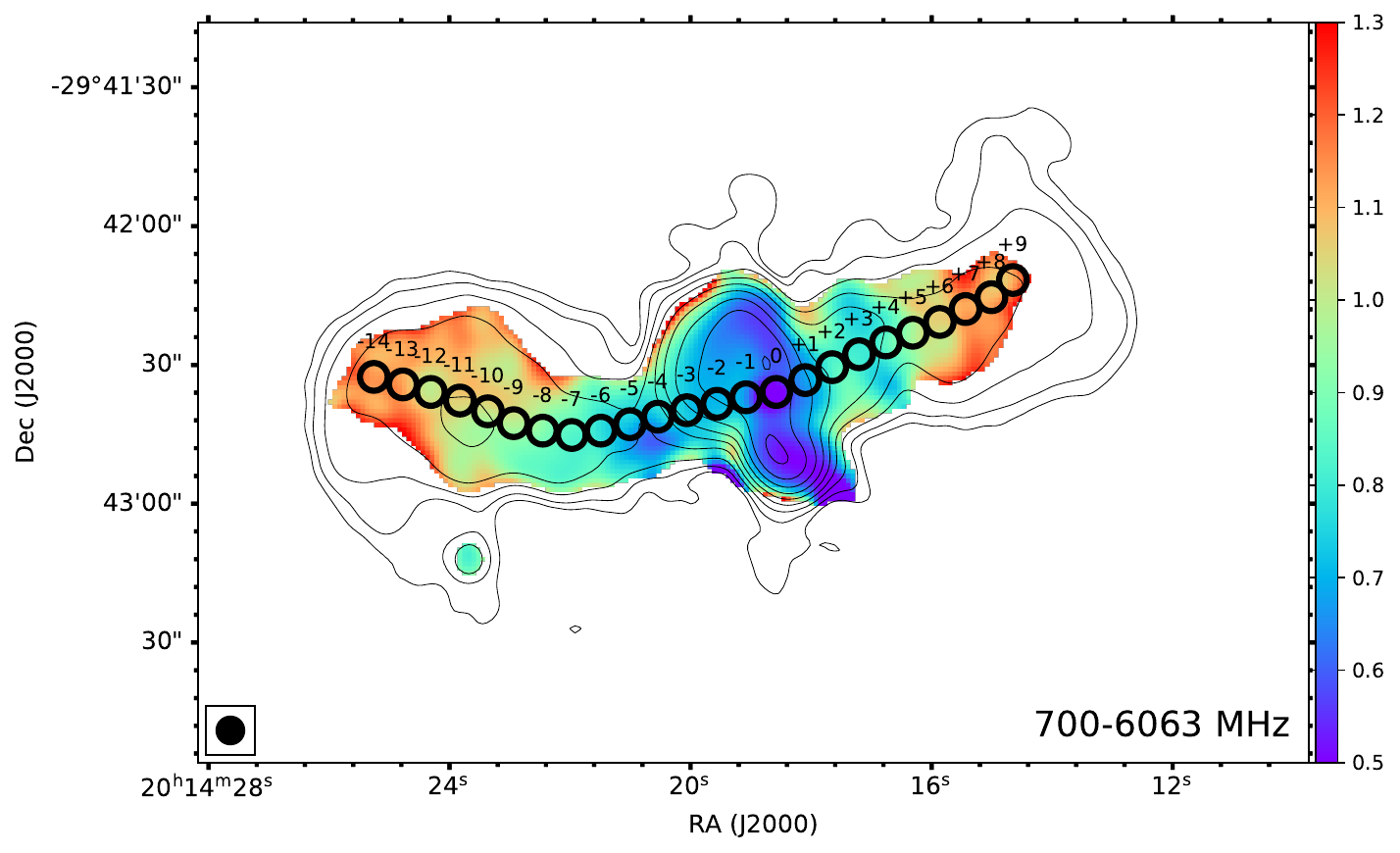} \\
\includegraphics[width=0.45\textwidth]{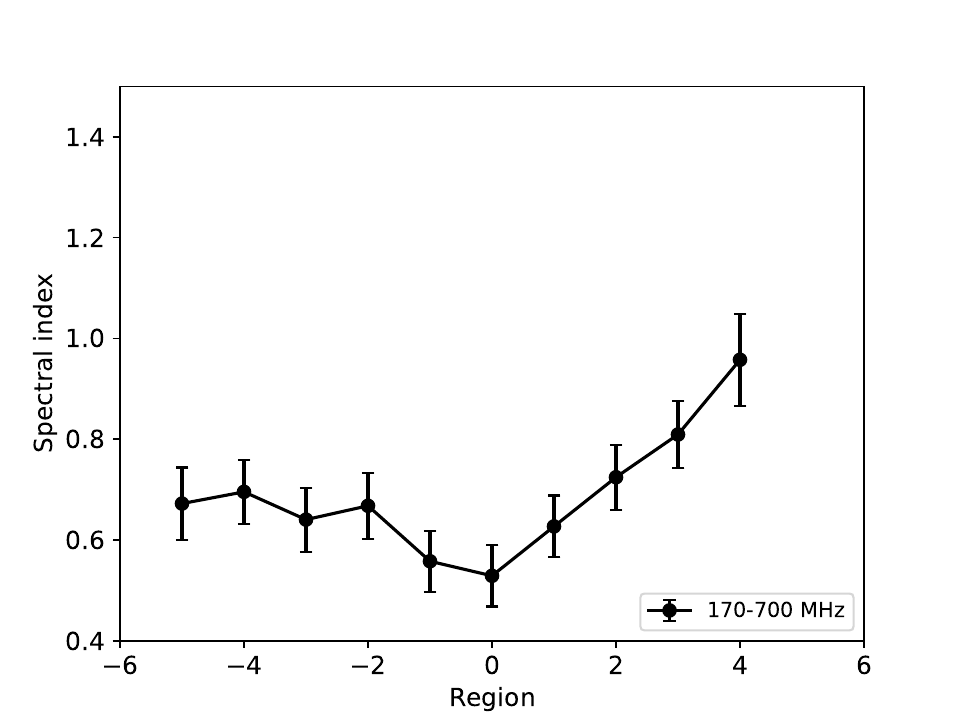} 
\includegraphics[width=0.45\textwidth]{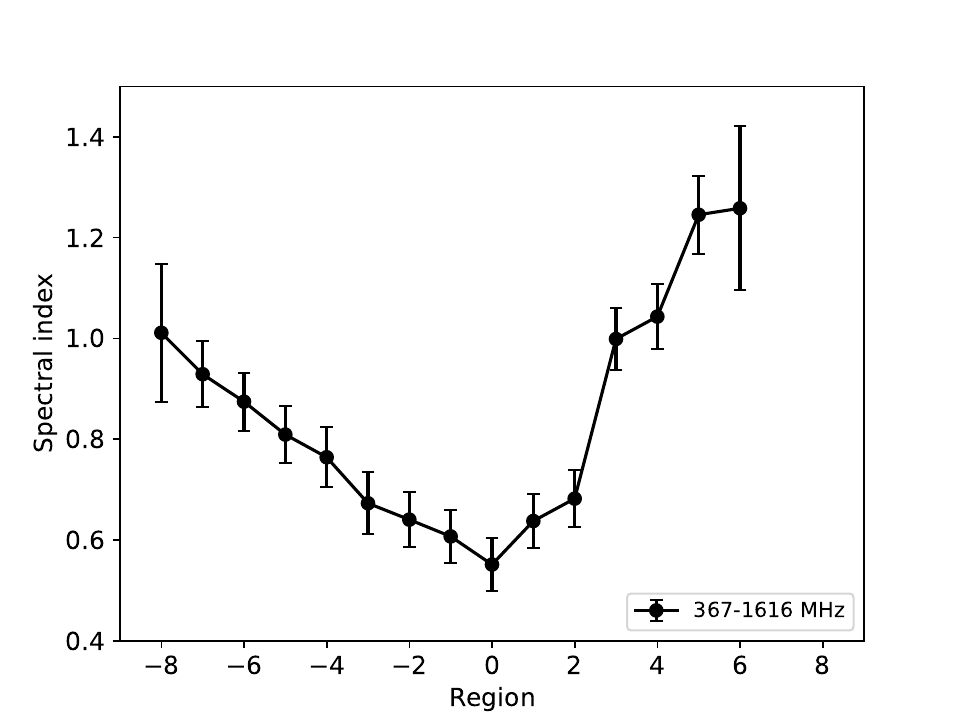} 
\includegraphics[width=0.45\textwidth]{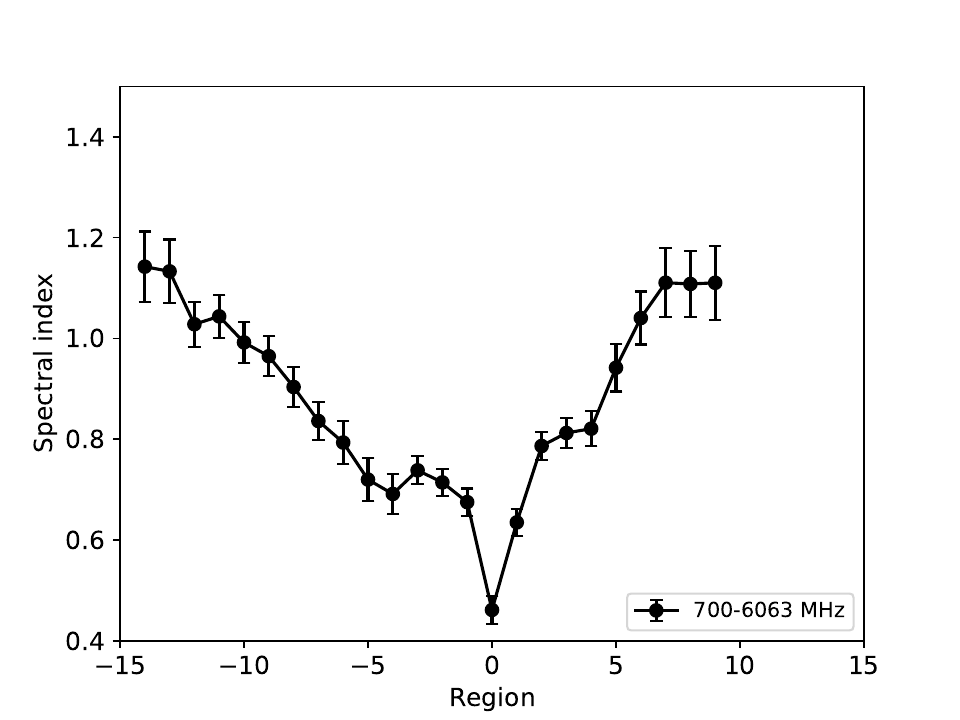} 
\includegraphics[width=0.45\textwidth]{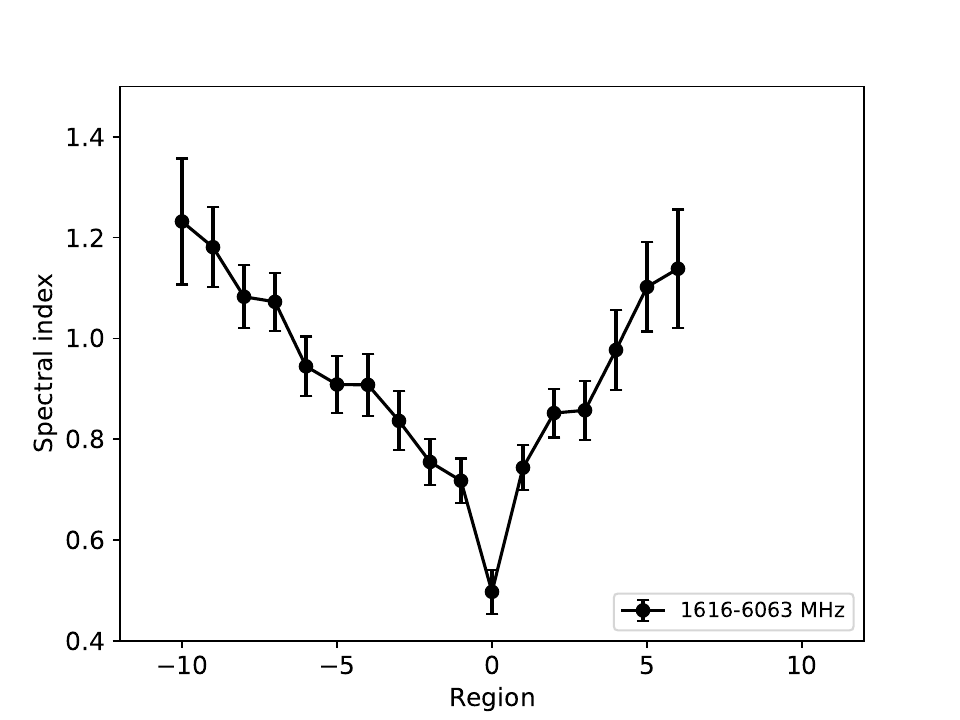} 
 \caption{Spectral index profiles along E-W direction computed from maps in Fig. \ref{fig: spixmap}. The considered spectral index maps are indicated in the legend of each profile. An example of the beam-size (Table \ref{tab: spixparam}) sampling circles is shown in the top panel for the 700-6063 MHz map only. The sampling regions are numbered from `0' (the location of the core), decrease towards east, and increase towards west.}
	\label{fig: spixprofile}
\end{figure*}

\begin{figure*}
	\centering

\includegraphics[width=0.32\textwidth]{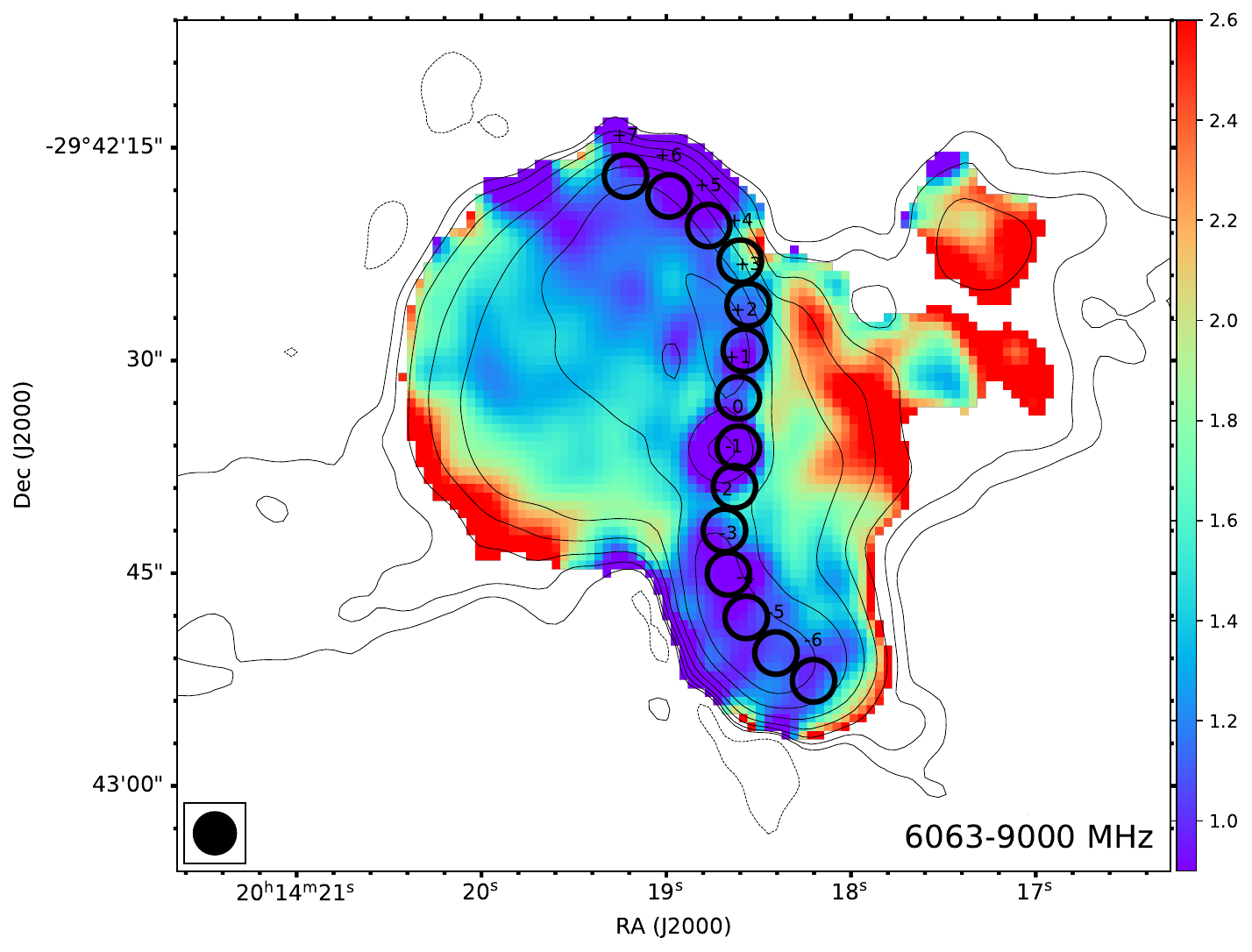} \\
\includegraphics[width=0.45\textwidth]{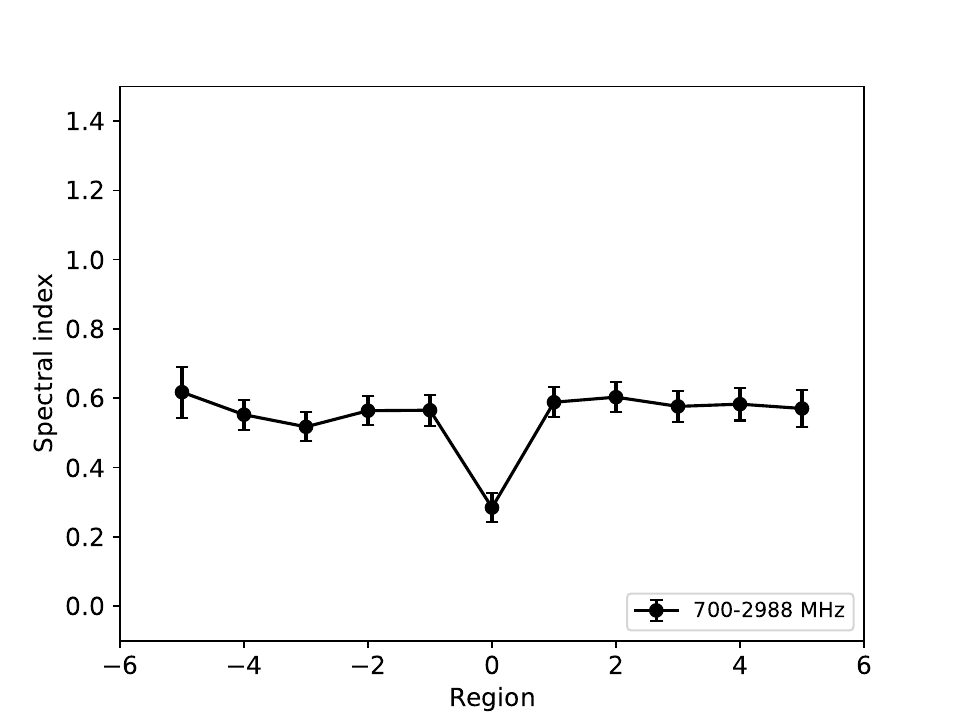} 
\includegraphics[width=0.45\textwidth]{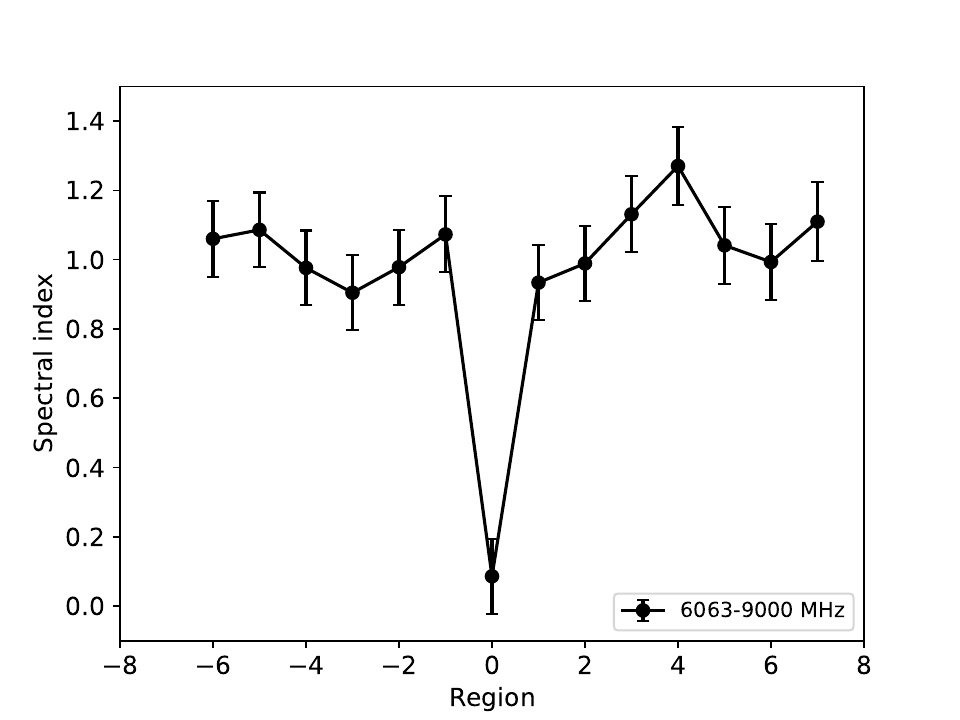}
	\caption{Spectral index profiles along N-S direction computed from maps in Fig. \ref{fig: spixmap}. The considered spectral index maps are indicated in the legend of each profile. An example of the beam-size (Table \ref{tab: spixparam}) sampling circles is shown in the top panel for the 6063-9000 MHz map only. The sampling regions are numbered from `0' (the location of the core), decrease towards south, and increase towards north.}
	\label{fig: spixprofile2}
\end{figure*} 

In this Section we present spectral index maps obtained by combining radio images at different frequencies, which have been produced with a common \textit{uv}-range and convolved to the same resolution. Pixels below a threshold of $5\sigma$ were masked. A summary of the imaging parameters is reported in Table \ref{tab: spixparam}. Our spectral index maps and corresponding errors are shown in Fig. \ref{fig: spixmap} and Fig. \ref{fig: errspixmap}, respectively.

In all the maps, the active lobes exhibit a flatter spectral index ($\alpha\sim 0.5-0.6$) than the wings ($\alpha\sim 0.7-1.5$), thus confirming the general conclusions from the analysis of the integrated spectra (Sect. \ref{sect: Integrated spectra}). Nevertheless, we found an arguable distribution of the spectral index for maps at 170-367 MHz and 367-700 MHz, which show regions having $\alpha\sim 0.3-0.4$ both in the lobes and wings, and alternating patches of flatter and steeper indices. We interpret this trend as driven by inadequate calibration of the shortest baselines of the band-3 dataset (see discussion in Sect. \ref{sect: uGMRT data}) rather than being physical and associated to particle re-acceleration. The comparison with our additional spectral index maps supports this interpretation, as the 170-700 MHz map is not consistent with the 170-367 MHz and 367-700 MHz maps, and the 367-1616 MHz map, which was produced by excluding baselines shorter than $900\lambda$, show more physically reasonable values of $\alpha$. Therefore, although band-3 data are still trustworthy in the context of integrated spectral measurements, the inaccurate calibration of short baselines has clear effects on our resolved spectral analysis (see also the high errors in Fig. \ref{fig: errspixmap} for maps including the 367 MHz data).

\cite{bruno19} reported on a progressive spectral steepening from the centre towards the wings at 1.7-6 GHz, albeit the relatively high associated errors prevented a strong constraint on the spectral index in the outermost regions of the wings. In the present work, we improved the quality of our images, and obtained spectral index maps that unambiguously highlight the spectral steepening with exceptional details. We computed the spectral index profiles that are shown in Figs. \ref{fig: spixprofile}, \ref{fig: spixprofile2} using beam-size (see Table \ref{tab: spixparam}) sampling circles. 

For the first time we confirm the spectral steepening along the wings over an unprecedentedly broad frequency range (Fig. \ref{fig: spixprofile}). Overall, spectral indices of the wings are in the range $\alpha\sim 0.7-1.4$, but the western wing exhibits a more pronounced steepening than the eastern wing, especially at low frequencies (170-700 MHz, 367-1616 MHz). This asymmetry may either be intrinsic, for example associated with electrons having different ages and/or magnetic field inhomogeneities, or arise from a combination of projection effects and resolution (see also Sect. \ref{sect: radiative age}). We notice that the spectral gradient is shallower at low frequencies than at high frequencies. Indeed, low frequency data allow us to trace low-energy particles, which can roughly maintain the initial (injection) spectral distribution for longer timescales.

The bottom panels of Fig. \ref{fig: spixmap} report the 700-2988 MHz and 6063-9000 MHz spectral index maps at $4''$ and $3''$, respectively. We used these maps to compute the spectral index profiles (Fig. \ref{fig: spixprofile2}) along the jets. The profile at 700-2988 MHz shows a constant value of $\alpha \sim 0.6$ for both the northern and southern jets. The average profile at 6063-9000 MHz is significantly steeper, as the spectral index is $\alpha \sim 1$, in line with the 5.5-9 GHz map in \cite{bruno19}; fluctuations of $\alpha$ along the jets are likely caused by projection effects or local variations of the magnetic field strength. At such high-resolution, the spectral distribution within the active lobes appears patchy, but this is likely caused by the different density of the \textit{uv}-coverage of the data.

\subsection{Radiative age}
\label{sect: radiative age}

\begin{figure*}
	\centering

\includegraphics[width=0.48\textwidth]{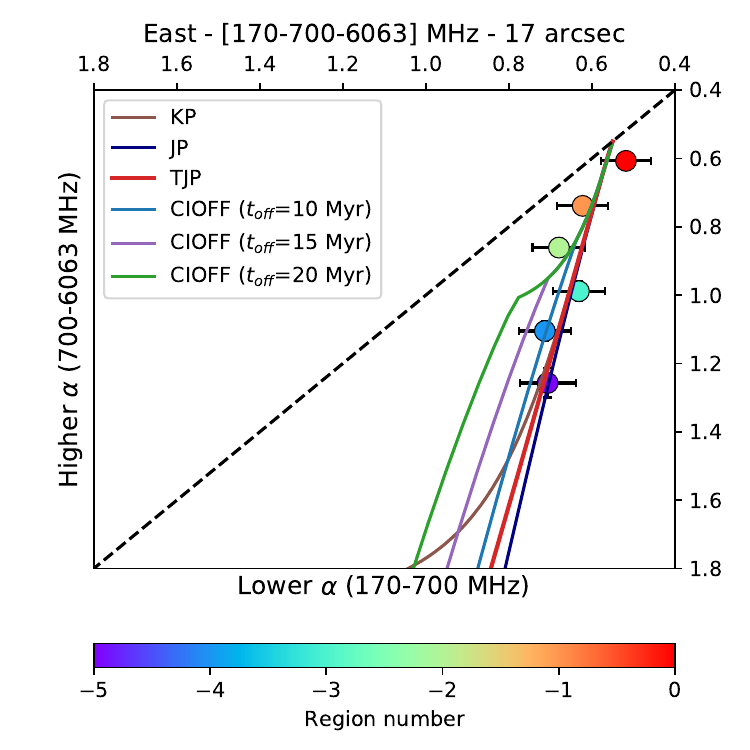}
\includegraphics[width=0.48\textwidth]{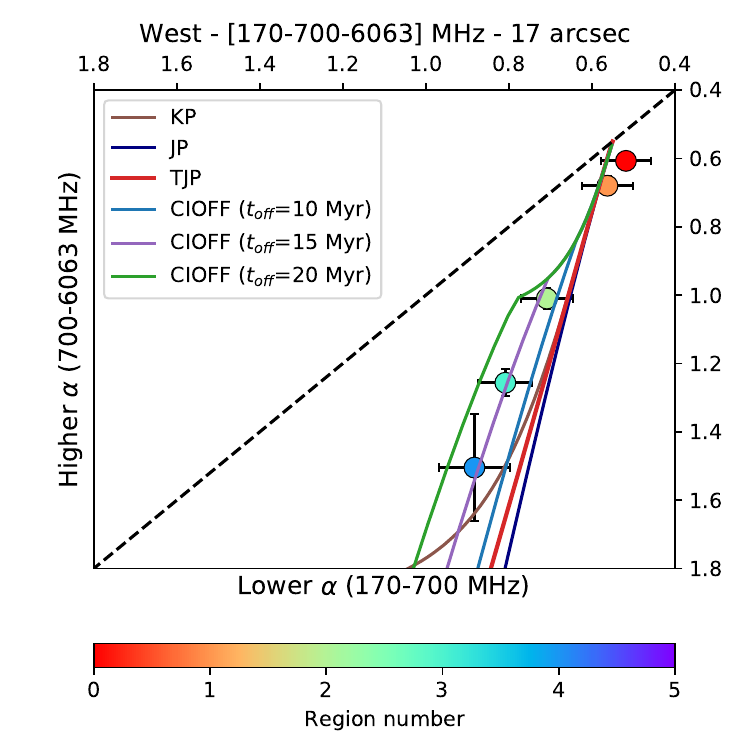}
\includegraphics[width=0.48\textwidth]{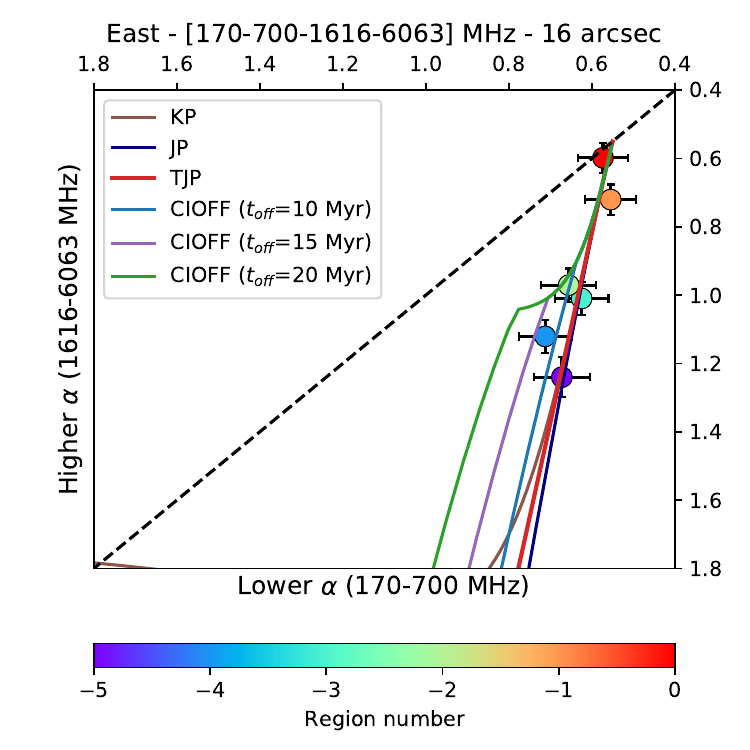} 
\includegraphics[width=0.48\textwidth]{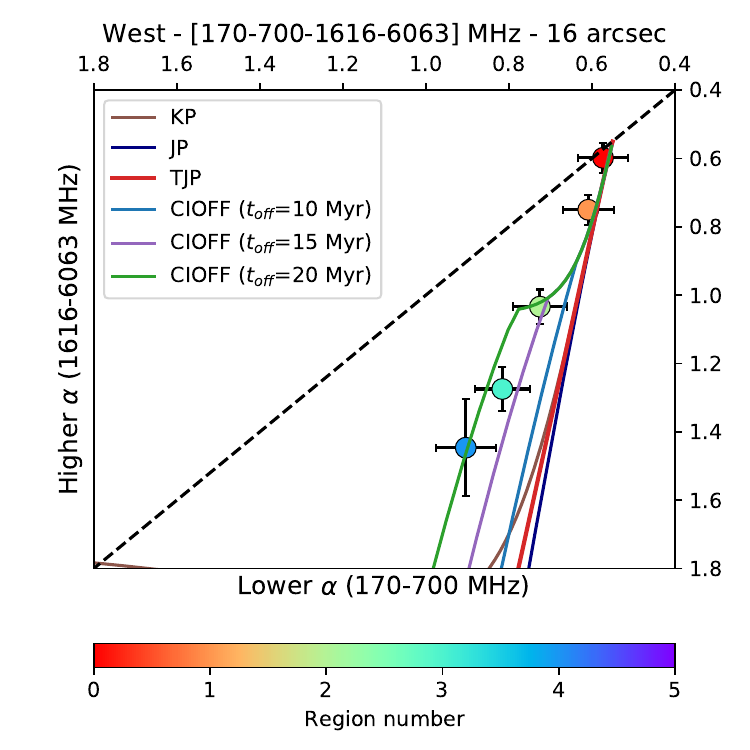} 
	\caption{Radio colour-colour diagrams. Data points are coloured based on the sampling (beam-size) region identifier as in Fig. \ref{fig: spixprofile}. The black dashed line (one-to-one line) indicates a power-law with $\alpha=\Gamma$. The solid lines are the theoretical KP (brown), JP (blue), TJP (red), CIOFF (cyan, purple, green for $t_{\rm off}=10,\; 15,\; 20$ Myr, respectively) ageing curves obtained with $\Gamma=0.55$ and $B_{\rm 0}=2.4 \; {\rm \mu G}$. The direction (eastward or westward) of the sampling with respect to region `0' (centred on the radio core), frequency pairs, and resolution are indicated on top of each panel. }
	\label{fig: ccp}
\end{figure*} 

\begin{figure*}
	\centering

\includegraphics[width=0.49\textwidth]{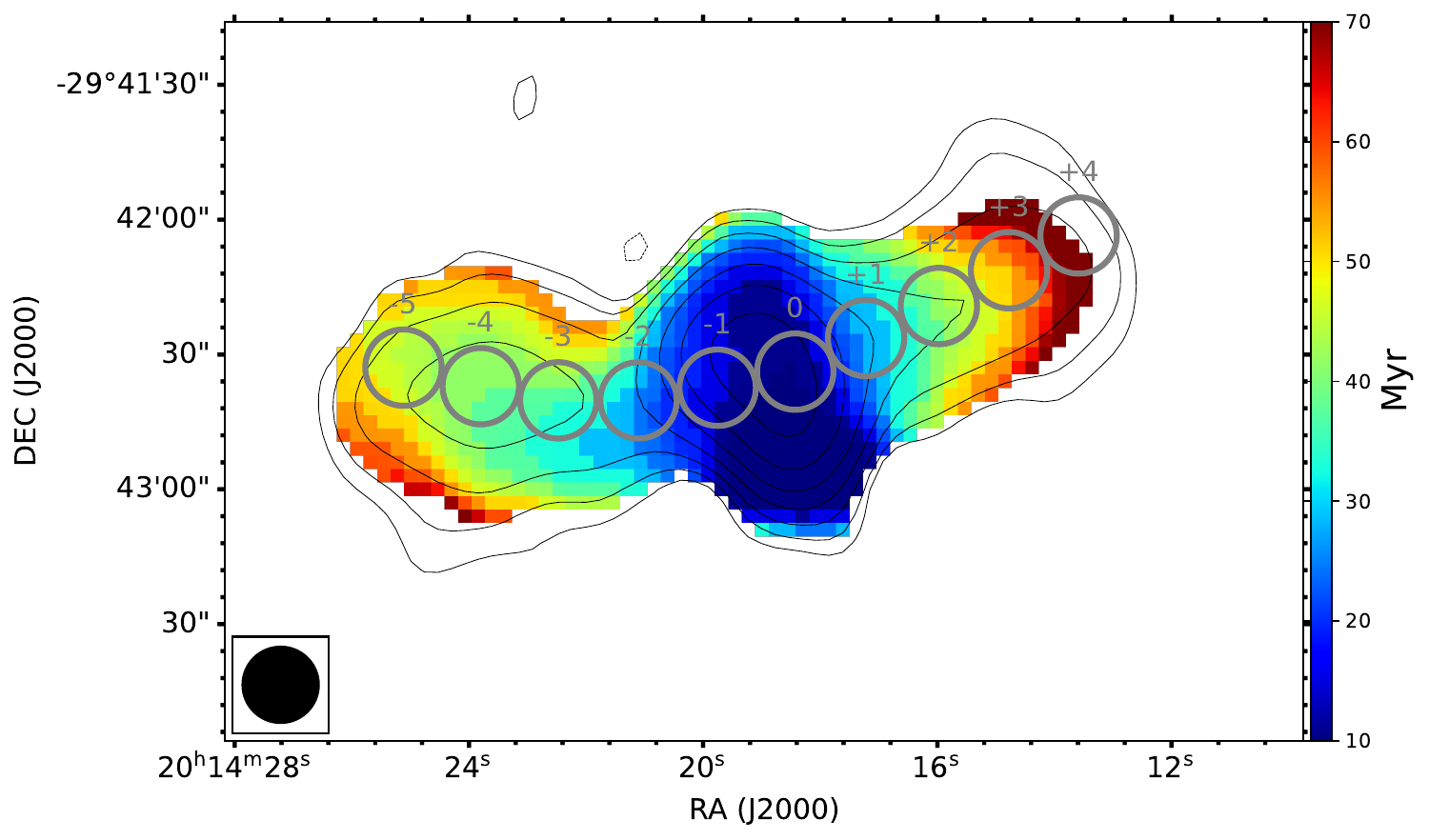}
\includegraphics[width=0.49\textwidth]{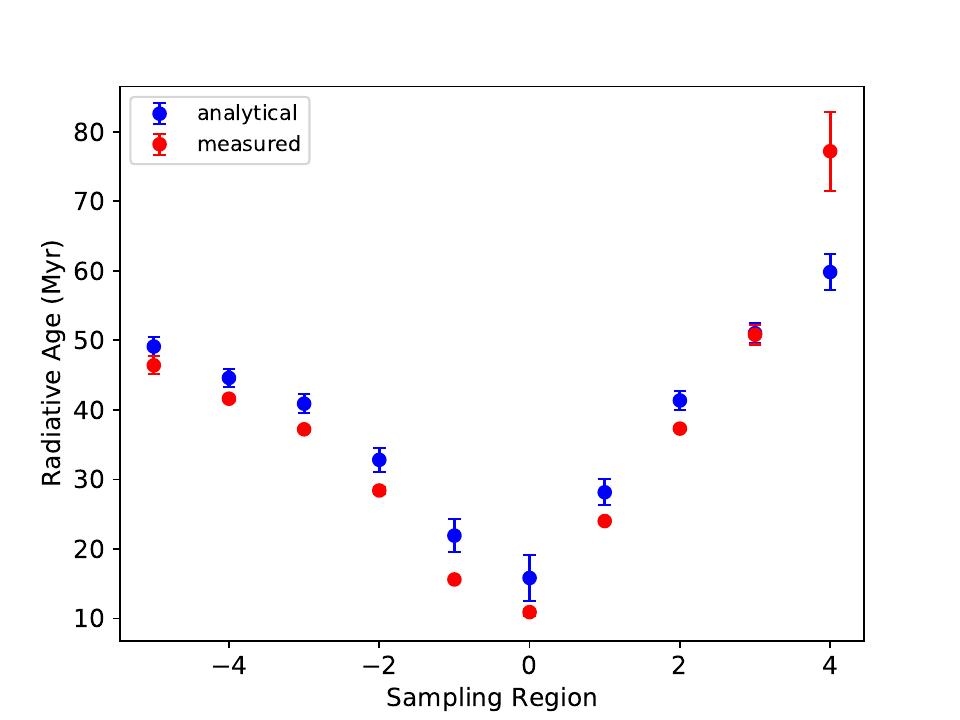}
	\caption{Radiative age of MRC 2011-298 at $17''$. \textit{Left}: Age map computed by fitting 170-700-1616-6063 MHz images with a TJP model ($\Gamma=0.55$, $B_{\rm 0}=2.4 \; {\rm \mu G}$). Typical errors are $\lesssim 5$ Myr. The overlaid contour levels are $[\pm5, \;10, \;20,\; 40,\; ...]\times \sigma$ from the 170 MHz image. \textit{Right}: Age profile obtained as the median within the grey circles in the left panel. Red data points are measured from the age map. Blue data points are obtained through Eq. \ref{eq: tcoolradio2} from the 700-6063 MHz spectral index profile.  }
	\label{fig: agemapTJP}
\end{figure*}

Relativistic particles ejected from the core of radio galaxies progressively age via synchrotron and inverse Compton losses. Their spectral evolution depends on the form of the initial electron energy distribution, that is $N(E)\propto E^{-(2\Gamma+1)}$, where $\Gamma$ is the injection spectral index (if Fermi I acceleration mechanism is assumed), and the spatial distribution of the magnetic field $B(r)$. Standard ageing models assuming a single injection event are the Kardashev-Pacholczyk \citep[KP;][]{kardashev62,pacholczyk70}, Jaffe-Perola \citep[JP;][]{jaffe73}, and Tribble-Jaffe-Perola \citep[TJP;][]{tribble93}. Both the KP and JP models assume a uniform magnetic field $B_0$. In the KP model, the pitch angle $\theta_{\rm p}$ is assumed to be constant and isotropic throughout the entire particle lifetime, thus allowing energetic electrons with small $\theta_{\rm p}$ to live indefinitely if emitting exclusively via synchrotron. In the JP model, $\theta_{\rm p}$ can be considered isotropic on short time scales only due to electron scattering, therefore a time-averaged pitch angle is assumed, which leads to an exponential cut-off at high energies in the electron distribution \citep[e.g.][]{hardcastle13}. The TJP model assumes a time-averaged $\theta_{\rm p}$ as for the JP model, but additionally considers Gaussian spatial fluctuations of the magnetic field around $B_{0}$; this setup produces a shallower high-energy cut-off, thus making electrons live longer for the TJP than for the JP model. In contrast to a single injection event, the Komissarov-Gubanov \citep[KGJP or CIOFF;][]{komissarov&gubanov94} model assumes the continuous injection of fresh particles for a prolonged time (up to the switch-off time $t_{\rm off}$), followed by a passive JP-like ageing; therefore, the CIOFF takes into account the mixing of particles with different ages.

In this Section we test the KP, JP, TJP, and CIOFF ageing models. We assume a value for the magnetic field that minimises the radiative losses and maximises the lifetime of the source, which is $B_0=B_{\rm CMB}/\sqrt{3}=2.4 \; \mu{\rm G}$, where $B_{\rm CMB}=3.25(1+z)^2 \; \mu{\rm G}$ is the equivalent magnetic field of the cosmic microwave background. We assume $\Gamma=0.55$ as derived in \cite{bruno19} through the {\tt findinject} task of the Broadband Radio Astronomy ToolS ({\tt BRATS\footnote{\url{https://www.askanastronomer.co.uk/brats/}};} \citealt{harwood13,harwood15}) software, which fits the observed radio spectrum and provides the injection index that minimises the $\chi^2$ distribution for the fitted model. For the CIOFF model, we test different values of the switch-off time, namely $t_{\rm off}=10,\; 15,\; 20$ Myr. We compared the ageing curves with the observed spectral distribution in radio colour-colour diagrams (RCCDs; \citealt{katz-stone93}). These are diagnostic plots to analyse the local shape of radio spectra computed from two pairs of frequencies (the spectral index in RCCDs is equivalent to a magnitude difference in optical colour-colour diagrams), regardless of the magnetic field and adiabatic expansion or compression. In RCCDs, the one-to-one line represents a power-law spectrum with $\alpha=\Gamma$, while data points lying below ($\alpha>\Gamma$) are subjected to ageing. We sampled the images discussed in Sect. \ref{sect: Integrated spectra} (170, 700, 6063 MHz from set A, and 170, 700, 1616, 6063 MHz from set B) with beam-size regions (similarly to Fig. \ref{fig: spixprofile}), and produced the RCCDs shown in Fig. \ref{fig: ccp} (for inspection purposes, we split the sampling towards east and west directions). 

The eastern spectral profiles can be reproduced by all single-injection ageing curves, whereas data points from the western wing slightly deviate from these models, and they are better described by the CIOFF curves. Such difference between the wings is not surprising, as in Sect. \ref{sect: Resolved spectral analysis} we found asymmetric spectral index profiles, and this is likely driven by projection effects. Specifically, the western wing is more inclined towards the observer, leading to the interception of multiple populations of electrons projected along the line of sight that are better described by a CIOFF model, whereas the eastern wing is closer to the plane of the sky and its spectral shape approaches to the single injection curves. We caution that the considered values of $t_{\rm off}$ should not be used to infer stringent timescales, as accurate model fitting to the spectrum would be necessary to this aim. We also stress that including or excluding L-band data provides roughly consistent results, thus suggesting that flux density losses are not a dominant issue, as outlined in Sect. \ref{sect: Constraining the break frequency}. In summary, we did not identify a preferential ageing model that better reproduces the observed spectral shape of the source. Therefore, we expect a marginal dependency of the estimate of the age (see below) on the considered model. In the following, we will consider the TJP as our reference model.

\cite{bruno19} fitted the spectrum of MRC 2011-298 pixel-per-pixel by using L-band, C-band, and X-band images with {\tt BRATS} to produce a radiative age map of the active lobes. Although fitting the age of the wings was not possible due to non-detection at X-band, the authors considered the following analytical formula, which is based on the spectral index computed between a pair of frequencies (see derivation in Appendix in \citealt{bruno19}):
\begin{equation}
t = \frac{1590}{(1+z)^\frac{1}{2}} \frac{{B}^\frac{1}{2}}{ B^2+B^2_{\rm CMB}}  \left(\frac{\ln{(\nu_{1})}-\ln{(\nu_{2})}}{\nu_{1}-\nu_{2}}\right)^\frac{1}{2} \left( \alpha - \Gamma \right)^\frac{1}{2} \; \; \; ,
\label{eq: tcoolradio_approx}
\end{equation}   
where the age, frequencies, and magnetic fields are expressed in units of Myr, GHz, and ${\rm \mu G}$, respectively. The present work benefits from sensitive low-$\nu$ data, which allow us to fit the spectrum of both lobes and wings, and measure their ages.

In Fig. \ref{fig: agemapTJP} we show the radiative age map (left panel) obtained by fitting images from set B with a TJP model (mean $\chi_{\rm red}^2 = 1.40$) and the corresponding profile (right panel, red data points). As expected based on the observed radial spectral steepening, the age progressively increases from the centre towards the wings. The age of the active lobes is $t\sim 20-30$ Myr, while the western and eastern wings reach maximum ages of $t \sim 80$ Myr and $t \sim 50$ Myr, respectively. This confirms that the wings are radiatively older than the active lobes. Specifically, we found differences of $\Delta t =27 \pm 1$ Myr between the western wing and lobes (regions `3' and `1', at $\sim120$ kpc and $\sim40$ kpc from the core\footnote{Data point `4' is inaccurate as it relies on a few pixels due to the adopted flux density threshold, and we thus excluded it in the age difference calculation.}), and $\Delta t =31 \pm 1$ Myr between the eastern wing and lobes (regions `-5' and `-1', at $\sim200$ kpc and $\sim40$ kpc from the core). Our present data allow us to probe larger distances (and older plasma) from the core, therefore these values are in line with the estimates obtained in \cite{bruno19} through Eq. \ref{eq: tcoolradio_approx} for inner regions of the wings.

The reliability of Eq. \ref{eq: tcoolradio_approx} is also confirmed by the age profile predicted from the 700-6063 MHz spectral index profile (blue data points in the right panel of Fig. \ref{fig: agemapTJP}). Overall, the analytical formula provides solid estimates of the age across the whole source, being higher by a mean factor of $\sim 1.1$ than the corresponding measured values. We caution that using the 170-700 MHz spectral index still provides useful upper limits to the age, but discrepancies with the measured age increase by a mean factor of $\sim 1.3$.  This is expected, as Eq. \ref{eq: tcoolradio2} is based on the assumption that $\nu_{1}$, $\nu_{2} \gtrsim \nu_{\rm b}$, which is not true at such low frequencies.

\section{Discussion}
\label{sect: Discussion}

Throughout this Section we discuss the results of our analysis. We first show how magnetic field variations impact the estimate of the spectral age. We then investigate whether precession of the jets can be responsible for their S-shape. Finally, we discuss literature formation models in the light of our findings for MRC 2011-298.

\subsection{Spectral age and magnetic field}
\label{sect: Spectral age and magnetic field}

\begin{figure}
	\centering
\includegraphics[width=0.48\textwidth]{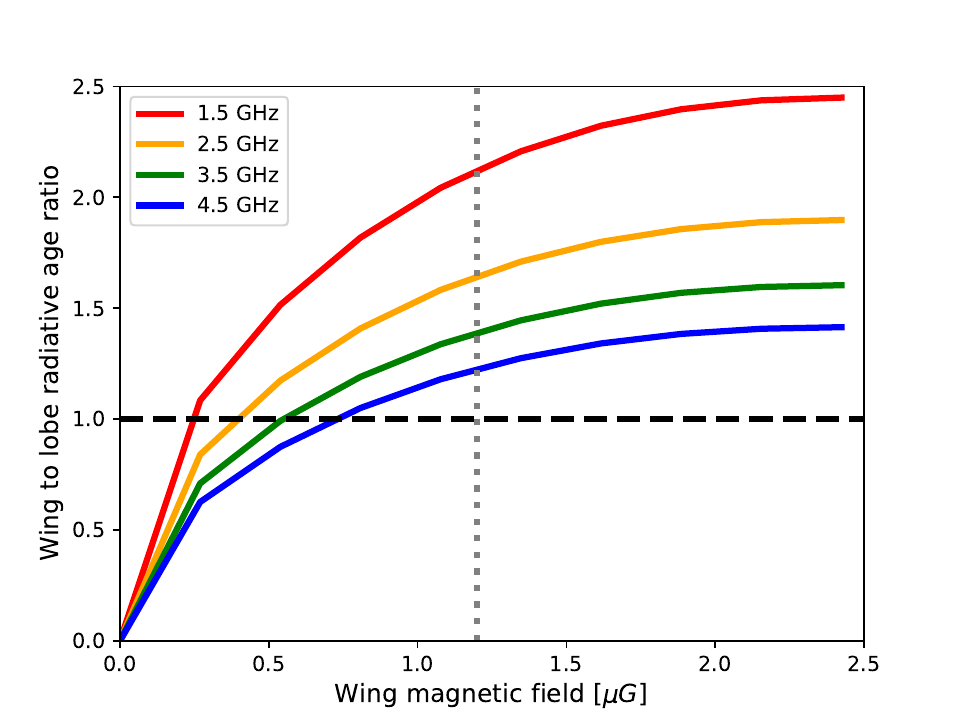}
	\caption{Ratio ($\tau$) of the radiative ages of the wings and lobes computed from Eq. \ref{eq: tcoolradio2} as a function of the average magnetic field of the wings ($B_{\rm W}$). Values of $B_{\rm L}=2.4 \; \mu {\rm G}$ and $\nu_{\rm b,L}= 9$ GHz are assumed for the active lobes. Different values of $\nu_{\rm b,W}$ are considered for the wings (see legend). The horizontal dashed line and vertical dotted line are drawn at $\tau=1$ and $B_{\rm W}=0.5B_{\rm L}$, respectively.}
	\label{fig: t-B}
\end{figure}

Understanding whether the active lobes are younger than the wings is key to testing theoretical formation scenarios for XRGs. In Sect. \ref{sect: radiative age} we constrained the radiative ages under the assumption of the TJP model, which is more refined than the JP and KP models in the treatment of $B$. However, the actual structure of $B$ in MRC 2011-298 is unknown, and variations of its strength across the source volume affect spectral age measurements.

The role of $B$ can be discussed based on simple assumptions. The radiative age is computed from the break frequency and magnetic field as \citep{miley80}:
\begin{equation}
t = 1590 \times \frac{{B}^{\frac{1}{2}}}{ B^2+B^2_{\rm CMB}} (1+z)^{-\frac{1}{2}} \nu_{\rm b}^{-\frac{1}{2}} \; \; \; {\rm [Myr]} \; \; \;,
\label{eq: tcoolradio}
\end{equation} 
where $B$ and $B_{\rm CMB}$ are expressed in $\mu {\rm G}$, and $\nu_{\rm b}$ in GHz. Based on Eq. \ref{eq: tcoolradio}, the wing to lobe age ratio is obtained as:
\begin{equation}
\tau = \frac{t_{\rm W}}{t_{\rm L}} = \left( \frac{B_{\rm L}^2+ B_{\rm CMB}^2}{B_{\rm W}^2+ B_{\rm CMB}^2} \right) \left( \frac{B_{\rm W}}{B_{\rm L}} \right)^{\frac{1}{2}} \left( \frac{\nu_{\rm b,W}}{\nu_{\rm b,L}} \right)^{-\frac{1}{2}} \; \; \;,
\label{eq: tcoolradio2}
\end{equation} 
where $B_{\rm L}$ and $B_{\rm W}$ are the average magnetic field strengths of the lobes and wings, respectively, and $\nu_{\rm b,L}\sim 9$ GHz and $\nu_{\rm b,W}\sim 1.4-4.7$ GHz are the estimated break frequencies (Sect. \ref{sect: Integrated spectra}). Reasonably, the magnetic field strength is higher across the active lobes than in the wings. In Eq. \ref{eq: tcoolradio2} we fixed $B_{\rm L}=2.4 \; \mu {\rm G}$ (Sect. \ref{sect: radiative age}) and $\nu_{\rm b,L}= 9$ GHz. The resulting  wing to lobe age ratio as a function of $B_{\rm W}$ is reported in Fig. \ref{fig: t-B} for different values of $\nu_{\rm b,W}$.

The age ratio is higher for more uniform magnetic fields across the whole source; for instance, at $B_{\rm W}=B_{\rm L}$, the maximum value of $\tau \sim 2.5$ is reached for $\nu_{\rm b,W}= 1.5$ GHz (red curve), while $\tau \sim 1.4$ for $\nu_{\rm b,W}= 4.5$ GHz (blue curve). However, Fig. \ref{fig: t-B} shows that a critical magnetic field value $B_{\rm c}$ exists for $\tau=1$ (dashed horizontal line), which means that the wings are younger than the lobes if $B_{\rm W}< B_{\rm c}$, regardless of $\nu_{\rm b,W}< \nu_{\rm b,L}$. For reference, $B_{\rm c}\sim 0.25 \; \mu {\rm G}\sim 0.1 B_{\rm L}$ for $\nu_{\rm b,W}= 1.5$ GHz and $B_{\rm c}\sim 0.8 \; \mu {\rm G}\sim 0.3 B_{\rm L}$ for $\nu_{\rm b,W}= 4.5$ GHz.

While based on simplistic considerations, as for example we did not include the temporal evolution of $B$, Fig. \ref{fig: t-B} indicates that a detailed and unbiased spectral analysis is necessary, but not sufficient to claim the relative age of morphologically complex radio galaxies. Several observational and numerical studies investigated the magnetic field distribution in XRGs, finding that the magnetic field lines are typically parallel to both the jet direction and wing length \citep[e.g.][]{hogbom79,black92,dennett-thorpe02,rottmann02,rossi17,cotton20,giri22}. Remarkably, the field strength was found to not vary dramatically from the active lobes to the wings, with typical average ratios $\lesssim 2$ for both XRGs with smaller and larger wings. Therefore, we can use this value as reference in Fig. \ref{fig: t-B} ($B_{\rm W}=1.2 \; \mu {\rm G}$, dotted vertical line), which yields $\tau\sim 2.1$ and $\tau\sim 1.2$ at 1.5 and 4.5 GHz, respectively. In summary, we can safely conclude that the wings are older than the lobes in MRC 2011-298, likely by a factor of $\tau> 2$ if $\nu_{\rm b,W} \sim 1.5$ GHz, which is in line with findings in Sect. \ref{sect: radiative age}.

We stress that by increasing $B_{\rm L}$, progressively smaller ratios $B_{\rm W}/B_{\rm L}$ would be necessary to match $\tau=1$. As remarkable examples, if $B_{\rm L}=B_{\rm CMB}=4.2 \; \mu {\rm G}$, then $B_{\rm c}\sim 0.07B_{\rm L}$ for $\nu_{\rm b,W} = 1.5$ GHz and $B_{\rm c}\sim 0.14B_{\rm L}$ for $\nu_{\rm b,W} = 4.5$ GHz; if $B_{\rm L}=10 \; \mu {\rm G}$, then $B_{\rm c}\sim 0.02B_{\rm L}$ for $\nu_{\rm b,W} = 1.5$ GHz and $B_{\rm c}\sim 0.04B_{\rm L}$ for $\nu_{\rm b,W} = 4.5$ GHz. In other words, if the actual magnetic field of the lobes were higher than the assumed $B_{\rm L}=2.4 \; \mu {\rm G}$, our conclusions on the younger age of the lobes would be further reinforced as $B_{\rm W} \ll 0.5 B_{\rm L} $ is unlikely.

\subsection{Jet precession}
\label{sect: Jet precession}

\begin{table}
\centering
   		\caption{List of free parameters for the optimisation algorithm  (Eq. \ref{eq: cost}) of the precession modelling of the jets (Eqs. \ref{eq: prec1}-\ref{eq: prec4}).}
    \label{tab: parametri precessione}   
\begin{tabular}{cccccc}
\hline
\noalign{\smallskip}
 & $P$ & $\psi$ & ${\rm v}$ \\
 & (Myr) & (deg) & ($10^{-2}c$) \\
\hline
\noalign{\smallskip}
S-band & $13.10 \pm 0.04$ & $15.13\pm 0.06$ & $2.20 \pm 0.01$ \\
C-band & $12.29 \pm 0.09$ & $9.62\pm 0.45$ & $2.59 \pm 0.01$ \\
X-band & $14.45 \pm 0.05$ & $13.26\pm 0.09$ & $2.13 \pm 0.01$ \\
Mean & $13.28 \pm 0.89$ & $12.67\pm 2.29$ & $2.31 \pm 0.20$ \\
\noalign{\smallskip}
\hline
\end{tabular}
   	  	\begin{tablenotes}
   	  	\centering
  	\item	{\small \textbf{Notes}. The fixed parameters are $s_{\rm rot}=-1$, $s_{\rm N,jet}=-1$, $s_{\rm S,jet}=+1$, $i=80$ deg, $\chi=260$ deg. The reported uncertainties are statistical for the fitted values of each band and systematical for the corresponding mean values (see details in Appendix \ref{sect: Precession modelling}). } 
   	\end{tablenotes}
   \end{table}

\begin{figure*}
	\centering
\includegraphics[width=0.33\textwidth]{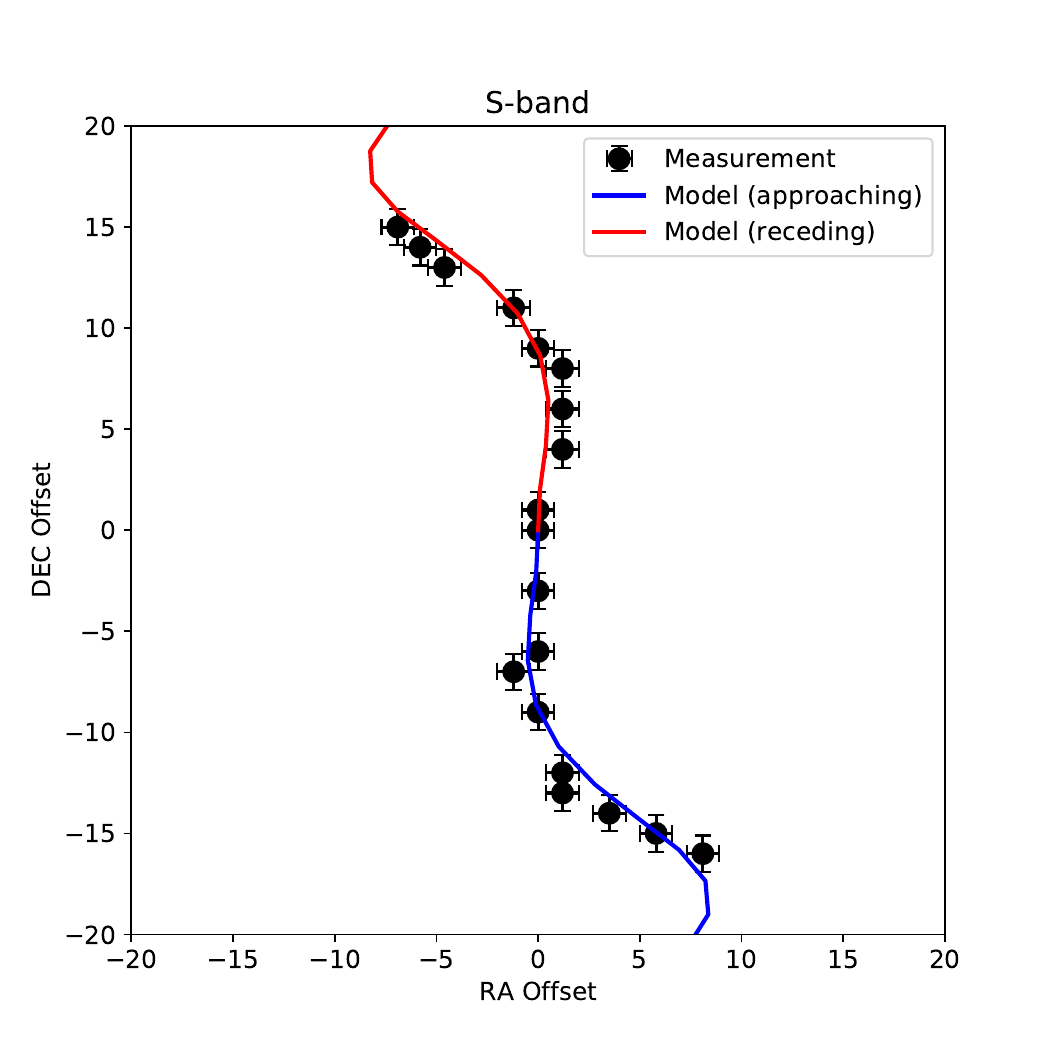}
\includegraphics[width=0.33\textwidth]{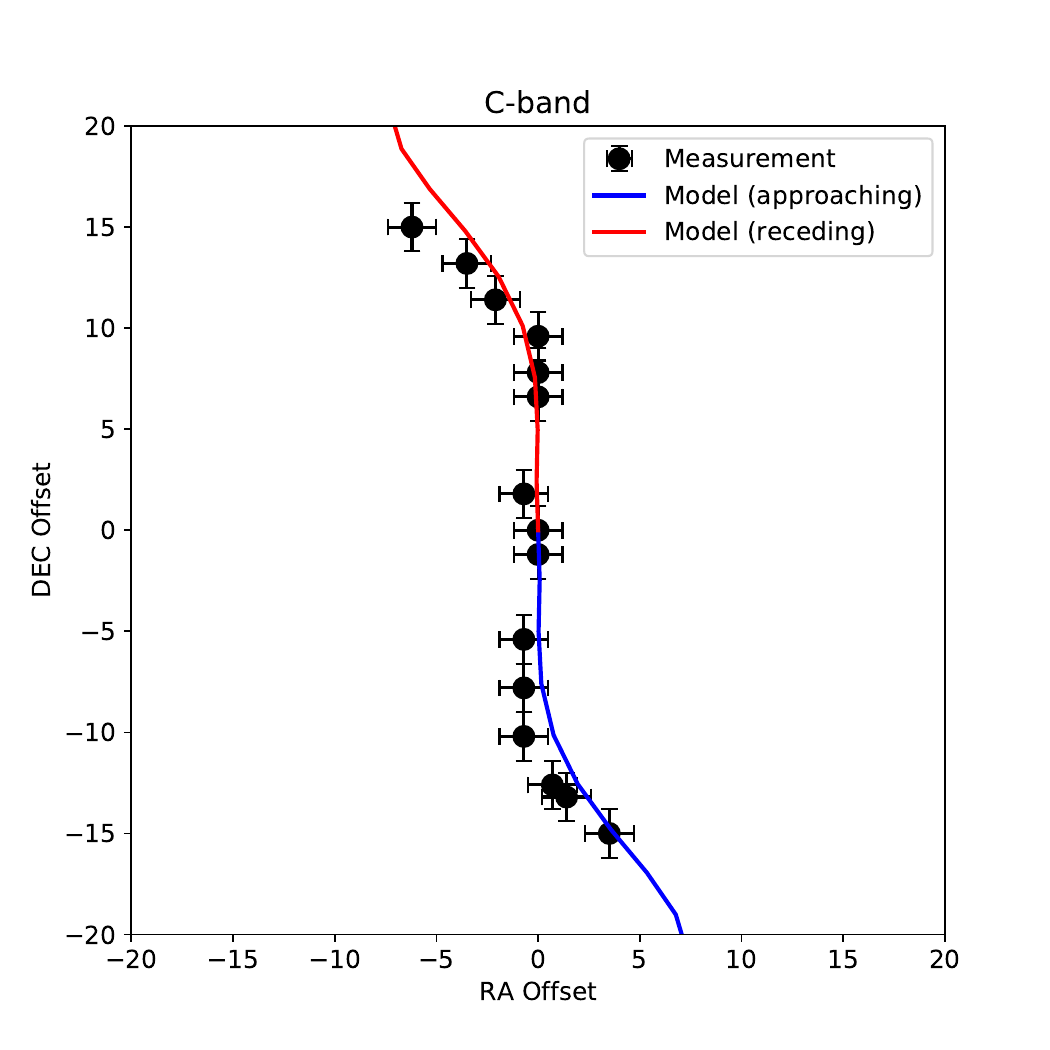}
\includegraphics[width=0.33\textwidth]{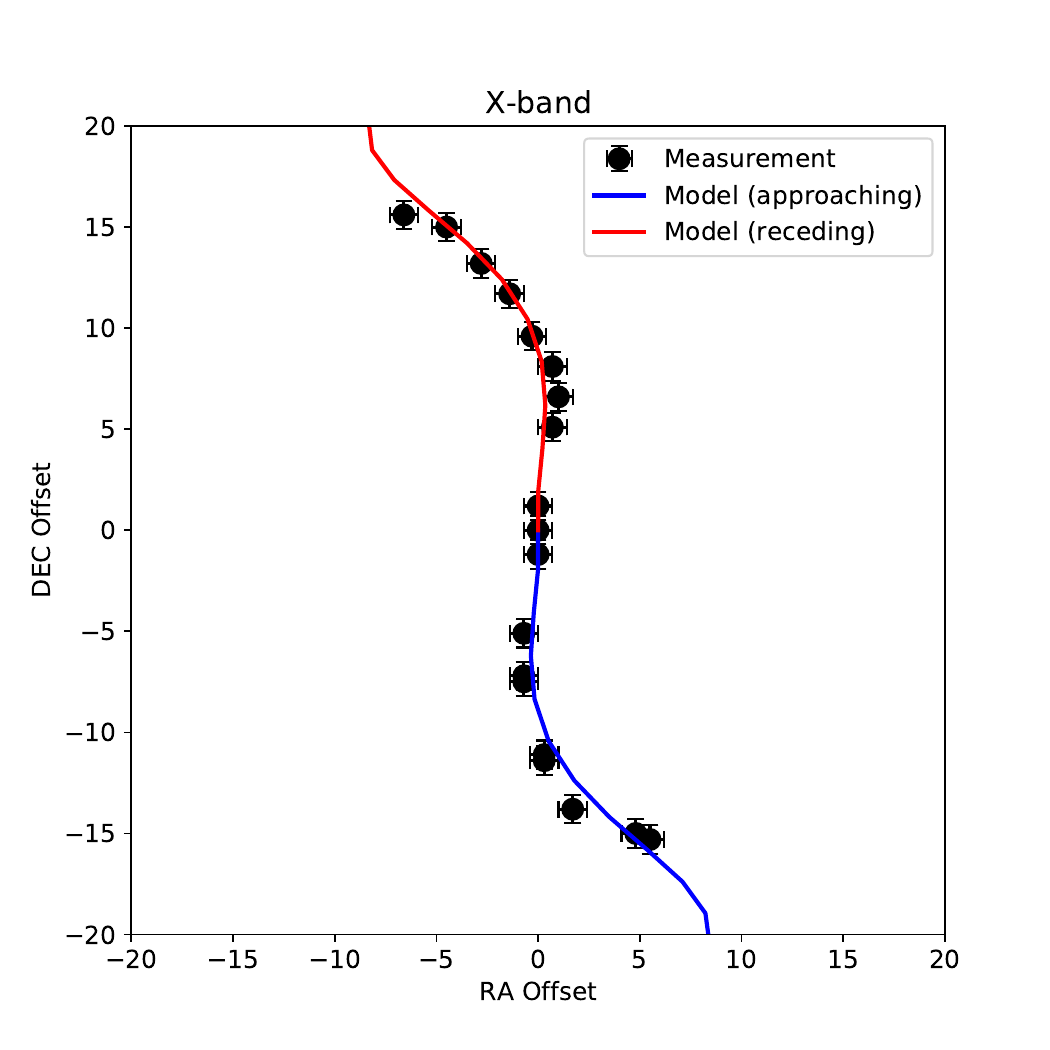}
	\caption{Results of the optimisation procedure (Appendix \ref{sect: Precession modelling}) for the precession of the jets at S-band (left), C-band (centre), and X-band (right). Black data points are the measured offset (in arcsec) from the core of the peak within each sampling region. Red and blue curves are the predicted precession paths for the receding and approaching jet, respectively. Fixed and optimised parameters are listed in Table \ref{tab: parametri precessione}.   }
	\label{fig: precession jet}
\end{figure*} 

Bent or twisted morphologies of the radio jets have been interpreted as the result of their precession \citep[e.g.][]{ekers78,gower82,morganti99,steenbrugge&blundell08,kun14,krause19,bruni21,ubertosi24}. In this Section, we test the hypothesis of precession for the S-shaped jets of MRC 2011-298. In Appendix \ref{sect: Precession modelling} we provide an overview of the theoretical framework (based on \citealt{hjellming&johnston81}), adopted procedures (based on \citealt{coriat19,ubertosi24}), and assumptions. 

We sampled S-band, C-band, and X-band radio images with beam-size regions following the emission of the jets (similarly to Fig. \ref{fig: profili brillanza}, top right panel). We derived the position of the peak within each region, and computed its distance from the core (region `0') in RA and Dec. Such measured offsets were used to optimise the parameters describing the projected motion of plasma bubbles along precessing jets (see details in Appendix \ref{sect: Precession modelling}). Specifically, we derived the period ($P$), half-opening precession angle ($\psi$), and jet velocity (${\rm v}$). In Fig. \ref{fig: precession jet} we report the measured offsets with overlaid the predicted jet path after optimisation. The optimised parameters at each frequency and their mean are listed in Table \ref{tab: parametri precessione}. We obtained values of the order of $P\sim 10$ Myr, $\psi\sim 10$ deg, and ${\rm v}\sim 0.02c$; we notice that this value of ${\rm v}$ is in line with typical velocities found for the jets of FRI galaxies on kiloparsec scales \citep[e.g.][]{bicknell94}. 

Our analysis shows that precession is a viable explanation to the bent morphology of the jets of MRC 2011-298. Possible phenomena inducing such precession are gravitational interactions between i) the optical nuclei A and B of the dumbbell system on large ($\sim 20$ kpc) scales \citep[e.g.][]{wirth82} or ii) two SMBHs, one of which active, within the primary core A on small ($\ll 0.1$ kpc) scales \citep[e.g.][]{komossa06}. For a binary system of total mass $M_{\rm tot}$ consisting of two equally-massive SMBHs in a circular orbit, we can obtain an upper limit to their separation based on the precession period as \citep[e.g.][]{krause19}
\begin{equation}
  d_{\rm BH} < 0.18 \times \left( \frac{P}{{\rm Myr}}\right)^\frac{2}{5}  \left( \frac{M_{\rm tot}}{{\rm 10^9 \; M_{\rm \odot}}}\right)^\frac{3}{5} \; \; \; [{\rm pc}] \; \; \;.   
  \label{eq: separation}
\end{equation}
By considering the estimate of the mass of the central SMBH derived in \cite{bruno19} ($M_{\rm BH}\sim 10^9 \; {\rm M_{\rm \odot}}$) as the total mass of a possible binary system, Eq. \ref{eq: separation} yields $d_{\rm BH} < 0.5$ pc. This value corresponds to $d_{\rm BH} < 0.2$ mas at the cluster redshift, and it is orders of magnitude lower than the resolution achieved by our present radio data ($\sim 2''$, corresponding to $\sim 5$ kpc); observations with the Very Long Baseline Interferometer (VLBI) are necessary to probe such small scales. 

We now discuss whether the wings of MRC 2011-298 could be produced as a consequence of the same precession of the jets. In this respect, \cite{nolting23} showed that XRGs can be originated in the case of long-term evolution of precessing jets with large $\psi\sim 30-45$ deg, but this is much higher than our constraints. Therefore, the formation of the overall X-shaped structure through precession seems unlikely. The observed morphology could be matched only if $\psi$ was larger in the past, but this hypothesis cannot be tested.

As mentioned in Sect. \ref{sect: Radio morphology}, uGMRT images reveal deflections of the wings at their edges, which provide an overall Z-shape for the wings that is reminiscent of precession. In the context of a rapid reorientation of the jets from the E-W to N-S axis (see also Sect. \ref{sect: On the origin of A3670}), we investigated the possibility that precession was already ongoing before the quick flip of the jet direction. We measured RA and Dec offsets along the wings from L-band and band-4 images and carried out an analogous optimisation procedure (not shown). However, the algorithm does not converge to stable solutions, thus preventing us from successfully reproduce the observed wing path. While this may suggest that the wings have not been subjected to precession in the past, many effects (such as ageing, expansion, projection effects, insufficient spatial resolution, inaccurate calibration) could alter the measured offsets and bias our procedure. Therefore, our analysis is inconclusive about the possible precession along the direction of the wings.

\subsection{Reorientation origin}
\label{sect: On the origin of A3670}

Throughout this Section we discuss formation scenarios proposed in the literature for XRGs in the light of the properties of MRC 2011-298. These are i) binary AGN, ii) jet-stellar shell diversion, iii) backflow, and iv) reorientation models \citep[e.g.][]{gopal-krishna12,giri24}, with the latter two being the most popular, although at present there is not a single scenario capable to reproduce all known XRGs. 

One of the proposed models considers the active lobes and wings as produced by a binary AGN system in perpendicular directions \citep{lal&rao07,lal19}, but the lack of XRGs with jetted wings is incompatible with this scenario. We exclude that the origin of the XRG in A3670 is associated with a binary AGN system, as it can account neither for the large age gaps between the wings and the lobes ($\sim 30$ Myr as found in Sect. \ref{sect: radiative age}) nor for the observed radio-optical alignments (Fig. \ref{fig: radio+panstarss}). We stress that this does not rule out the presence of a possible binary SMBH system, in which only one member is currently active. 

\cite{bruno19} showed that the properties of MRC 2011-298 are not in contrast with the jet-stellar shell interaction model \citep{gopal-krishna12}. According to this scenario, the jets propagating along the major axis of the host galaxy are deflected perpendicularly by the encounter with a system of gaseous stellar shells \citep[][]{gopal-krishna&saripalli84,gopal-krishna03,hodgeskluck10b,almeida11}, which are produced as a consequence of a recent wet merger, thus forming the wings along the minor axis. Although galaxy mergers are a favourable channel for the formation of BCGs, and the dumbbell nature of our target supports this hypothesis, we do not have signatures that could confirm the presence of shells in the BCG in A3670. For example, no evidence of recent mergers is provided by optical images (see Fig. \ref{fig: radio+panstarss}) and spectrum \citep{6dFGS09}, which shows absorption lines as typical of old stellar populations in early-type galaxies (such as neutral Mg and Na). Moreover, the large age difference between the lobes and wings would imply a long interaction with the shells, which is unlikely. \cite{bruno19} suggested that the S-shape of the jets could be induced by the deflection with rotating shells, but in Sect. \ref{sect: Jet precession} we showed that this may be reasonably driven by precession. Therefore, introducing the hypothesis of stellar shells is not necessary, and our present results disfavour this model.

Hydrodynamical models \citep[e.g.][]{leahy&williams84, kraft05,capetti02,hodgeskluck11,rossi17,cotton20,giri23} are promising to explain the formation of the majority of known XRGs. These rely on the assumption of the alignments between radio, optical, and X-ray axes, as indeed typically observed. Nonetheless, the origin of the wings from diverted backflowed plasma is unlikely for MRC 2011-298. Indeed, the formation of prominent wings such as those of MRC 2011-298 is notably challenging in the framework of hydrodynamical scenarios \citep[e.g.][]{giri24}. Most importantly, low-power FRI-type jets can hardly induce backflow from interaction with the ambient medium. To overcome this issue, \cite{saripalli&subrahmanyan09} proposed that wings in FRI-type XRGs have been formed via backflow from FRII-type jets during a previous AGN outburst phase. This scenario implies the evolution of FRII into FRI radio galaxies, which however is speculative and not supported by observations (e.g. \citealt{hardcastle&croston20}, for a review on the FRI-FRII dichotomy). Therefore, we confidently exclude the backflow origin for MRC 2011-298.    

Very extended, steep-spectrum, and old wings can naturally arise in the context of reorientation models. Radio jets can slowly reorient as a consequence of long-term precession \citep[e.g.][]{dennett-thorpe02,nolting23} or almost instantaneously change their propagation direction by large angles as a consequence of the coalescence of SMBHs or inhomogeneous mass accretion \citep[e.g.][]{dennett-thorpe02,merritt&ekers02,liu04,gergely&biermann09,lalakos22}. \cite{bruno19} disfavoured the precession scenario for the formation of the wings, which is expected to produce Z-shaped (i.e. large-distance offset with respect to the radio core) rather than X-shaped structures. This is further supported by our present analysis on the precession of the jets (Sect. \ref{sect: Radio morphology}), as we obtained a narrow ($\psi \sim 10$ deg) precessing cone that is incompatible with the formation of large wings. Based on the inefficient accretion that is thought to be associated with FRI radio galaxies, \cite{liu04} argued that the interaction of the SMBH with the accretion disk is negligible, therefore it is unlikely that such interaction has driven the spin-flip in out target. On the other hand, spin-flip caused by coalescing SMBHs is still plausible for MRC 2011-298, even though \cite{bruno19} stressed that this model could not account for the observed radio-optical alignments and high ellipticity of the host. 

\cite{giri23} carried out 3D relativistic magnetohydrodynamic simulations aimed at probing the formation of XRGs through backflow and reorientation in various conditions. Interestingly, two of their simulated XRGs (labelled as `qr90\_XY', `sr90\_XY' in \citealt{giri23}) are surprisingly similar to MRC 2011-298. These are the result of quick (instant) and slow (5 Myr) reorientation\footnote{As discussed by \cite{giri23}, a slow reorientation of 5 Myr is still fast enough to produce indistinguishable morphologies as for instant spin-flip.} of the jets by $90$ deg from the minor axis to the major axis of the host galaxy. This configuration, which leaves fossil plasma in the form of prominent wings along the minor axis and fresh active lobes along the major axis, is exactly what we observe in our target. The thermal gas pressure gradient plays a key role in the simulation, as it favours not only the growth in size, but also the lateral expansion of the wings. Overall, this scenario solves the caveats raised in \cite{bruno19} by showing that jets initially propagating along the minor axis of an asymmetric medium are able to produce wings through the orthogonal spin-flip. It is worth noticing that the X-ray atmosphere simulated by \citealt{giri23} has an ellipticity of $\varepsilon=0.3$, which is consistent with the ellipticity measured in the optical for the BCG in A3670 ($\varepsilon=0.28$, \citealt{makarov14}). In summary, both the observed properties and simulations support the spin-flip origin for the XRG in A3670 and the additional role of the ambient medium for its evolution.

\section{Summary and conclusions}
\label{sect: Summary and conclusions}

The brightest cluster galaxy in A3670 hosts MRC 2011-298 (Figs. \ref{fig: radio+panstarss}, \ref{fig: mappefullres}), a prominent X-shaped radio galaxy that has been first studied by \cite{bruno19} at gigahertz frequencies. In the present work, we followed-up the target with deep radio data at lower frequencies to constrain its origin by investigating the spectral evolution of the active lobes and wings. Specifically, we combined uGMRT and JVLA radio data spanning wide ranges in frequency (from 120 MHz to 10 GHz) and spatial resolution (down to $\sim 5$ kpc), which provided us a detailed view of the source.

MRC 2011-298 is characterised by large wings, extending along E-W for $\sim 450\; {\rm kpc}$, and orthogonal active lobes, extending along N-S for $\sim 150\; {\rm kpc}$. Therefore, the total wing to lobe length ratio is $\sim 3$, which is impressively large if compared with typical XRGs. Moreover, as usual for the majority of XRGs, the wings and the lobes are aligned with the minor and major optical axis of the host, respectively. The active lobes host S-shaped radio jets of FRI-type. It is well established that XRGs exhibiting FRI-type jets are uncommon. We demonstrated that the S-shaped bending could be the result of precession of the jets (Sect. \ref{sect: Jet precession}, Fig. \ref{fig: precession jet}) on a period of $P= 13.3\pm 0.9$ Myr. Possibly, a companion SMBH within a binary system may have induced such jet precession.

We measured an average spectral index $\alpha=0.69\pm 0.02$ for the whole source (Sect. \ref{sect: Integrated spectra}, Fig. \ref{fig: spettro integrato}), as typical of classical radio galaxies. However, the spectral distribution is not uniform, as the spectral index is steeper in the wings than in the lobes, with a steepening trend outwards along the E-W direction (Sect. \ref{sect: Resolved spectral analysis}, Figs. \ref{fig: spixmap}, \ref{fig: spixprofile}). The sensitivity of our data allowed us to constrain the break frequencies of each component (Sect. \ref{sect: Constraining the break frequency}), finding $\nu_{\rm b,L}\sim 9$ GHz for the lobes and $\nu_{\rm b,W}\sim 1.4-4.7$ GHz for the wings. These values provided us fundamental constraints on the radiative age of the lobes and wings. Indeed, we produced a radiative age map at sufficient resolution ($\sim 40$ kpc, Fig. \ref{fig: agemapTJP}) to study the age trend across the source and confirm that the wings are significantly older than the lobes (by $\Delta t\gtrsim 30$ Myr).  
 
An essential part of our work was devoted to unbias the spectral analysis, by accurately investigating the role of sparse and inhomogeneous sampling of the \textit{uv}-coverage (Sects. \ref{sect: Integrated spectra}, \ref{sect: Constraining the break frequency}), imperfect calibration (Sect. \ref{sect: Resolved spectral analysis}), and spatial variations of the magnetic field when lacking direct polarisation measurements (Sect. \ref{sect: Spectral age and magnetic field}). We showed that ignoring these effects could easily lead to wrong conclusions, especially when dealing with data covering a limited frequency range. Our detailed approach yielded solid measurements that we used to discuss the origin of the target based on theoretical models and simulations in the literature (Sect. \ref{sect: On the origin of A3670}).

We confidently confirm that the reorientation of the jets is responsible for the formation of the X-shaped structure of MRC 2011-298. In this framework, the jets reoriented from E-W to N-S, leaving fossil plasma in the wings and injecting fresh particles within the present active lobes. We stress that this reorientation is thought to be driven by the abrupt (on a timescale $ t \sim 0-5$ Myr) flip of the spin of the central SMBH by $\sim 90$ deg, which is thus a distinct process from the precession mentioned above. The phenomenon triggering such spin-flip is debated, but it may be possibly associated with the coalescence of two SMBHs.

In conclusion, current data and simulations indicate that different mechanisms can lead to the formation of XRGs. Upcoming radio surveys, such as those from the Square Kilometer Array (SKA), the LOw Frequency ARray (LOFAR) 2.0, and the next-generation Very Large Array (ngVLA), will scrutiny the sky with unprecedented capabilities, enabling  multi-frequency analysis at various resolutions of large samples of XRGs. These could be potentially distinguished into sub-populations based on their observed radio properties and auxiliary information on their host galaxy and environment. Therefore, identifying XRGs with a well constrained origin to be used as reference is still of primary relevance. Our analysis of MRC 2011-298 demonstrated that spin-flip is a valid scenario to explain XRGs with extended and old wings, without the necessity of introducing additional complexity, such as the presence of stellar shells or evolution of the jets from FRII-type to FRI-type.

Although we unveiled the origin of the XRG in A3670, many open questions are still to be addressed. First, X-ray data are essential to probe the environmental medium, and thus confirm the role of pressure gradients in the evolution of the wings subsequent to the jet reorientation, as suggested by the observed radio-optical alignments and ellipticity of the host, and supported by simulations. Second, polarisation radio data are necessary to firmly determine the magnetic field structure and strength, providing tighter constraints on the spectral age and the time scales for the reorientation. Finally, radio data at sub-arcsec resolution could provide valuable information on the central engine, possible binary SMBHs, and jet precession below parsec scales. 

\begin{acknowledgements}
We thank the referee for their comments and suggestions, that have improved the presentation of the manuscript. LB thanks L. Gregorini for useful advice and discussion. MB acknowledges support from the Next Generation EU funds within the National Recovery and Resilience Plan (PNRR), Mission 4 - Education and Research, Component 2 - From Research to Business (M4C2), Investment Line 3.1 - Strengthening and creation of Research Infrastructures, Project IR0000034 – 'STILES - Strengthening the Italian Leadership in ELT and SKA'. The National Radio Astronomy Observatory is a facility of the National Science Foundation operated under cooperative agreement by Associated Universities, Inc. We thank the staff of the GMRT that made these observations possible. GMRT is run by the National Centre for Radio Astrophysics of the Tata Institute of Fundamental Research. The Pan-STARRS1 Surveys (PS1) and the PS1 public science archive have been made possible through contributions by the Institute for Astronomy, the University of Hawaii, the Pan-STARRS Project Office, the Max-Planck Society and its participating institutes, the Max Planck Institute for Astronomy, Heidelberg and the Max Planck Institute for Extraterrestrial Physics, Garching, The Johns Hopkins University, Durham University, the University of Edinburgh, the Queen's University Belfast, the Harvard-Smithsonian Center for Astrophysics, the Las Cumbres Observatory Global Telescope Network Incorporated, the National Central University of Taiwan, the Space Telescope Science Institute, the National Aeronautics and Space Administration under Grant No. NNX08AR22G issued through the Planetary Science Division of the NASA Science Mission Directorate, the National Science Foundation Grant No. AST–1238877, the University of Maryland, Eotvos Lorand University (ELTE), the Los Alamos National Laboratory, and the Gordon and Betty Moore Foundation. This research has made use of SAOImageDS9, developed by Smithsonian Astrophysical Observatory \citep{ds9}. This research has made use of the VizieR catalogue access tool, CDS, Strasbourg Astronomical Observatory, France (DOI: 10.26093/cds/vizier). This research made use of APLpy, an open-source plotting package for Python \citep{robitaille&bressert12APLPY}, Astropy, a community-developed core Python package for Astronomy \citep{astropycollaboration13,astropycollaboration18}, Matplotlib \citep{hunter07MATPLOTLIB}, Numpy \citep{harris20NUMPY}, SciPy \citep{scipy}.

\end{acknowledgements}

\bibliographystyle{aa}
\bibliography{bibliografia}

\begin{appendix}

\section{Sampling of \textit{uv}-plane}
\label{sect: uvplane}

\begin{figure*}
	\centering

\includegraphics[width=0.32\textwidth]{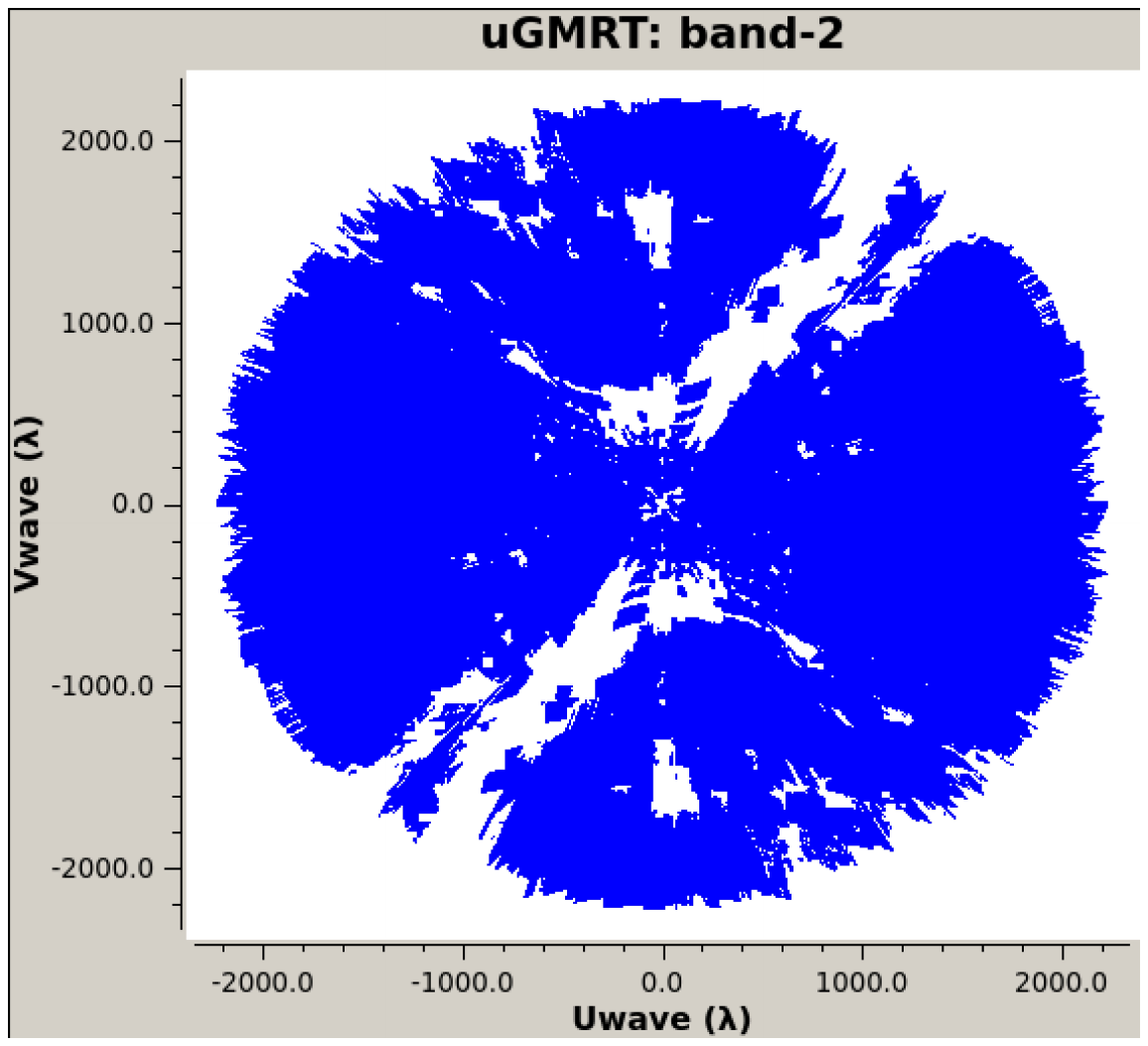}
\includegraphics[width=0.32\textwidth]{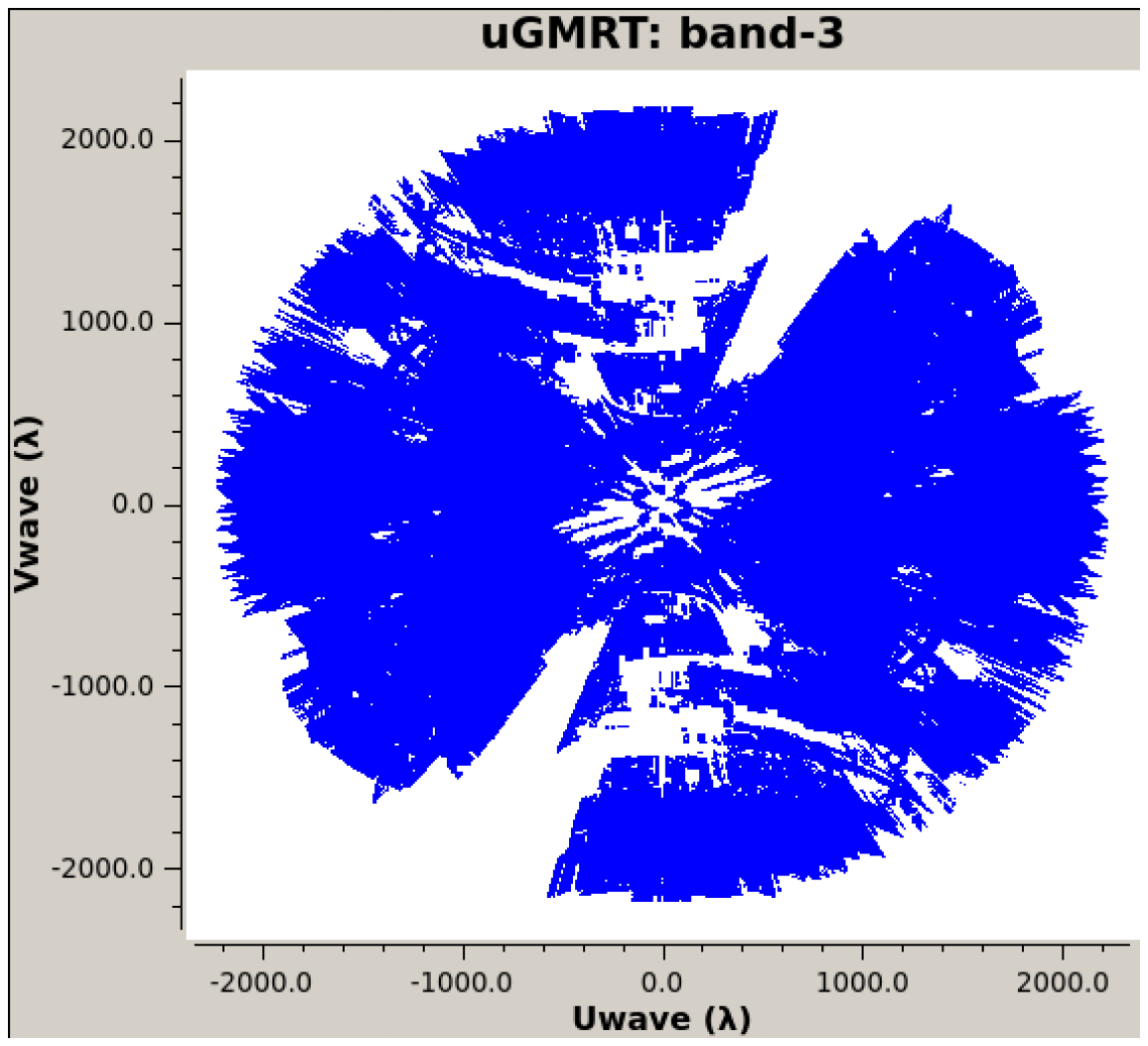}
\includegraphics[width=0.32\textwidth]{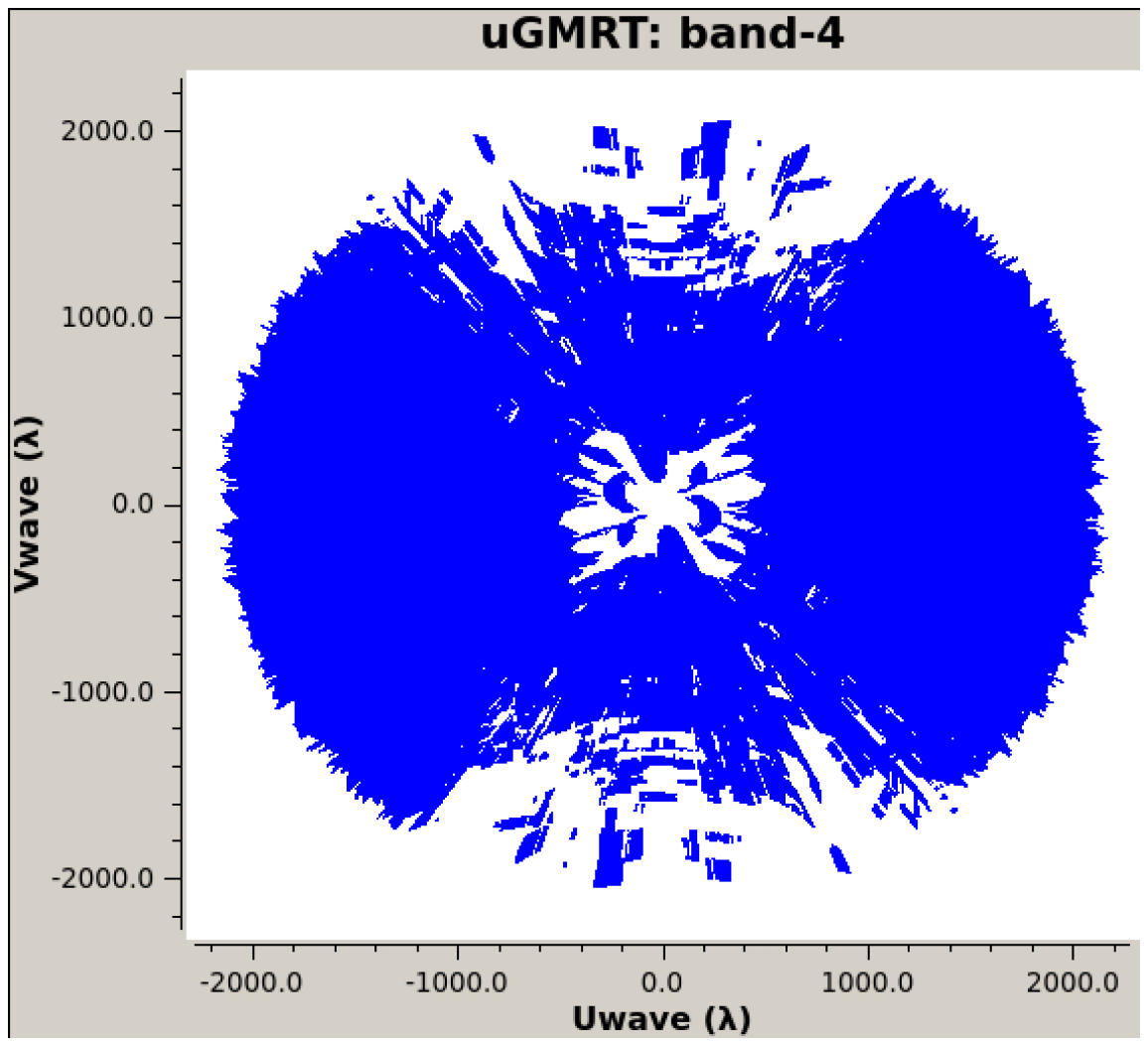}
\includegraphics[width=0.32\textwidth]{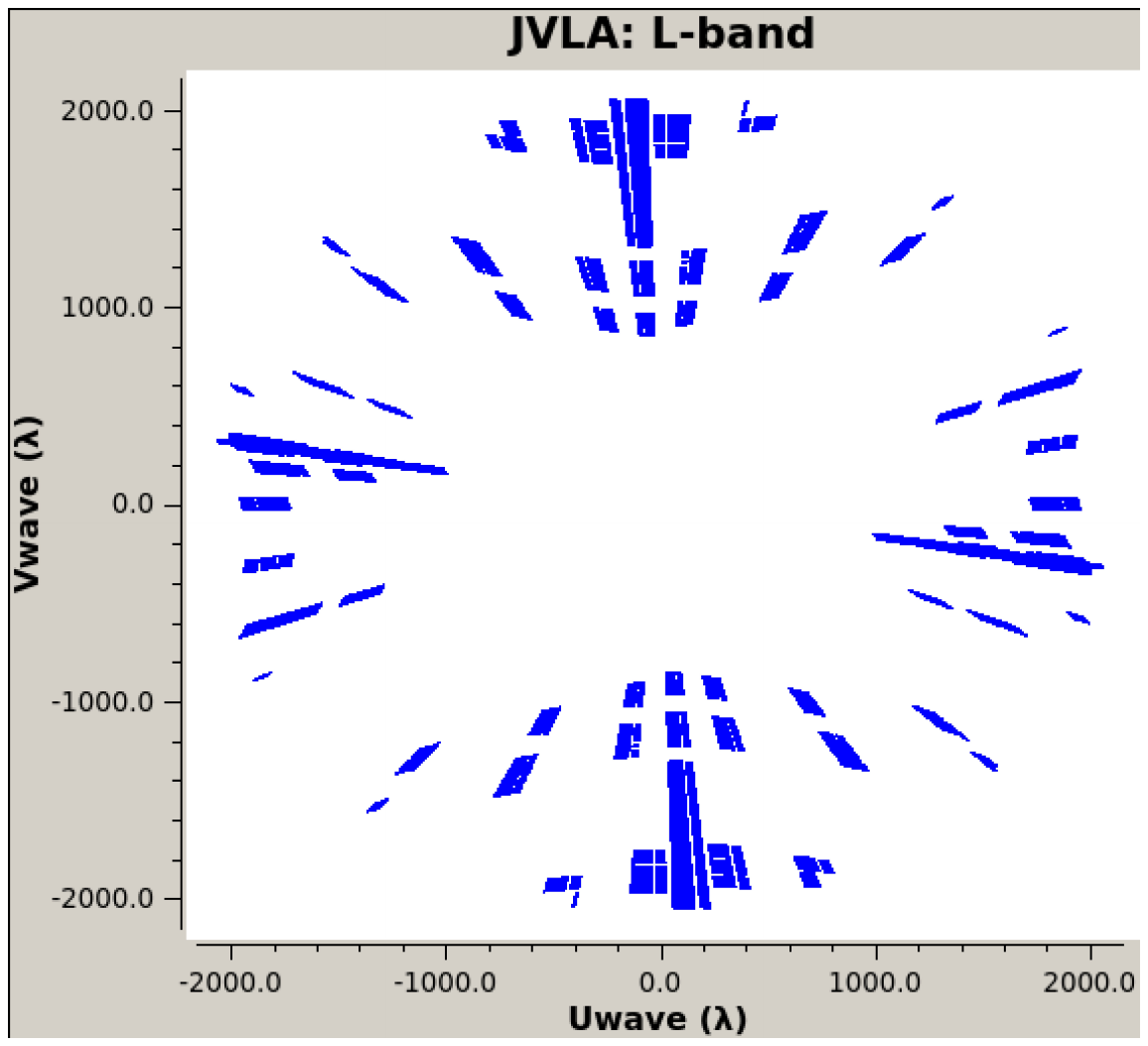}
\includegraphics[width=0.32\textwidth]{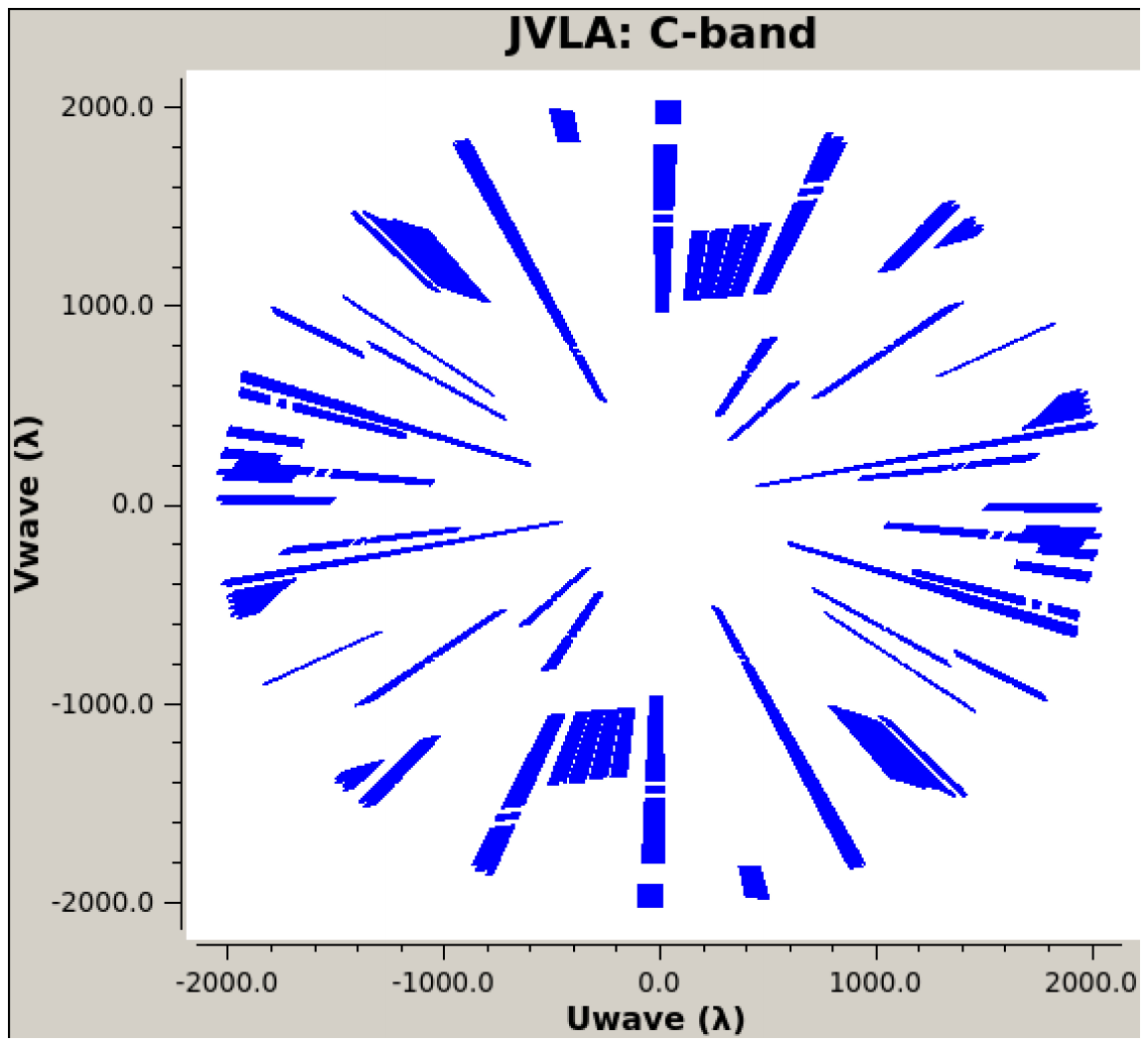}

	\caption{Inner \textit{uv}-plane of our observations. Only baseline lengths $<2 \; {\rm k}\lambda$ are displayed for clearer inspection. \textit{From top left to bottom right}: uGMRT band-2, uGMRT band-3, uGMRT band-4, JVLA L-band, JVLA C-band.}
	\label{fig: uvplane}
\end{figure*} 

In Fig. \ref{fig: uvplane} we report the (inner) \textit{uv}-planes for our uGMRT (band-2, band-3, band-4) and JVLA (L-band, C-band) observations. For inspection purposes, only baseline lengths $<2 \; {\rm k}\lambda$ are displayed.

\section{Spectral index error maps}
\label{sect: errspixmap}

\begin{figure*}
	\centering

\includegraphics[width=0.49\textwidth]{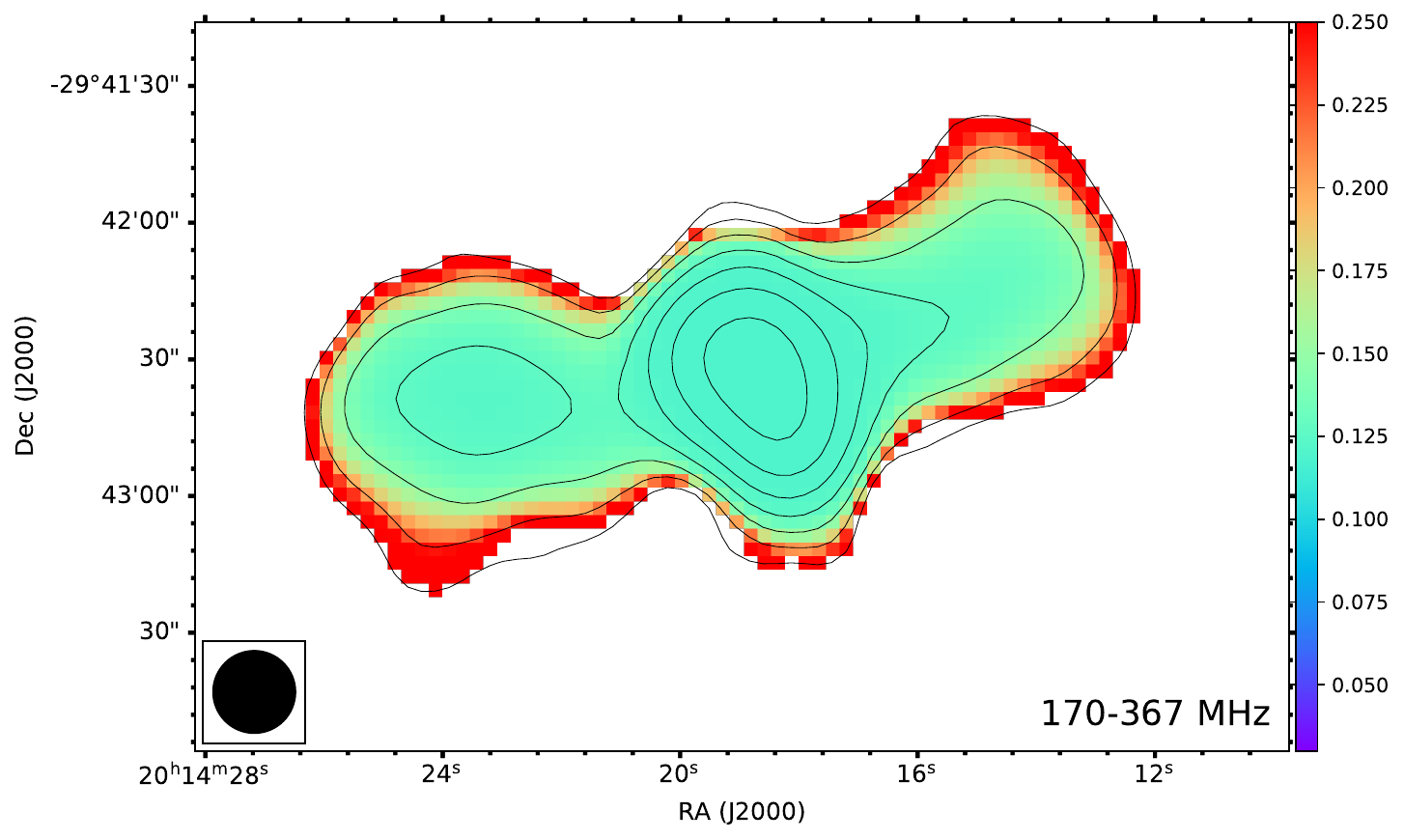}
\includegraphics[width=0.49\textwidth]{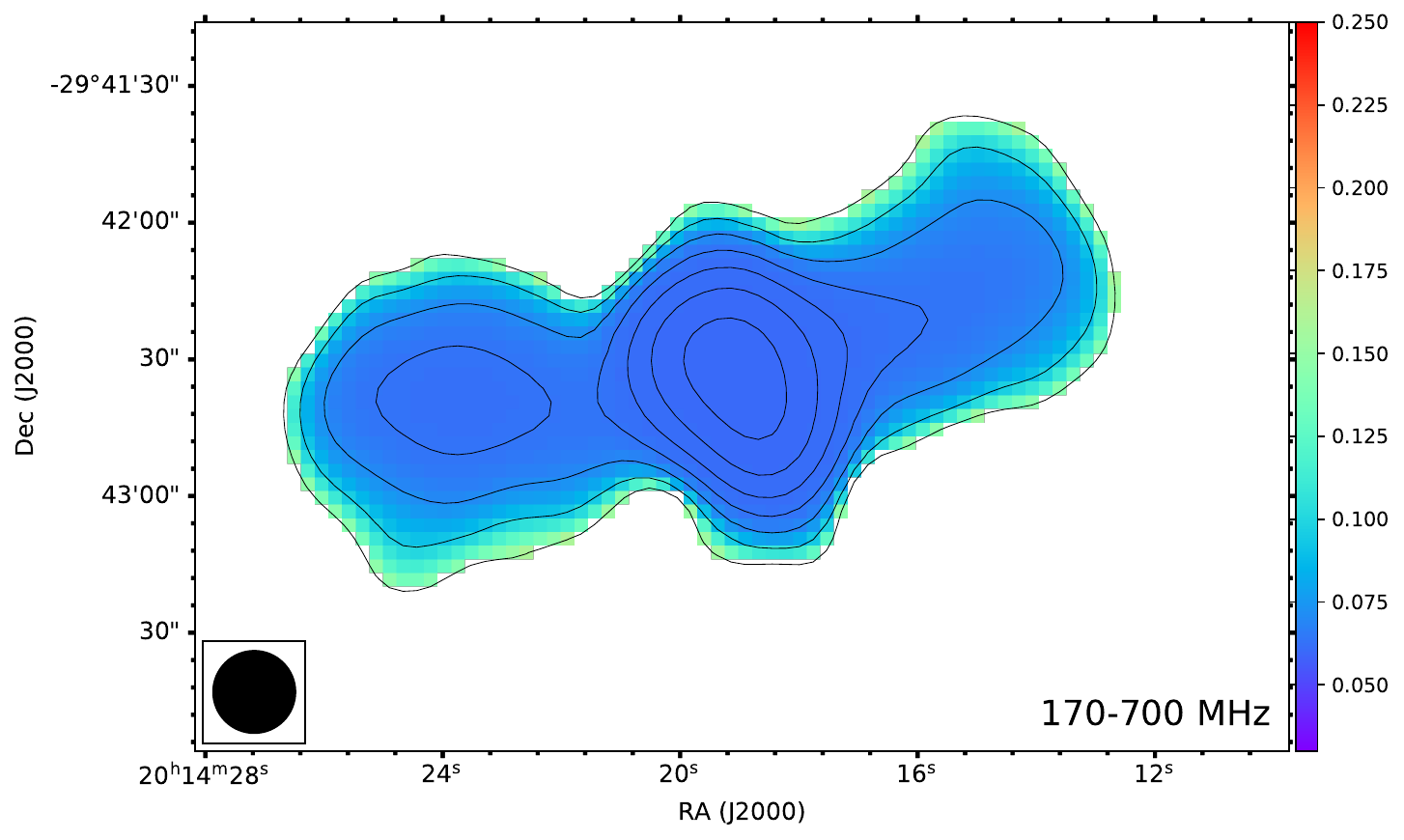}
\includegraphics[width=0.49\textwidth]{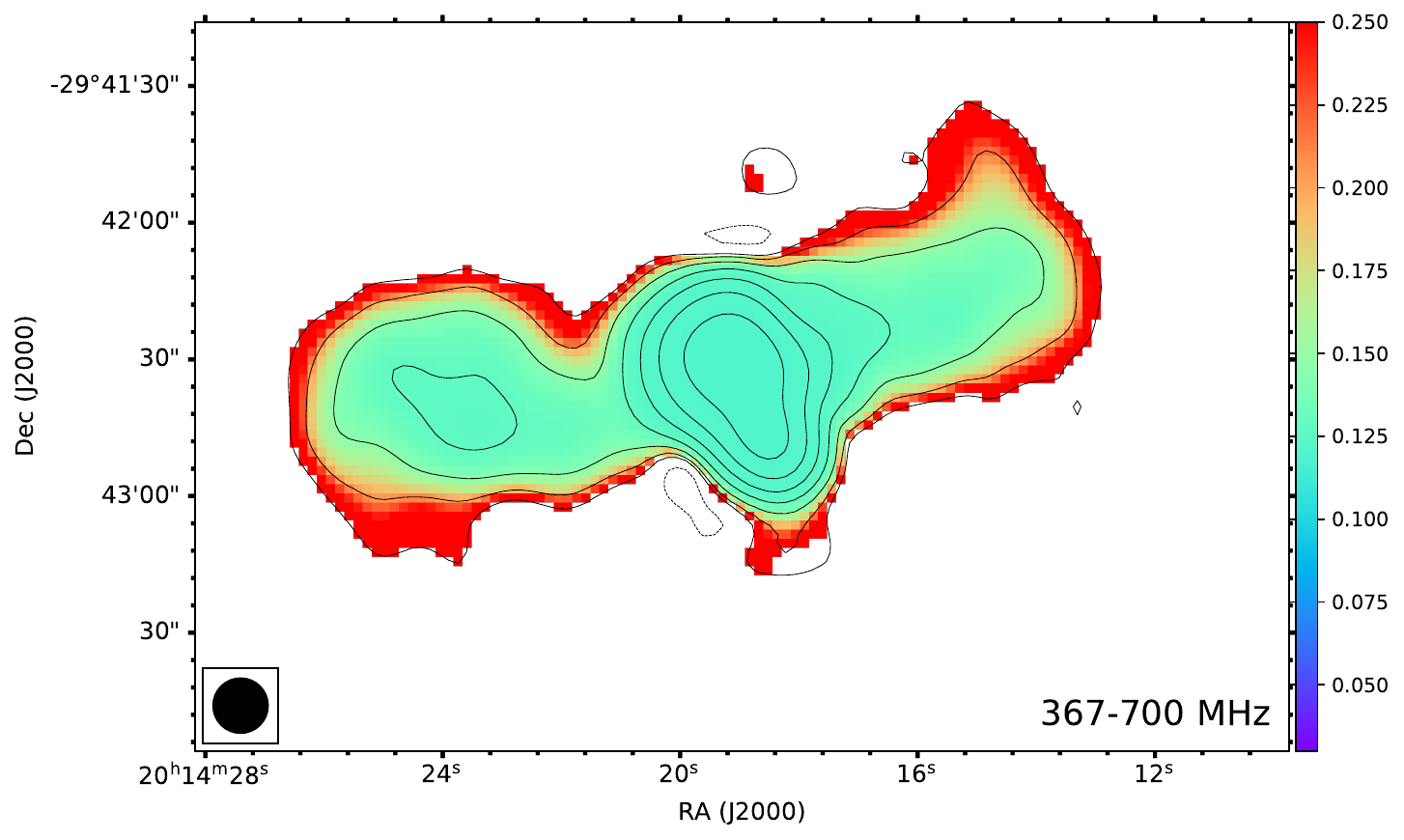}
\includegraphics[width=0.49\textwidth]{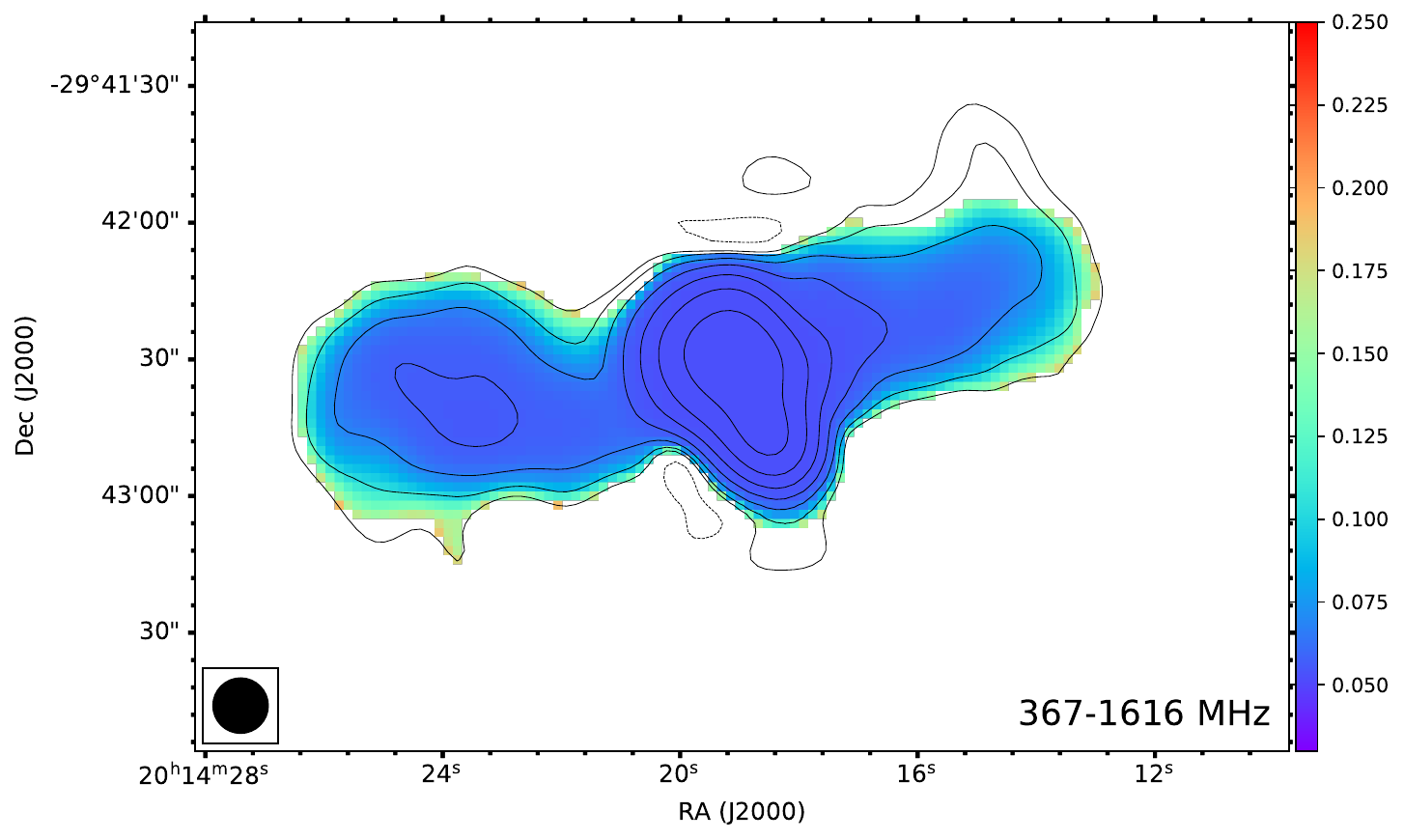}
\includegraphics[width=0.49\textwidth]{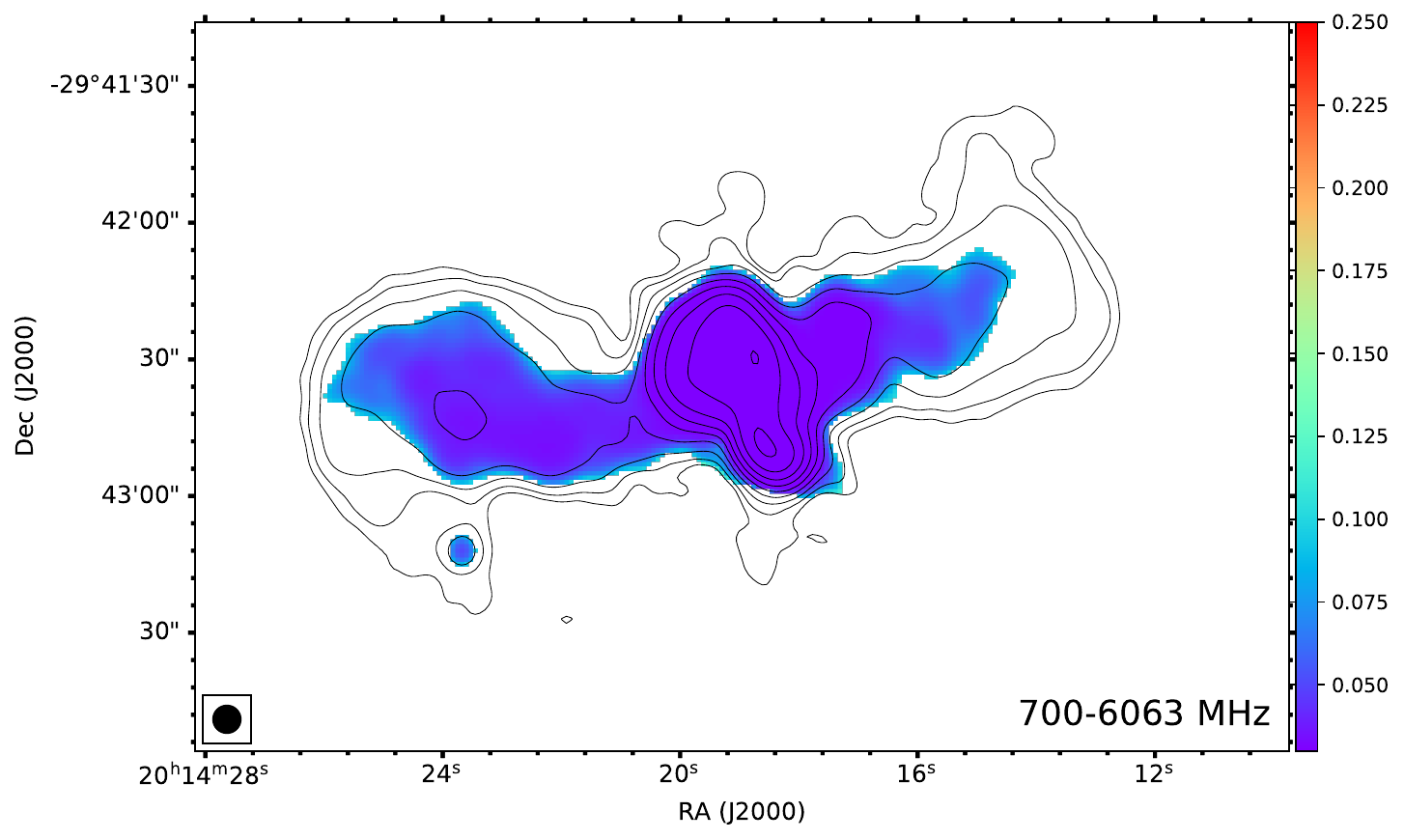}
\includegraphics[width=0.49\textwidth]{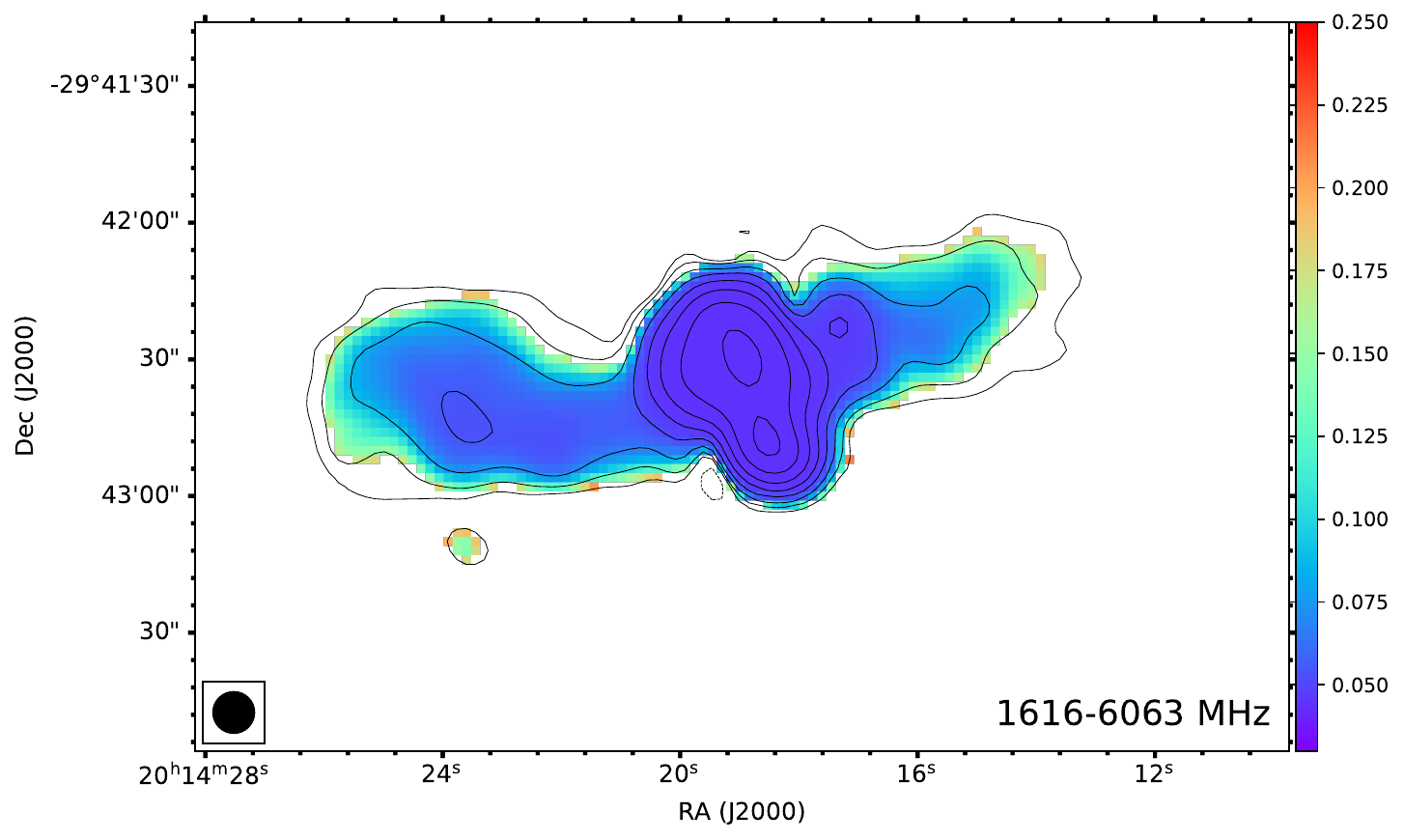}
\includegraphics[width=0.40\textwidth]{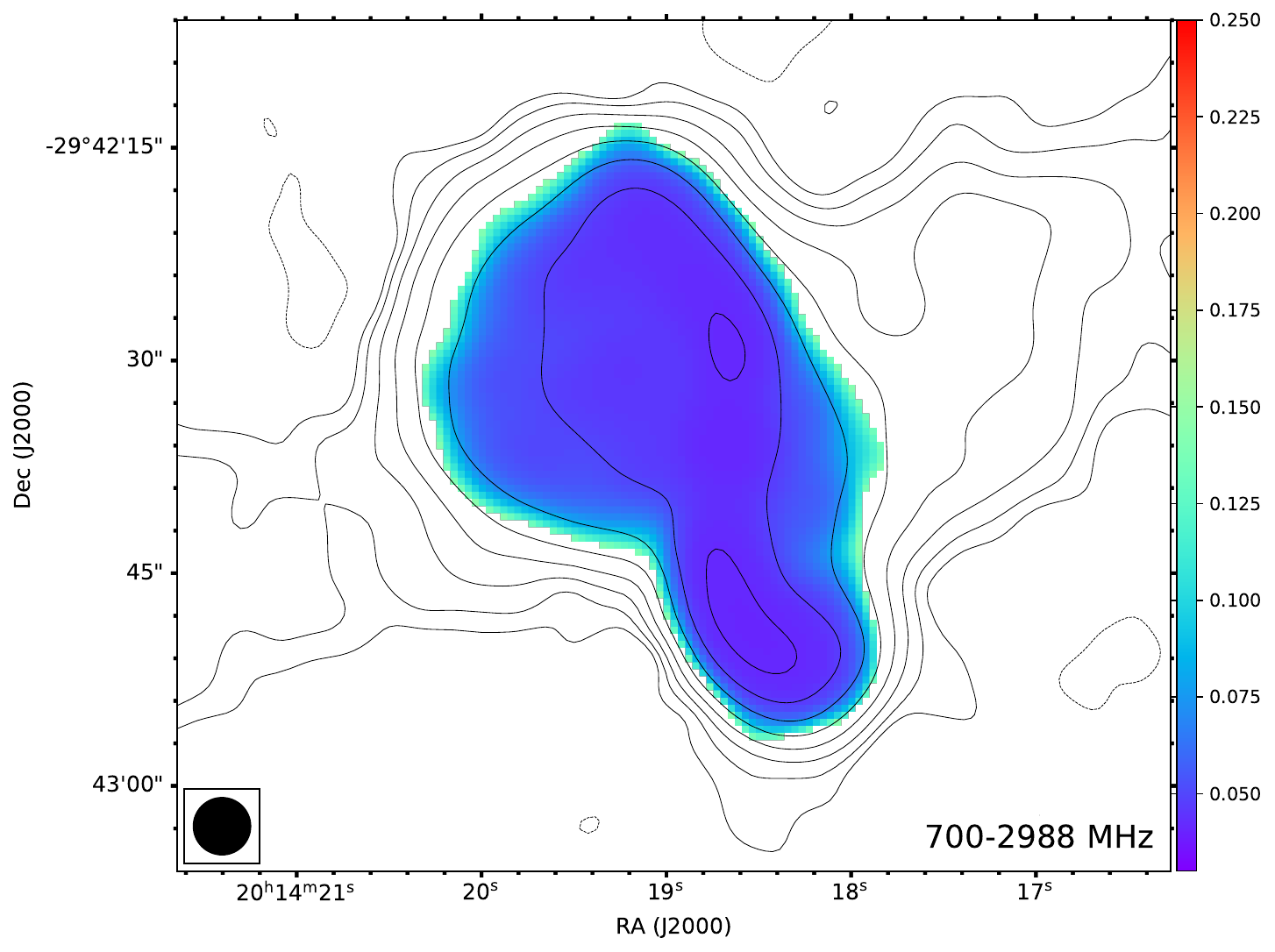}
\includegraphics[width=0.40\textwidth]{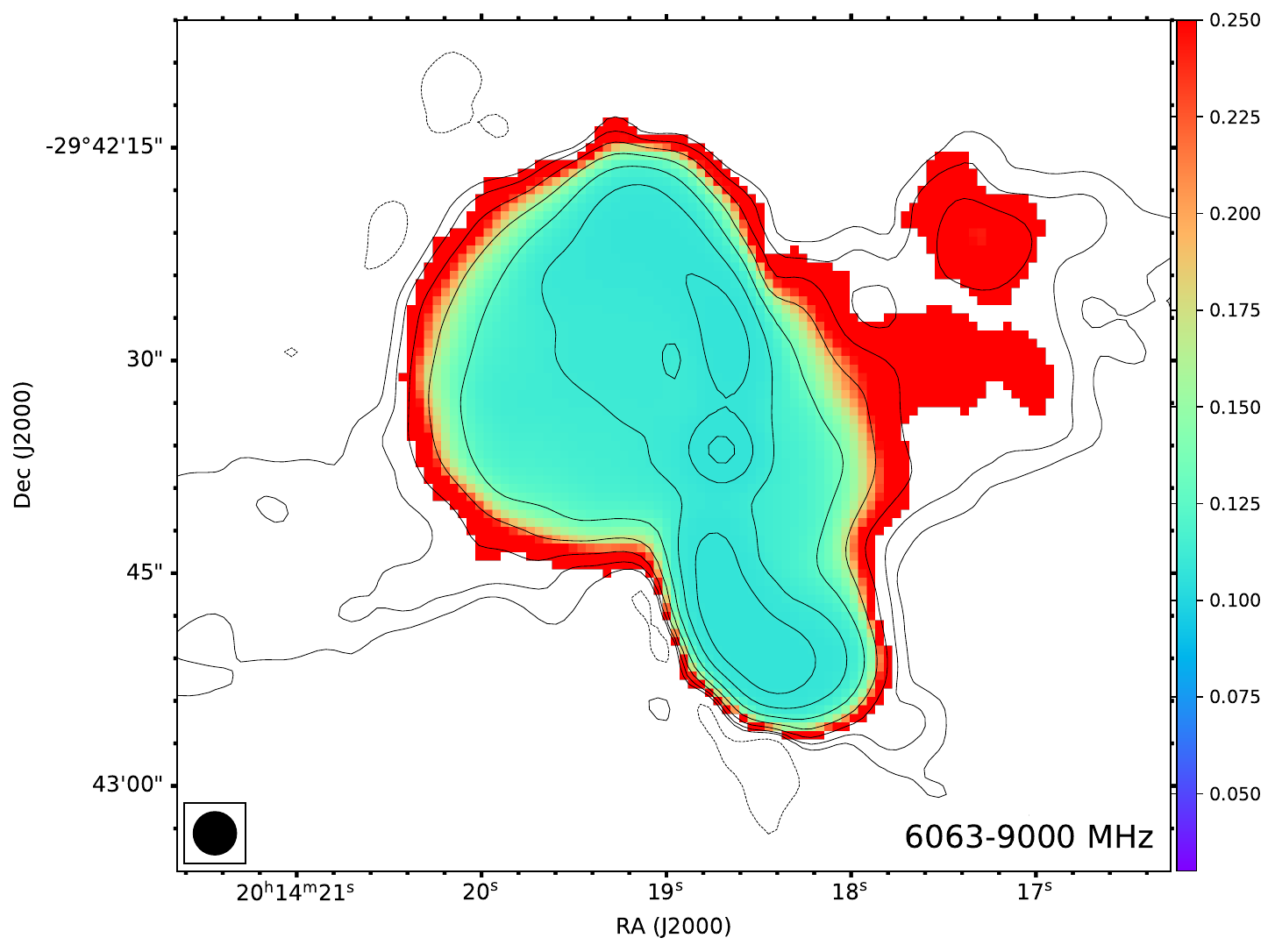}
	\caption{Spectral index error maps corresponding to the maps reported in Fig. \ref{fig: spixmap}.}
	\label{fig: errspixmap}
\end{figure*}

In Fig. \ref{fig: errspixmap} we report the error maps associated with the spectral index maps shown in Fig. \ref{fig: spixmap}. Errors are obtained as 
\begin{equation}
\Delta \alpha =  \left\lvert \frac{1}{\ln{ \left( \frac{\nu_{\rm 1}}{\nu_{\rm 2} } \right) }}\right\lvert \sqrt{ \left( \frac{\Delta S_{\rm 1}}{S_{\rm 1}}\right)^2 + \left( \frac{\Delta S_{\rm 2}}{S_{\rm 2}}\right)^2 } \; \; \; ,
\label{eq: spectralindexerrorformula}
\end{equation}
where $\Delta S$ are computed as in Eq. \ref{eq: erroronflux}.

\section{Precession modelling}
\label{sect: Precession modelling}

In this Section we summarise the theoretical framework describing the precession of relativistic jets, which is detailed in \cite{hjellming&johnston81}. We outline below the main equations describing the path of two plasma bubbles that are ejected in opposite directions from the core at a time $t_{\rm ej}$ with a velocity ${\rm v}$ (assumed to be equal to the jet advance speed), while the jets precess around the ejection axis. By evolving the system from $t_{\rm ej}$ to the fixed present time $t_{\rm obs}$ with a $t_{\rm ref}$ step, such equations predict the position of the bubbles through a series of snapshots, which can be then compared to the actual jet morphology observed in radio images. 

The angular velocity of the precessing jets is 
\begin{equation}
    \Omega = \frac{2 \pi s_{\rm rot} }{P} \; \; \; ,
    \label{eq: prec1}
\end{equation}
where $P$ is the precession period and $s_{\rm rot}$ is the direction of rotation ($s_{\rm rot}=1$ is anticlockwise, $s_{\rm rot}=-1$ is clockwise). Through a suitable rotation of the reference system (see Fig. 1 in \citealt{hjellming&johnston81}), the components of the velocity can be expressed in polar coordinates in a new Cartesian system having the $x$-axis aligned with the line of sight, yielding:  
\begin{equation}
\begin{cases}
     {\rm v}_{\rm x} = s_{\rm jet}  {\rm v} [\sin{\psi} \sin{i} \cos{(\Omega (t_{\rm ej}-t_{\rm ref}))} + \cos{\psi} \cos{i} ] \\ {\rm v}_{\rm y} = s_{\rm jet}  {\rm v} \sin{\psi}  \sin{(\Omega (t_{\rm ej}-t_{\rm ref}))} \; \; \; \; \; \; \; \; \; \; \; \; \; \; \; \; \; \; \; \; \; \; \; \; \; \; \; \; \; , \\ {\rm v}_{\rm z} = s_{\rm jet}  {\rm v} [\cos{\psi} \sin{i} - \cos{i} \sin{\psi} \cos{(\Omega (t_{\rm ej}-t_{\rm ref})]}
\end{cases}
\label{eq: prec2}
\end{equation}
where $s_{\rm jet}=-1$ for the redshifted (receding) jet and $s_{\rm jet}=+1$ for the blueshifted (approaching) jet, $\psi$ is the half-opening angle of the precession cone, and $i$ is the viewing angle between the source and the observer. By further rotating the Cartesian coordinate system by an angle $\chi$, which thus represents the position angle of the jets measured from the west axis, the velocity components projected into the sky (i.e. Right Ascension $\alpha$ and Declination $\delta$) system are obtained as:
\begin{equation}
    \begin{cases}
        {\rm v}_{\rm \alpha} = \sin{\chi} {\rm v}_{\rm y} + \cos{\chi} {\rm v}_{\rm z} \\ 
    {\rm v}_{\rm \delta} = \cos{\chi} {\rm v}_{\rm y} - \sin{\chi} {\rm v}_{\rm z}
    \end{cases}
    \label{eq: prec3}
\end{equation}
After a time $\Delta t = t_{\rm obs} - t_{\rm ej} $, the bubbles will be offset with respect to the position of the core. This offset is expressed in terms of proper motions as:
\begin{equation}
    \begin{cases}
         \mu_{\rm \alpha} = \frac{{\rm v}_{\rm \alpha}  (t_{\rm obs} - t_{\rm ej})}{ D_{L}\left(1-\frac{{\rm v}_{\rm x}}{c}\right)} \\
         \mu_{\rm \delta} = \frac{{\rm v}_{\rm \delta}  (t_{\rm obs} - t_{\rm ej})}{D_{L}\left(1-\frac{{\rm v}_{\rm x}}{c}\right)}
    \end{cases}
    \label{eq: prec4}
\end{equation}
where $c$ is the light speed and $D_{L}$ is the luminosity distance to the source. 

Following the optimisation strategy detailed in \cite{coriat19} and \cite{ubertosi24}, we used Eqs. \ref{eq: prec1}-\ref{eq: prec4} to minimise the difference between predicted and observed proper motions, and thus constrain the set of parameters that better reproduce the data. To this aim, we need to minimise the following (`cost') function 
\begin{equation}
\mathcal{C} = \Upsilon +  \sum_{j}^{} \left[ \left( \frac{\hat{\mu}_{\rm \alpha, j} - \mu_{\rm \alpha, j}}{\hat{\sigma}_{\rm \alpha, j}} \right)^2 + \left( \frac{\hat{\mu}_{\rm \delta, j} - \mu_{\rm \delta, j}}{\hat{\sigma}_{\rm \delta, j}}  \right)^2 \right] \; \; \; ,
\label{eq: cost}
\end{equation}
where the coordinates $(\mu_{\rm \alpha, j} ; \mu_{\rm \delta, j})$ and $(\hat{\mu}_{\rm \alpha, j} ; \hat{\mu}_{\rm \delta, j})$ are the predicted (from Eq. \ref{eq: prec4}) and measured (from radio images) proper motions, respectively. Each difference is weighted by the radio beam area, with $(\hat{\sigma}_{\rm \delta, j};  \hat{\sigma}_{\rm \alpha, j})$ being the beam semi-axes. The term $\Upsilon=\Upsilon({\rm v}, i, \alpha)$ is the penalty function defined in \cite{coriat19} to exclude combinations of ${\rm v}$ and $i$ yielding unrealistic flux density ratios of the approaching and receding jets. 

The minimisation of $\mathcal{C}$ was performed through the {\tt differential\_evolution} \citep{storn&price97} function within the {\tt scipy.optimize}\footnote{\url{https://docs.scipy.org/doc/scipy/tutorial/optimize.html}} package. The required model parameters are $s_{\rm rot}$, $P$, $s_{\rm jet}$, ${\rm v}$, $\psi$, $i$, $\chi$. We fixed $s_{\rm jet}=1$ and $s_{\rm jet}=-1$ for the southern and northern jet, respectively. By testing both $s_{\rm rot}=\pm 1$ we constrained the rotation direction to be clockwise ($s_{\rm rot}= -1$) for MRC 2011-298.  As we were not able to optimise the viewing angle and the jet position angle (meaning that the cost is highly sensitive to small variations of these parameters and they cannot be properly constrained), we tested reasonable sets of values and finally fixed them to $i=80$ deg and $\chi=260$ deg, which provide slightly lower values of $\mathcal{C}$. The remaining parameters ($P$, $\psi$, ${\rm v}$) were derived by optimisation of Eq. \ref{eq: cost}, and we obtained estimates of statistical and systematical uncertainties as in \cite{ubertosi24}. Following a bootstrap approach, we repeated the optimisation 100 times, and considered the standard deviation of the parameter distributions as the statistical error. By running the optimisation for S-band, C-band, and X-band images, we computed the systematical error as the standard deviation of the optimised mean values at each frequency. 

The results of our procedure are reported in Table \ref{tab: parametri precessione}. We found negligible statistical errors ($< 5\%$), which indicate that the optimisation algorithm is stable and all parameters are well determined at each frequency. The systematical errors are higher, but still relatively small ($\lesssim 20\%$ at most), highlighting the consistency of the results at different frequencies. We stress that varying $i$ and/or $\chi$ by $\sim 10$ deg does not meaningfully affect the fitted parameters.

\end{appendix}

\end{document}